\documentclass[twocolumn]{aastex63}

\usepackage{enumerate}
\usepackage{graphicx}
\usepackage{longtable,booktabs}
\usepackage{multirow}
\usepackage{appendix}
\usepackage{float}
\usepackage{amssymb}
\usepackage{lineno}
\usepackage{threeparttable}
\usepackage{soul}

\submitjournal{APJ}

\accepted{Nov. 1, 2022}

\shorttitle{The H-poor Superluminous Supernovae from ZTF-I}
\shortauthors{Chen et al.}
\watermark{Draft}
\graphicspath{{./}{figures/}}
\begin{document}

\title{The Hydrogen-Poor Superluminous Supernovae from the Zwicky Transient Facility Phase-I Survey: I. Light Curves and Measurements}

\author[0000-0001-5175-4652]{Z.~H.~Chen}
\affil{Physics Department and Tsinghua Center for Astrophysics (THCA), Tsinghua University, Beijing, 100084, China}

\author[0000-0003-1710-9339]{Lin~Yan}
\affil{Caltech Optical Observatories,
California Institute of Technology,
Pasadena, CA 91125, USA}

\author[0000-0002-5477-0217]{T.~Kangas}
\affil{KTH Royal Institute of Technology, Stockholm, Sweden}

\author{R.~Lunnan}
\affil{The Oskar Klein Centre, Department of Astronomy, Stockholm University, AlbaNova, SE-106 91 Stockholm, Sweden}

\author{S.~Schulze}
\affil{The Oskar Klein Centre, Department of Physics, Stockholm University, AlbaNova, SE-106 91 Stockholm, Sweden}

\author[0000-0003-1546-6615]{J.~Sollerman}
\affil{The Oskar Klein Centre, Department of Astronomy, Stockholm University, AlbaNova, SE-106 91 Stockholm, Sweden}

\author{D.~A.~Perley}
\affil{Astrophysics Research Institute, Liverpool John Moores University, 146 Brownlow Hill, Liverpool L3 5RF, UK}

\author[0000-0002-1066-6098]{T.-W.~Chen}
\affil{The Oskar Klein Centre, Department of Astronomy, Stockholm University, AlbaNova, SE-106 91 Stockholm, Sweden}

\author{K.~Taggart}
\affil{Department of Astronomy and Astrophysics, University of California, Santa Cruz, CA 95064, USA}

\author{K.~R.~Hinds}
\affiliation{Astrophysics Research Institute, Liverpool John Moores University, 146 Brownlow Hill, Liverpool L3 5RF, UK}

\author{A.~Gal-Yam}
\affil{Department of particle physics and astrophysics, Weizmann Institute of Science, 76100 Rehovot, Israel}

\author[0000-0002-7334-2357]{X.~F.~Wang}
\affil{Physics Department and Tsinghua Center for Astrophysics (THCA), Tsinghua University, Beijing, 100084, China}
\affil{Beijing Planetarium, Beijing Academy of Sciences and Technology, Beijing, 100044, China}

\author{I.~Andreoni}
\affil{Cahill Center for Astrophysics, California Institute of Technology, 1200 E. California Blvd. Pasadena, CA 91125, USA}
\author[0000-0001-8018-5348]{E.~Bellm}
\affiliation{DIRAC Institute, Department of Astronomy, University of Washington, 3910 15th Avenue NE, Seattle, WA 98195, USA}
\author[0000-0002-7777-216X]{J.~S.~Bloom}
\affiliation{Department of Astrophysics, University of California, Berkeley, CA 94720-3411, USA}
\affiliation{Lawrence Berkeley National Laboratory, 1 Cyclotron Road, MS 50B-4206, Berkeley, CA 94720, USA}

\author{K.~Burdge}
\affil{Division of Physics, Mathematics, and Astronomy, California Institute of Technology, Pasadena, CA 91125, USA}
\author{A.~Burgos}
\affil{Instituto de Astrofísica de Canarias, E-38 200 La Laguna, Tenerife, Spain}
\author{D.~Cook}
\affil{IPAC, California Institute of Technology, 1200 E. California Blvd, Pasadena, CA 91125, USA}
\author{A.~Dahiwale}
\affil{Division of Physics, Mathematics, and Astronomy, California Institute of Technology, Pasadena, CA 91125, USA}
\author{K.~De}
\affil{Cahill Center for Astrophysics, California Institute of Technology, 1200 E. California Blvd. Pasadena, CA 91125, USA}
\author[0000-0002-5884-7867]{R.~Dekany}
\affil{Caltech Optical Observatories, California Institute of Technology, Pasadena, CA 91125, USA}
\author{A.~Dugas}
\affil{Department of Physics and Astronomy, Watanabe 416, 2505 Correa Road, Honolulu, HI 96822, USA}
\author{S.~Frederik}
\affil{Department of Astronomy, University of Maryland, MD 20742-2421, USA}
\author{C.~Fremling}
\affil{Division of Physics, Mathematics, and Astronomy, California Institute of Technology, Pasadena, CA 91125, USA}
\author{M.~Graham}
\affil{Cahill Center for Astrophysics, California Institute of Technology, 1200 E. California Blvd. Pasadena, CA 91125, USA}

\author{M.~Hankins}
\affil{Arkansas Tech University, Russellville, AR 72801, USA}
\author{A.~Ho}
\affil{Department of Astronomy and Miller Institute for Basic Research in Science, University of California, Berkeley, CA, USA}
\author{J.~Jencson}
\affil{Steward Observatory, University of Arizona, 933 North Cherry Avenue, Tucson, AZ 85721-0065, USA}
\author{V.~Karambelkar}
\affil{Cahill Center for Astrophysics, California Institute of Technology, 1200 E. California Blvd. Pasadena, CA 91125, USA}
\author{M.~Kasliwal}
\affil{Division of Physics, Mathematics, and Astronomy, California Institute of Technology, Pasadena, CA 91125, USA}
\author{S.~Kulkarni}
\affil{Division of Physics, Mathematics, and Astronomy, California Institute of Technology, Pasadena, CA 91125, USA}
\author[0000-0003-2451-5482]{R.~Laher}
\affiliation{IPAC, California Institute of Technology, 1200 E. California Blvd, Pasadena, CA 91125, USA}
\author[0000-0001-7648-4142]{B.~Rusholme}
\affiliation{IPAC, California Institute of Technology, 1200 E. California Blvd, Pasadena, CA 91125, USA}
\author{Y.~Sharma}
\affil{Division of Physics, Mathematics, and Astronomy, California Institute of Technology, Pasadena, CA 91125, USA}
\author{F.~Taddia}
\affil{Department of Physics and Astronomy, Aarhus University, Ny Munkegade 120, DK-8000 Aarhus C, Denmark}
\author{L.~Tartaglia}
\affil{INAF-Osservatorio Astronomico di Padova, Vicolo dell’Osservatorio 5, I-35122 Padova, Italy}
\author{B.~P.~Thomas}
\affil{Department of Astronomy, University of Texas at Austin, 2515 Speedway, Stop C1400 Austin, TX 78712-1205, USA}
\author{A.~Tzanidakis}
\affil{Division of Physics, Mathematics, and Astronomy, California Institute of Technology, Pasadena, CA 91125, USA}
\author{J.~Van~Roestel}
\affil{Division of Physics, Mathematics, and Astronomy, California Institute of Technology, Pasadena, CA 91125, USA}
\author{R.~Walter}
\affil{Cahill Center for Astrophysics, California Institute of Technology, 1200 E. California Blvd. Pasadena, CA 91125, USA}
\author{Y.~Yang}
\affil{Department of Astronomy, University of California, Berkeley, CA 94720-3411, USA}
\author{Y.~H.~Yao}
\affil{Cahill Center for Astrophysics, California Institute of Technology, 1200 E. California Blvd. Pasadena, CA 91125, USA}
\author{O.~Yaron}
\affil{Department of particle physics and astrophysics, Weizmann Institute of Science, 76100 Rehovot, Israel}





\correspondingauthor{Z.~H.~Chen, Lin~Yan}
\email{chenzh18@mails.tsinghua.edu.cn, lyan@caltech.edu}

\begin{abstract}
During the Zwicky Transient Facility (ZTF) Phase-I operation, 78 hydrogen-poor superluminous supernovae (SLSNe-I) were discovered in less than three years, making up the largest sample from a single survey. This paper (Paper I) presents the data, including the optical/ultraviolet light curves and classification spectra, while Paper II in this series will focus on the detailed analysis of the light curves and modeling. Our photometry is primarily taken by the ZTF in the $g,r,i$ bands, and with additional data from other ground-based facilities and {\it Swift}. The events of our sample cover a redshift range of $z = 0.06 - 0.67$, with a median and $1\sigma$ error (16\% and 84\% percentiles) $z_{\rm med} = 0.265^{+0.143}_{-0.135}$. The peak luminosity covers $-22.8\,{\rm mag} \leq M_{g,\rm peak} \leq -19.8$\,mag, with a median value of $-21.48^{+1.13}_{-0.61}$\,mag. Their light curves evolve slowly with the mean rest-frame rise time of $t_{\rm rise} = 41.9\pm17.8$\,days. The luminosity and time scale distributions suggest that low luminosity SLSNe-I with peak luminosity $\sim -20$\,mag or extremely fast rising events ($<10$\,days) exist but are rare. We confirm previous findings that slowly rising SLSNe-I also tend to fade slowly. The rest-frame color and temperature evolution show large scatters, suggesting that the SLSN-I population may have diverse spectral energy distributions. The peak rest-frame color shows a moderate correlation with the peak absolute magnitude, {\it i.e.} brighter SLSNe-I tend to have bluer colors. With optical and ultraviolet photometry, we construct bolometric luminosity and derive a bolometric correction relation generally applicable for converting $g,r$-band photometry to bolometric luminosity for SLSNe-I. 

\end{abstract}


\section{Introduction} \label{sec:intro}
Superluminous supernovae (SLSNe) are a rare class of stellar explosions first discovered over fifteen years ago \citep[{\it i.e.} SN\,2005ap,][]{Quimby2007}. Their peak luminosities ($10^{43 - 44}\,{\rm erg\, s^{-1}}$) are 10 -- 100 times higher than those of normal Type Ia and core-collapse supernovae (SNe). Their light curves (LCs) usually evolve rather slowly, with rise times of $\sim 20 - 100$\,days. The combination of these two features can not be explained by conventional SN models, {\it i.e.} standard radioactive decay \citep{Kasen2017}. With the discovery of the first several SLSNe (SN\,2005ap, SN\,2006gy, SN\,2007bi \&\ SN\,2008es), it was quickly recognized that, like normal SNe, SLSNe can be divided into two spectroscopic subclasses, one with hydrogen emission lines \citep[SLSNe-II,][]{Miller2009,Gezari2009,Inserra_2018b} and the other without hydrogen \citep[SLSNe-I,][]{Quimby2007, Ofek2007, Gal-Yam2009, Smith2007, Gal-Yam2012}. In recent years, the subclass of H-poor but Helium-rich SLSNe (SLSNe-Ib) were first identified by \citet{Quimby_2018}, and later a sample by \citep{Yan2020}.

Three popular models have been proposed to explain the extraordinary radiative power of SLSNe. One involves energy injection from a central engine, such as the spin-down of a fast-rotating neutron star \citep[magnetar,][]{Kasen2010, Woosley2010}. Alternatively, the interaction between the SN ejecta and dense circumstellar material (CSM) can efficiently convert kinetic energy into radiation \citep{Chevalier2011,Chatzopoulos2013}. Finally, some SLSNe could be powered by massive amounts of $^{56}$Ni synthesised in a pair-instability supernova explosion (PISN) of low-metallicity stars with masses $> 140 \, M_{\odot}$ \citep{Woosley2007,Kasen2011}.  It is commonly accepted that most SLSNe-II are analogous to Type IIn SNe \citep{Schlegal1990, Filippenko1997}, primarily powered by ejecta interactions with dense CSM \citep{Ofek2007, Miller2010, Chevalier2011}, while a small fraction show broad H$\alpha$ features with no signs of strong interaction in spectra, {\it e.g.} SN\,2008es, SN\,2013hx, PS15br \citep{Miller2009,Gezari2009,Inserra_2018b}, but interactions are still likely required in these events \citep{Kangas2022}.

Between 2005 and 2009, a handful of SLSNe were discovered by several untargeted transient surveys which were not specifically targeting bright nearby galaxies. This small number of luminous events sparked a flurry of studies in both theory and observation of SLSNe. The next big advance in this field came between 2009 and 2016 when large area, untargeted transient surveys started operating. For example, the Palomar Transient Factory \citep[PTF,][]{Law2009}, the Pan-STARRS1 Medium Deep Survey \citep[PS1 MDS,][]{Chambers_2016}, the Catalina Real-time Transient Survey \citep[CRTS,][]{Drake2009}, the All-Sky Automated Survey for SuperNovae \citep[ASAS-SN,][]{Shappee2014}, the Dark Energy Survey \citep[DES,][]{DES_2005} and the Gaia Photometric Science Alerts \citep[Gaia,][]{Hodgkin_2021} made major contributions to the discoveries of several dozen SLSNe at both low ($z\sim0.2$) and high redshift ($z\sim1$) \citep{Nicholl2015b, DeCia2018, Quimby_2018, Lunnan2018, Angus_2019}. However, with over $\sim90$ discovered SLSNe-I by the end of 2017, many questions regarding their physical nature still remained unclear. For example, the SLSN volumetric rates are poorly constrained, with only estimates from small SLSN-I samples \citep{Quimby2013,McCrum2015,Prajs2017,Frohmaier2021}. Attempts to examine the statistical distributions, such as luminosity functions, were also quite limited due to small number statistics. 

Assembling a large sample of low-$z$ SLSNe with a well-defined survey volume and cadence is one of the goals of the Zwicky Transient Facility \citep[ZTF,][]{Graham2019,Bellm2019,Masci2019}. ZTF utilizes a 600-megapixel camera mounted on the Palomar Samuel Oschin 48\,inch Schmidt telescope to reach a 47 deg$^2$ field of view \citep{Dekany2020}. ZTF can cover the full Northern Sky in 3 days down to a $5\sigma$ limiting magnitude of $20.5 - 20.8$\,magnitude, which is about 3.5 and 0.5 magnitudes deeper than that of ASAS-SN \citep{Shappee2014}, and the Asteroid Terrestrial-impact Last Alert System \citep[ATLAS,][]{Tonry2018}, respectively. The ZTF survey offers several advantages for discovering rare transient events such as SLSNe. It is an untargeted, all sky, and moderately high cadence survey, probing large volumes with its large area coverage and deep sensitivity limits. Its well-defined survey strategy -- area coverage and cadence -- also makes it possible to quantify the survey efficiency.

ZTF conducted several surveys with different cadences (ranging from minutes to days) and area coverage during the phase-I operation \citep{Bellm2019b}. Among them, a particularly important one for extragalactic transient studies is the Northern Sky Public Survey. ZTF covered roughly the entire northern sky accessible from Palomar, corresponding to a total sky area of $\sim23,675$\,deg$^2$. In every 3 days, each field was observed once in $g$ band and once in $r$ band, with an interval of at least 30 minutes between observations.

Between March 17, 2018 and October 31, 2020, ZTF Phase-I \footnote{The ZTF public survey transitioned to a 2-day cadence on October 1, 2020, however the official start of the ZTF Phase II is December 1, 2020. We picked our date range for the convenience of the sources analysed in our sample.} discovered and spectroscopically confirmed 85 SLSNe-I, 6 SLSNe-I.5 (classified as SLSNe-I but showing H lines after the peak) and 61 SLSNe-II (defined as SNe\,II with peak magnitudes brighter than $-20.0$\,mag). 
The number of SLSNe discovered by ZTF from 2018 to 2020 (about 60 per year) is roughly 5 -- 7 times higher than that detected in any previous year. 
The SLSN-I sample will be the focus of a series of three papers. Paper I (this paper) presents the observational data and analysis of the overall observational properties. Paper II \citep{Chen_2022} discusses the LC modeling and analysis of the LC morphology. Paper III (Yan et al. in preparation) will present the derived SLSN-I volumetric rates and luminosity functions at $z \leq 0.7$. Several additional papers based on some individual SLSNe discovered during the ZTF Phase-I operation have been recently published. \citet{Lunnan2020} showcased the first four SLSNe-I discovered by ZTF during its science commissioning phase. \citet{Yan2020} presented the discovery of six He-rich SLSNe-I (SLSNe-Ib), revealing additional constraints on the progenitor mass-loss history. 

This paper is organized as follows. Section~\ref{sec:sample} introduces the selection and classification of this sample. Section~\ref{sec:data} presents the photometry from ZTF and other facilities.  Section~\ref{sec:results} discusses our methodology with various photometric corrections and the calculations of peak absolute magnitudes.  The measurements of time scales, colors, black-body temperatures and bolometric luminosities are presented in Section~\ref{sec:measure}. Section~\ref{sec:summary} summarizes the conclusions. Throughout the paper, all magnitudes are in the AB system unless explicitly noted otherwise. 
We adopt a $\Lambda$CDM cosmology with H$_0 = 70.0$\,km\,s$^{-1}$\,Mpc$^{-1}$, $\Omega_M = 0.3$ and $\Omega_{\Lambda} = 0.7$. 

\section{The SLSN-I sample from ZTF-I} 
\label{sec:sample} 
During the phase-I survey, ZTF discovered 85 SLSNe-I. This paper focuses on 78 of these 85 events whose LCs have turned over from the peak by October 31, 2020, enabling better LC modeling. Of these 78 SLSNe-I, seven can be classified as He-rich SLSNe-Ib, including six published by \citet{Yan2020} and one by \citet{Terreran2020}. For completeness, this sample paper also includes the four sources published in \citet{Lunnan2020}. In addition, Schulze et al. (in preparation) will focus on an extremely slow and peculiar SLSN-I, ZTF18acenqto (SN\,2018ibb), and provided the derived parameters to include in our catalog.

\begin{figure*}[htp]
\centering
\includegraphics[width=\textwidth]{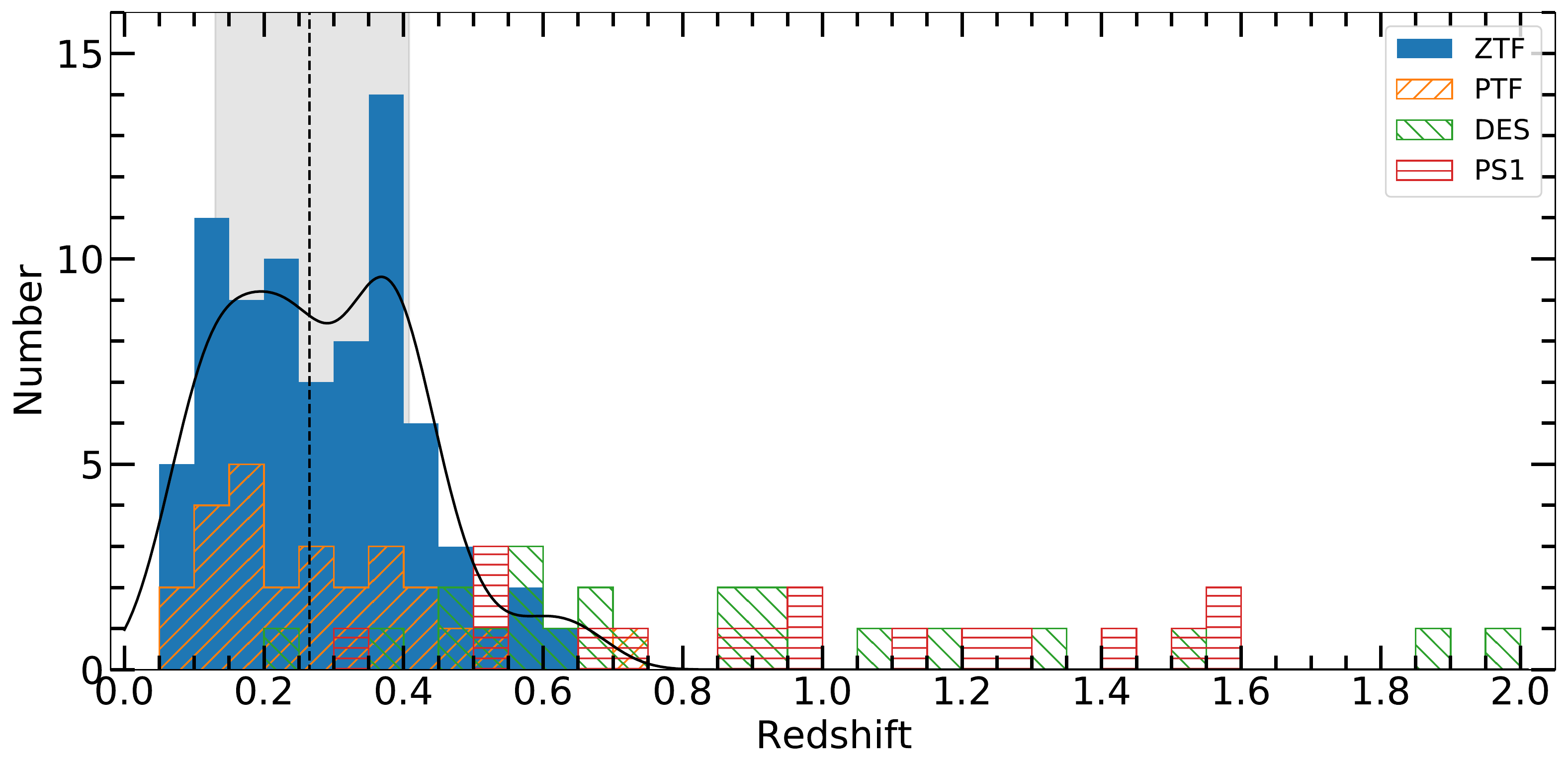}
\caption{Distribution of redshifts for the sample of 78 SLSNe-I presented in this paper.  Other SLSN-I samples are plotted as hollow bars for comparison. The dashed line and the shaded area mark the median value and $1\sigma$ error (16\% and 84\% percentiles) of ZTF
sample, $z_{\rm med} = 0.265^{+0.143}_{-0.135}$. The black solid line shows the kernel density estimation of the distribution.}
\label{fig:redshift}
\end{figure*}

Table~\ref{tab:basicinfo} compiles the metadata for each of the targets, including the internal ZTF name, IAU name, right ascension (RA), declination (DEC), redshift, Galactic extinction $E(B-V)$, discovery group, and additional information on spectral classification. Our sample covers the redshift of $z \sim 0.06 - 0.67$ with a median of $z_{\rm med} = 0.265^{+0.143}_{-0.135}$. All redshifts in our sample are determined using the narrow emission lines from the host galaxy, except for nine events without host lines. The redshifts for these nine events are estimated from template matching by running \textit{superfit} \citep{Howell2006} over a range of $z$ values. These redshifts are less accurate and marked with $\star$ in Table~\ref{tab:basicinfo}. There are two additional events, ZTF19abcvwrz (SN\,2019aamx) and ZTF19aawsqsc (SN\,2019hno), which also have less accurate redshifts because of the low signal-to-noise ratios of Mg II $\lambda\lambda\,2796,2803$ absorption lines in their host-galaxy spectra. Figure~\ref{fig:redshift} displays the redshift distribution of the full sample, including those derived from template fitting. To avoid possible biases caused by the choice of histogram grids, we apply kernel density estimation on all the histograms in this paper using a Gaussian kernel offered by the machine learning package Scikit-learn \citep{scikit-learn}, as shown in Figure~\ref{fig:redshift}.

Several SLSN-I samples from different surveys, including PS1, DES and PTF, as well as samples collected from the literature by \citet{Nicholl2015b} and \citet{Inserra_2018a}, have revealed many important properties of SLSNe-I. Compared with these previous samples, our sample size is significantly larger and the observing cadence is also better, as shown in Table~\ref{tab:previous}. These two key features allow us to investigate the LC properties of SLSNe with much better statistics.

\begin{center}
\begin{longtable*}{lcccl}
\caption{SLSN samples}\\
\toprule

\multirow{2}{*}{Source} & \multirow{2}{*}{Candidates} & \multirow{2}{*}{Redshift range} & Observing cadence$^{\rm a}$ & \multirow{2}{*}{Reference} \\
& & & (rest-frame days) & \\

\midrule
Literature & 25$^{\rm b}$(14) & 0.10 - 1.19 & - & \citet{Nicholl2015b,Inserra_2018a} \\ 
PS1 & 17 & 0.32 - 1.57 & 2.30 & \citet{Lunnan2018} \\ 
DES & 22 & 0.22 - 2.00 & 3.65 & \citet{Angus_2019} \\ 
PTF & 26 & 0.06 - 0.74 & 2.26 & \citet{DeCia2018} \\ 
ZTF & 78 & 0.06 - 0.67 & 1.45 & This paper \\
\bottomrule

\hspace*{\fill} \\
\multicolumn{5}{l}{$^{\rm a}$The median value of the observing cadence in the rest frame.}\\
\multicolumn{5}{l}{$^{\rm b}$Including 11 SLSNe-I from PTF and PS1 samples and 14 independent SLSNe-I.}\\
\label{tab:previous}
\end{longtable*}
\end{center}

\subsection{Photometric Selection} \label{sec:filter}
Here we briefly describe the photometric selection system of our SLSN-I candidates. Daily ZTF alerts are ingested into a dynamic science portal called the GROWTH Marshal \citep{Kasliwal2019}. A filter is implemented within the Marshal to select SLSN candidates.  The candidates passing the filter do not get automatically saved; instead, each week, a human scanner has to visually examine the LC of each candidate selected by the filter and make a decision if it is worth being saved. The subsequent spectral follow up is based on the human-saved candidates. This filter adopts several cutoff conditions, including: {\bf [1]} Not moving; the same alert is detected in two consecutive epochs with $\Delta t > 0.02$\,days. {\bf [2]} Not a star, based on the SDSS star-galaxy scores \citep{Tachibana2018}. {\bf [3]} Excluding bogus alerts based on scores constructed by \citet{Duev2019}. {\bf [4]} Not in the galactic plane, with Galactic latitude $\rm |b| > 7^\circ$. {\bf [5]} Variability was not detected at this location more than a year prior to the alert.

In addition, we assign numerical scores to several properties, including: {\bf[1]} slowly rising events; {\bf[2]} faint blue hosts; {\bf[3]} spatial location of the transient relative to the center of the host (against nuclear transient); {\bf[4]} brightness of the candidate relative to the host brightness (against faint transients); {\bf[5]} the time interval between the first detection and the last one (against very long-lived transients like active galactic nucleus). This candidate filter gives high scores (thus preferences) to slowly rising events with faint blue hosts. For example, a candidate rising at least 20 - 25 days with a faint host galaxy will have a high score and pass the filter and be also saved by the human scanner.

We note that the ZTF collaboration has multiple transient groups which also perform daily alert stream scanning, and most transient candidates can be saved by multiple groups, for example, the ZTF Bright Transient Survey \citep[BTS,][]{Fremling2020}, the ZTF Census of the Local Universe \citep[CLU,][]{De2020}, the fast transient group, the stripped envelop SN group and the ZTF Nuclear Transient group. Therefore, the classification of a specific SLSN-I candidate can also be drawn from other classification efforts, most noticeably the BTS. A few classifications are taken from the Transient Name Server (TNS) reported by external groups. But almost all of these externally-classified sources were also identified as candidates by our filter. The filtering method will be described in detail in the forthcoming work. 

With the above selection criteria, on the order of 50 candidates per week are saved and get another round of vetting when spectral classification observations are planned. Any candidates brighter than 18.5 mag are classified by the ZTF BTS. Their classification efficiency at $\leq18.5$ mag is very high, close to $95\%$ \citep{Fremling2020, Perley2020}. 
We anticipate that our catalog is almost entirely complete to SLSNe within the ZTF footprint peaking at magnitudes brighter than 18.5 mag, and highly complete to 19.0 mag.  For magnitudes fainter than 19.0 (44 out of 78 events in total), there may be biases primarily related to the nature of the host and/or the rising phase of the light curve that will be examined in future work. The detailed structure of the light curve, {\it e.g.} the presence of bumps, was not a crucial factor in selection or triggering follow up.

Our effort for spectroscopic classification is primarily focused on SLSN candidates fainter than 18.5 mag, using the Spectral Energy Distribution Machine \citep[SEDM,][]{Blagorodnova2018} and the Double Beam Spectrograph \citep[DBSP,][]{Oke1983} mounted on the Palomar 60\,inch (P60) and 200\,inch (P200) telescope, respectively. Additional facilities include the Low Resolution Imaging Spectrometer \citep[LRIS,][]{Oke1995} on the Keck I telescope, the Alhambra Faint Object Spectrograph and Camera (ALFOSC) on the 2.56m Nordic Optical Telescope (NOT), SPectrograph for the Rapid Acquisition of Transients (SPRAT) on the 2m Liverpool Telescope (LT), and the Intermediate-dispersion Spectrograph and Imaging System (ISIS) on the 4.2~m William Herschel Telescope (WHT). The basic information of classification spectra is listed in Table~\ref{tab:spec}. 
The spectral reduction is performed using various standard reduction pipelines. This includes the 
SEDM automated pipeline \citep{Rigault2019}, the pyraf-dbsp package \citep{Bellm2016} and DBSP\_DRP \citep{Roberson2022} pipelines for the DBSP data, and the LPipe package \citep{Perley_2019} for the LRIS data.

Finally, we note that the SLSN project has at least 0.5 to 1 night of DBSP time on P200 per month for spectral classification (PI: Yan). More quantitative discussions on the completeness of spectral classification will be included in Paper III. 

\subsection{Spectral Classification} \label{subsec:identify}

Every event in our sample has at least one spectrum and most have multi-epoch spectra. The hallmark spectral features for SLSNe-I are the five O\,II absorption features in the wavelength range of $3737 - 4650$~\AA~ in pre-peak and/or near-peak optical spectra. These were identified and discussed in \citet{Quimby2011,Quimby_2018}. We utilize the large SLSN-I spectral template library assembled by \citet{Quimby_2018}, and update the library by adding the missing phase information to some of the templates. Our classification relies on matching with the spectral templates using SNID \citep{Blondin2007} and \textit{superfit}.

To determine the best matched spectral templates, we run both \textit{superfit} and SNID on the smoothed spectra with host-galaxy emission lines removed and with redshifts fixed for most sources. If the spectrum of a SLSN-I has a good match with that of a SN\,Ic but its $g$-band peak luminosity is higher than $-20.0$\,mag, we classify this candidate as a SLSN-I since many SLSNe-I develop spectra similar to those of SNe\,Ic after the peak \citep{Pastorello_2010,Quimby_2018,Gal-Yam_2019}. 

In Appendix A, Figures~\ref{fig:classification_0} -- \ref{fig:classification_2} present the best-fit spectral template for each event of our sample, with the information of event name, phase and template information labeled after the spectra. The phases are measured relative to the rest-frame $g$-band peak in this paper. We note that the phase differences between the observed spectra of our SLSNe-I and the templates in the library are not zero, but generally within 30\,days (rest frame). This is not surprising since SLSNe-I can have similar photospheric spectra but different LC evolution time scales \citep[see][]{Kangas_2017,Quimby_2018}. 

The above classification procedure can sometimes give ambiguous results for a small number of events, {\it i.e.} two events for our sample. In these cases, we rely on additional LC information, such as rise time and peak luminosity to break the degeneracy. For example, ZTF19aacxrab (SN\,2019J) and ZTF19aaqrime (SN\,2019kwt) are almost equally well-matched with the spectral templates of SN\,Ia and SLSN-I. In the case of SN\,2019J, although the overall spectral features broadly match with those of SNe\,Ia, its spectrum lacks \ion{S}{2}$\,\lambda\lambda\,5433,5606$ commonly seen in SNe\,Ia. In addition, because of their slow rising LCs and high peak luminosities ({\it i.e.} SN\,2019kwt has $M_g \sim -22.8$\,mag), we adopt the SLSN-I classifications, corresponding to the template spectra of SN\,2007bi and PTF09cnd, respectively. 

In summary, the 78 sources listed in Table~\ref{tab:basicinfo} can be classified as SLSNe-I according to their features of spectra and light curves. The classification spectra are made available to the public as part of electronic data at the Journal website and will be uploaded to the Weizmann Interactive Supernova Data Repository \citep[WISeREP,][]{Yaron2012}\footnote{\url{https://www.wiserep.org/}}.

\section{Observations and Data} \label{sec:data}

\subsection{ZTF data and forced photometry} \label{sec:forcedphot}

The bulk of the photometric data comes from ZTF, including the data from the public survey with a 3-day cadence, the ZTF partnership and Caltech surveys with a faster cadence ($\leq 2$\,days) over smaller areas \citep{Bellm2019b}. 

The IPAC ZTF pipeline produces reference-subtracted images using the ZOGY algorithm \citep{Zackay2016} and aperture photometry for all transients detected at $\geq 5\sigma$. However, there are two issues with the LCs produced by this pipeline. One is that the reference images built in 2018 may contain signals from transients that exploded during the same period, for which the photometric offset needs to be corrected. The second issue is that the upper limits prior to the first detection are aperture photometry and based on the image noise measured over the entire quadrant. This can significantly underestimate transient signals.
 
To fix these two issues, we perform forced PSF photometry using the software provided by IPAC\footnote{ \url{http://web.ipac.caltech.edu/staff/fmasci/ztf/forcedphot.pdf}}.  With a code from \citet{Yuhan2019}, we refine the astrometric position of each event by using only the images around the peak phase. The very early and late-time forced photometry without transient signals allow us to compute baseline offsets to the LCs. In addition, we reject bad-quality data 
if: {\bf [1]} the image processing and instrumental calibration fail to meet predefined quality criteria;
{\bf [2]} the robust estimate of $1\sigma$ value of the spatial noise per pixel in the image is over 25;
{\bf [3]} the seeing is larger than $5^{''}$, similar to what was used in \citet{Yuhan2019}. The photometry of the same transient observed with different CCD quadrants can have a systematic offset. This problem gets fixed by our photometric reprocessing as well. For the final photometric collections, a detection is defined as $4\sigma$ ($3\sigma$ for {\it Swift} data) above the background and an upper limit is computed at $3\sigma$. 

The ZTF astrometric and photometric systems are calibrated using the {\it Gaia} DR1 data and the Pan-STARRS 1 catalogs, respectively \citep[for details see][]{Masci2019}. The output magnitudes are in the AB system. The airmass and color term corrections are included when the photometry is calibrated to the PS1 system. The color information for color corrections ($g-r$ for $g$, $r$ bands and $r-i$ for $i$ band) is obtained from the $g$, $r$, $i$ photometry taken at the same epoch or neighboring observations. 

\subsection{Supplemental photometry} \label{sec:swift}

\subsubsection{P60, LT and P200 photometry}

When ZTF is not able to take observations due to scheduling conflicts, bad weather, or when the transient is fainter than the ZTF detection limits, additional data are taken with SEDM on the P60 \citep{Blagorodnova2018}, the Optical imager of the Infrared-Optical suite of instruments (IO:O) on the LT \citep{Steele04} or the Wafer-Scale camera for Prime (WASP)\footnote{\url{https://sites.astro.caltech.edu/palomar/observer/200inchResources/waspmanual.html}} on the P200. All of the P60, LT and P200 photometric results are obtained via PSF fitting, which are calibrated with the PS1 and SDSS (for $u$-band and part of P60 data only) standard stars. For the P60 data, the systematic errors between PS1 and SDSS systems may introduce an additional uncertainty of 0.01 mag in r and i bands and a larger error (up to 0.25 mag) in g band. The airmass and color-term corrections are included in all the P60/LT/P200 data. 
The P60 data are processed using the software Fpipe \citep{Fremling2016} and image subtraction uses the SDSS references. While LT photometric data are processed with the software specifically built for IO:O \citep[][Taggart et al. 2022 in prep.]{Steele04,Fremling2016}. The IO:O collects data in SDSS $u$, $g$, $r$, $i$ and $z$ bands and image subtraction is performed using reference images from the PS1 or the SDSS (for $u$-band only). The P200 data are processed using the software AutoPhOT \citep{Brennan2022} and the PS1 reference images are used for image subtractions before performing photometry. 

\subsubsection{{\it Swift} data}
27 of 78 events in our sample have observations from the UV/Optical Telescope \citep[UVOT,][]{Roming2005} aboard the \textit{Neil Gehrels Swift Observatory} \citep{Gehrels04}. We retrieved the UVOT data from the NASA \textit{Swift} Data Archive\footnote{\url{https://heasarc.gsfc.nasa.gov/cgi-bin/W3Browse/swift.pl}}, and used the standard UVOT data analysis software distributed with {\tt HEASoft} version 6.19\footnote{\url{https://heasarc.gsfc.nasa.gov/lheasoft/download.html}} \citep{HEAsoft2014}, along with the standard calibration data. We manually define sky background apertures devoid of any sources. Four events in our sample, namely ZTF18aavrmcg (SN\,2018bgv), ZTF18acslpji (SN\,2018hti), ZTF19aawfbtg (SN\,2019hge) and ZTF19abpbopt (SN\,2019neq), have bright host galaxies. We therefore requested host-galaxy images at very late phases when the SNe had faded. The photometry for these four objects are obtained by subtracting out the host galaxy fluxes. For most of the other events, their host galaxies are very faint and contribute no more than $10\%$ of the SN fluxes in the UV-bands, so no host-galaxy subtractions are applied. 

All the photometry data are listed in Table~\ref{tab:photometry}, and are available to the public in electronic format at the journal website. 

\section{The Observed Light Curves} \label{sec:results}

The observed LCs for the full sample are presented in Figures~\ref{fig:lc_0} - \ref{fig:lc_5}, where the left Y-axis is apparent magnitude and the X-axis is the rest-frame days relative to the $g$-band peak phase. For a small number of events without a $g$-band peak phase, we use the $r$-band peak phase as the reference. 
Table~\ref{tab:photometry} contains the complete upper limits over much wider time ranges (early and late times) than those shown in the plots. For better display, we plot only a small number of photometric upper limits.

One striking feature apparent in Figures~\ref{fig:lc_0} - \ref{fig:lc_5} is that a large fraction of the LCs in our sample show significant undulations, and some have multiple peaks. In Paper II, we find $18-44\%$ of the well-sampled SLSNe-I have LC undulations. Among a small sub-set of SLSNe-I with early phase coverage, $6-44\%$ show early double peak LCs. The quantitative analysis of the LCs and the discussions of these features are included in Paper II.

\subsection{Magnitude corrections} \label{subsec:correction}

\subsubsection{Extinction corrections}
The Galactic reddening $E(B-V)$ is taken from the \citet{Schlafly_2011} dust map using the NASA/IPAC InfraRed Science Archive (IRSA) database. The extinction corrections at different wavelengths are computed using the empirical dust extinction laws of \citet{Fitzpatrick_2007} with $R_V=3.1$.

Previous studies have shown that SLSN-I hosts are mostly low-mass, metal-poor dwarf galaxies \citep{Lunnan2014,Lunnan2015,Leloudas2015,Perley2016,Angus2016,Chen2017a,Schulze2018}. This is the case for most of the events in our sample. Because their LCs do not show particularly red colors at pre-peak phases, we do not make any host-galaxy reddening corrections to these events. However, a small fraction of SLSN-I hosts are massive \citep[$10^{9-10}M_\odot$,][]{Perley2016,Chen2017}, and the host-galaxy reddening in such cases can be large. We find that seven events in our sample (Table~\ref{tab:hostextinction}) have non-negligible host-galaxy reddening. They either reside in bright hosts ($M_r \lesssim -18.5$\,mag) or have redder $(g-r)$ colors at peak than average (see Figure~\ref{fig:peakcolor}), which is presumably due to host-galaxy reddening.

\begin{center}
\begin{longtable}{lcc}
\caption{Host-galaxy Reddening}\\
\toprule
Name & $E(B-V)_{\rm host}$ (mag) & Method \\
\midrule
SN\,2018don & 0.4$^a$ & spec.temp \\
SN\,2018kyt & 0.11 & line ratio \\
SN\,2019kwt & 0.22 & line ratio \\
SN\,2019aamx & 0.07 & spec.temp \\
SN\,2019aamu & 0.23 & spec.temp \\
SN\,2019stc & 0.18 & line ratio \\
SN\,2020xkv & 0.24 & spec.temp \\
\bottomrule
\hspace*{\fill} \\
\multicolumn{3}{l}{$^{\rm a}$Estimated in \citet{Lunnan2020}.}
\label{tab:hostextinction}
\end{longtable}
\end{center}

The host-galaxy reddening is difficult to measure accurately, but can be roughly estimated using two different methods. The first one uses Balmer line ratio measured from the host-galaxy spectra. Without dust attenuation, the predicted line ratio between H$\alpha$ and H$\beta$ remains constant in a wide range of temperature and electron density in the nebular regions \citep[{\it e.g.} ${ \rm H\alpha/H\beta}\approx2.86$,][]{Osterbrock_2006}. The attenuation caused by dust scattering is wavelength-dependent and stronger at shorter wavelengths. This method is affected by both stellar and supernova absorption features. We remove the stellar Balmer absorption using {\it FIREFLY} \citep{Wilkinson_2017} before we measure the line flux, but ignore the influence of supernova features. This method is applied to the host spectra of ZTF18acyxnyw (SN\,2018kyt) and ZTF19acbonaa (SN\,2019stc). Note that the SN location may not coincide with the \ion{H}{2} region responsible for the Balmer lines, so the real extinction at the SN site could be different from the inferred value. But such an estimation is at least an indicator of whether the galaxy has significant dust. 

The second method is to infer the host-galaxy reddening by matching the observed spectrum (including spectral continuum slope and absorption features) with a SLSN-I template with negligible extinction and at a similar phase. For this method, the assumption is that the observed red spectral color is due to the host-galaxy reddening. The results from this method are very uncertain, but offer crude extinction estimates for those events without host emission lines. This method is applied to ZTF18aajqcue (SN\,2018don), SN\,2019aamx and ZTF19acvxquk (SN\,2019aamu), whose spectra lack distinct Balmer emission lines from host galaxies. 
For example, to match the near-peak spectrum of SN\,2018don, the $t\sim+54$\,day template spectrum of SN\,2007bi is required to be reddened by $E(B-V) \sim 0.4$\,mag \citep{Lunnan2020}. Similarly, the $+9$\,day spectrum of SN\,2019aamu matches well with the $-7$\,day template spectrum of PTF12gty after considering  a host-galaxy reddening of $E(B-V) = 0.23$\,mag. And the $+13$\,day spectrum of SN\,2019aamx matches well with the $+22$\,day template spectrum of PTF09cwl with $E(B-V) = 0.07$\,mag (shown in Figures~\ref{fig:classification_0} -- \ref{fig:classification_2}). Note that both PTF12gty and PTF09cwl have a faint, dwarf host, so their host-galaxy reddenings are assumed to be negligible \citep{Quimby_2018,Perley2016}.

For SN\,2019kwt and ZTF20abzaacf (SN\,2020xkv), we applied both of the above methods, since their spectra contain strong supernova signals and we are not able to remove the stellar absorption nor supernova features. The derived $E(B-V)$ for SN\,2019kwt is $0.22$\,mag from the host-galaxy spectrum and $0.18$\,mag from template matching, which appear to be consistent. In the case of SN\,2020xkv, the $E(B-V)$ derived from the spectral matching ranges from $0.22 - 0.26$\,mag, in contrast to the lower value ($0.17$\,mag) inferred from the Balmer decrement. Its spectrum has very low signal-to-noise ratio at H$\beta$, and with both highly uncertain measurements, we adopt the average value of $0.24$\,mag from the template matching method.

\subsubsection{The K-corrections}
One of the key parameters of SLSNe-I is the absolute $g$-band peak magnitudes. Proper estimate of this value requires proper K-corrections to derive reliable rest-frame values, either in $g$-band for sources at $z \leq 0.17$ or in $r$-band for sources at $z>0.17$. This switch is chosen because the observed $r$-band LC at $z>0.17$ is closer to the rest-frame $g$-band in wavelength.

Most events in our sample have at least one spectrum around the LC peak ({\it i.e.} $\lesssim 20$\,days in rest frame), which allow us to compute the K-corrections using the formula described in \citet{Hogg2002}. For those without spectra near the peak phases, we apply a constant K-correction of $-2.5\times {\rm log}(1+z)$. This approximation is not too far from the spectral estimates, as shown below.

Figure~\ref{fig:kcorrection} compares the calculated K-corrections based on observed spectra (blue dots) and an assumed $10^4$\,K blackbody spectral energy distribution (SED) (green line) with the constant correction (black line). The constant value of $-2.5\times {\rm log}(1+z)$ appears to capture most of the corrections within a range of $\pm0.1$\,mag (shaded region) for SLSNe-I at $0.06 < z < 0.67$. We list the K-corrections for each event in Table~\ref{tab:LCproperties}.

\begin{figure}[htp]
\includegraphics[width=0.5\textwidth]{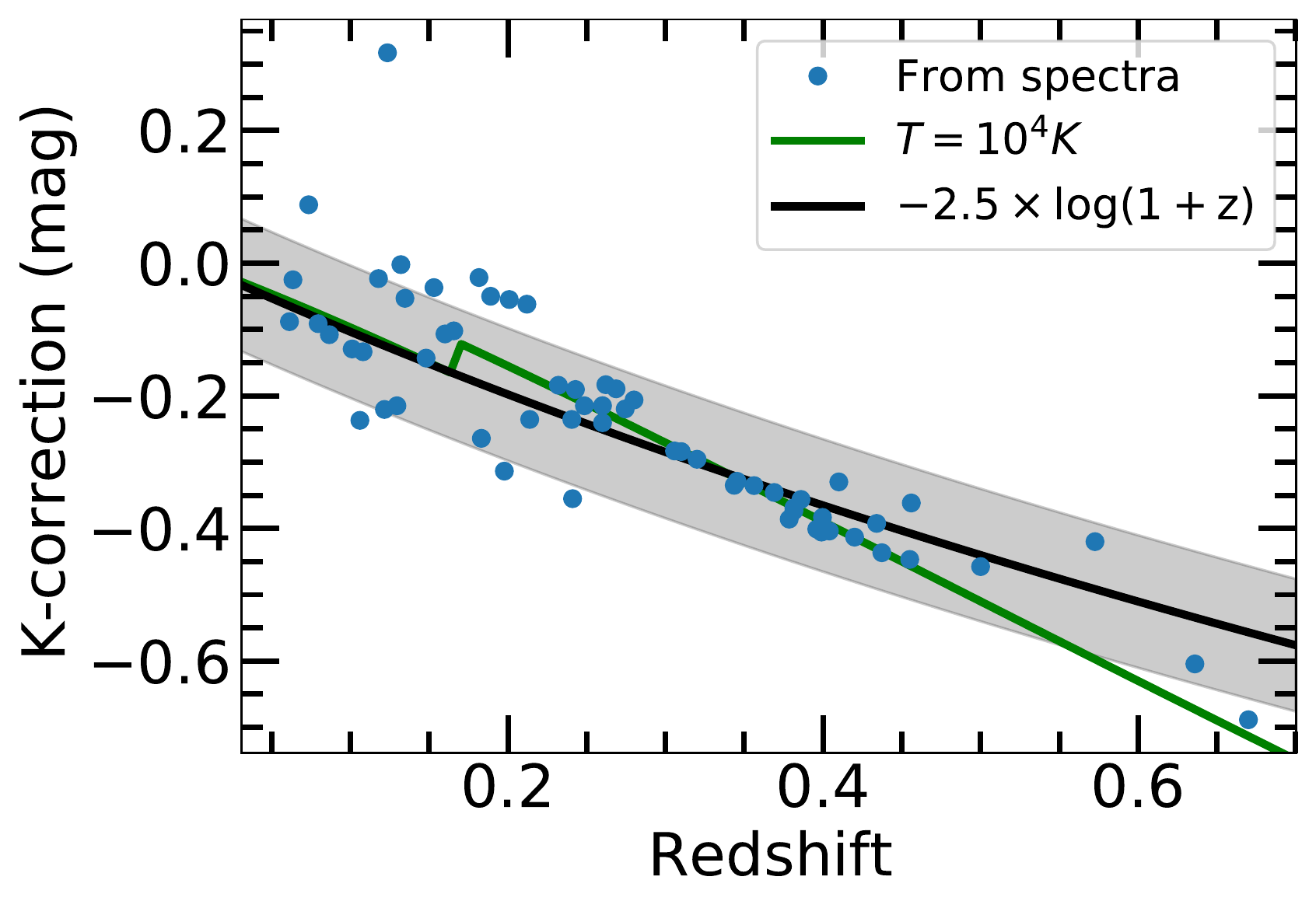}
\caption{K-corrections from the observed band to the rest-frame $g$ band for SLSNe-I around the peak. Blue points represent the corrections computed from the observed spectra. The black line shows the constant correction of $-2.5\times {\rm log}(1+z)$ with an uncertainty of 0.1\,mag shown by the gray shaded area. The green line shows the correction calculated assuming a blackbody spectrum with a temperature of $10^4$\,K. The break at $z=0.17$ is caused by the change of observed band (from $g$ to $r$ band).}
\label{fig:kcorrection}
\end{figure}

\subsubsection{The S-corrections} \label{subsubsec:S-correction}

As the $g$-, $r$- and $i$-band photometry of SLSNe-I presented in this work are obtained from four different telescopes, the errors caused by filter differences need to be evaluated.
Since most of the data are from the ZTF it self, we choose to correct the LT/P200/P60 photometry to that of the ZTF filters. Following the method of \citet{Stritzinger2002,Stritzinger2005}, \citet{Pignata2008} and \citet{Wang2009}, 
the correction between different instruments (known as the S-correction) can be computed using
\begin{equation}
Sc_{\rm \lambda 1} = M_{\rm \lambda 1} - m_{\rm \lambda 1} - CT_{\rm \lambda 1} \left(m_{\rm \lambda 1} - m_{\rm \lambda 2}\right) - ZP_{\rm \lambda 1},
\end{equation}
where, $M_{\rm \lambda 1}$ is the SN synthetic magnitude computed with the response functions of ZTF $\lambda 1$ filter, and $m_{\rm \lambda 1}$ and $m_{\rm \lambda 2}$ are the SN synthetic magnitudes computed with the LT/P200/P60 filters. $CT_{\rm \lambda 1}$ is the color term and $ZP_{\rm \lambda 1}$ is a zero point, which are measured by convolving the ZTF filters with a large sample of spectrophotometric Landolt standard stars from \citet{Stritzinger2005}. The response function above includes filter transmission, detector quantum efficiency and atmospheric transmission.

As S-corrections are spectrum-dependent and accurate measurements at any given epoch require far more SLSN-I spectra than we have in this work. We use the S-correction values computed from the spectra as an additional $1\sigma$ error in photometry, and add them to the existing errors of LT/P200/P60 photometry in quadrature. The additional $1\sigma$ errors due to the S-corrections are listed in Table~\ref{tab:Sc}.

\begin{center}
\begin{longtable}[htbp!]{lccc}
\caption{The estimated error caused by the S-correction.}\\
\toprule
\multirow{2}{*}{Telescopes} & \multicolumn{3}{c}{Filters} \\
                           & $g$\,(mag)     & $r$\,(mag)     & $i$\,(mag)     \\
\midrule
LT & 0.018 & 0.044 & 0.030 \\
P60 & 0.042 & 0.030 & 0.026 \\
P200 & 0.023 & 0.030 & 0.025 \\
\bottomrule
\label{tab:Sc}
\end{longtable}
\end{center}



\subsection{Empirical LC Fitting Method} \label{sec:lcfitting}

The LC analysis requires estimates of various parameters such as peak magnitude, rise time and rise rate, which all involve numerical interpolation and fitting. We adopt a machine learning algorithm, Gaussian Process (GP) regression which has various kernel functions. The GP interpolation can reduce the influence of outliers and gives robust error estimates. We use a composite kernel which is the sum of Mat{\'e}rn kernel with a white noise kernel. We tested Mat{\'e}rn kernels with $\nu$ parameter of 3/2 (Mat{\'e}rn 3/2) \citep{Inserra_2018a,Angus_2019,Lunnan2020} as well as $\nu = 5/2$ (Mat{\'e}rn 5/2). When we compute peak magnitudes and phases, GP fit with Mat{\'e}rn 3/2 is performed in the flux space. The Python package {\tt george} \citep{Ambikasaran2014} and the Scikit-learn give comparable results. 

\subsection{LCs in Absolute g-band magnitudes}
\label{subsec:Photmetry} 

To set an approximate luminosity scale, we also plot the absolute magnitude on the right Y-axis of Figures~\ref{fig:lc_0} - \ref{fig:lc_5}. This is calculated by assuming a constant K-correction of $-2.5\times {\rm log}(1+z)$. In these figures, the LCs have been corrected for the Galactic and host-galaxy reddening whenever possible. 

\begin{figure}[htp]
\centering
\includegraphics[width=0.5\textwidth]{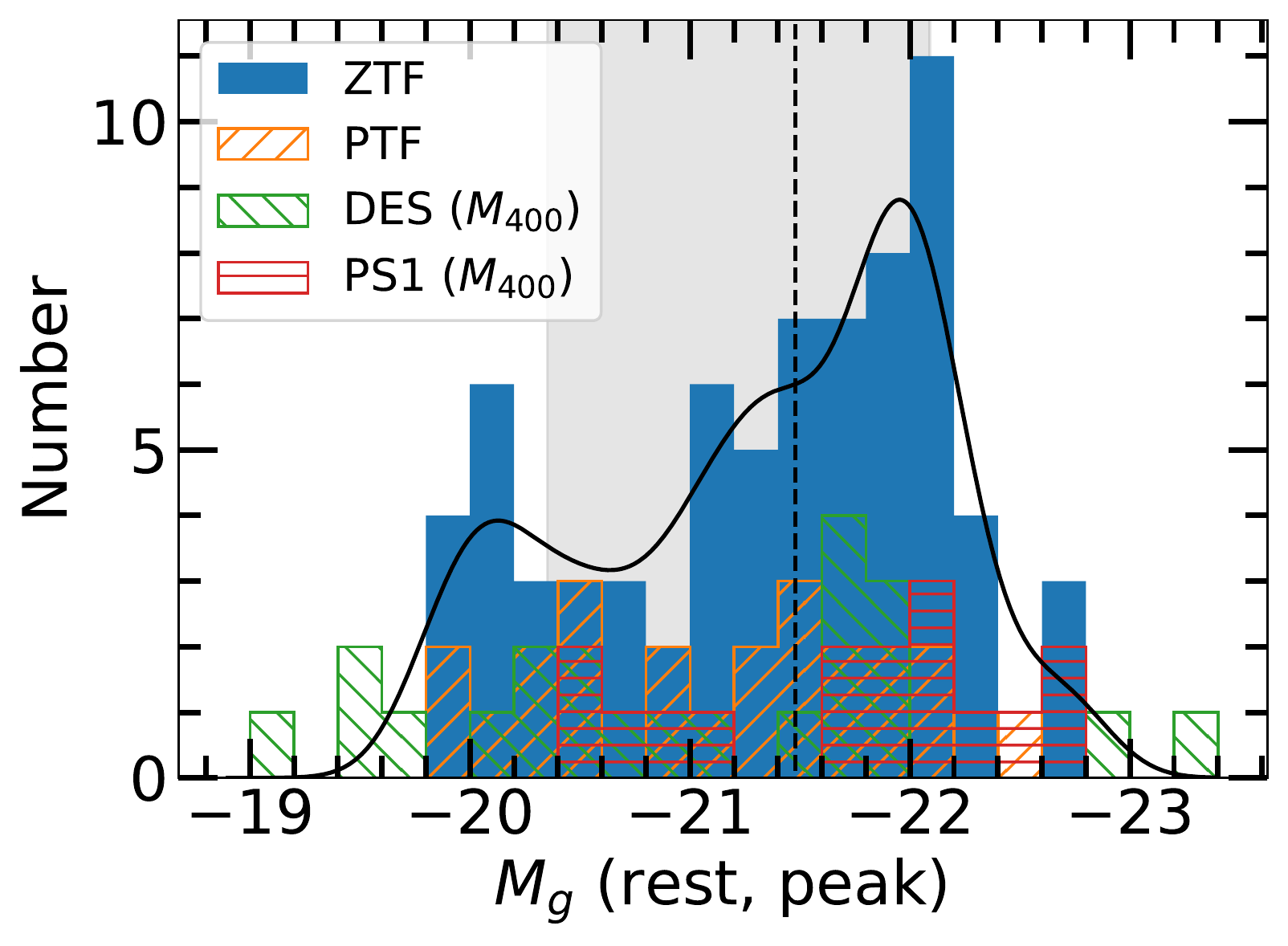}
\caption{The distribution of the rest-frame $g$-band absolute peak magnitudes. The dashed line and shaded region mark the median value and the $1\sigma$ dispersion, $-21.48^{+1.13}_{-0.61}$\,mag. The black solid line shows the kernel density estimation of the distribution. Other SLSN-I samples are plotted in different hatching patterns for comparison. DES and PS1 samples are measured at 400\,nm, which is bluer than the $g$ band (472\,nm).}
\label{fig:peak}
\end{figure}

To calculate the apparent peak magnitude and phase, we run the GP to interpolate the LCs in flux space. The errors are shown as $1\sigma$ uncertainty. The $g$-band absolute peak magnitudes are derived using $M_g = m_g - \mu - KC$, where $\mu$ is the distance modulus, $KC$ is the K-correction, and $m_g$ is the apparent $g$-band magnitudes corrected for both the host (in rest frame) and the Galactic extinction (in observed frame). The rest-frame $g$-band absolute peak magnitudes and the peak dates are tabulated in Table~\ref{tab:LCproperties}. The peaks of some events are not well constrained due to the poor sampling around the peak. Figure~\ref{fig:peak} shows the distribution of the $g$-band absolute peak magnitudes ($M_g$) for the ZTF sample, ranging from $-19.8$\,mag to $-22.8$\,mag with a median and $1\sigma$ error of $M_{g,\rm med} = -21.48^{+1.13}_{-0.61}$\,mag.  Both our median value and dispersion are consistent with those of previous samples.

The weak bimodal distribution in the peak absolute magnitudes should not be over-interpreted because this is the raw distribution function without applying corrections for various selection biases ({\it e.g.} Malmquist bias, preference of our selection system for slowly-rising events). Selection biases could also have a strong impact on the distribution of the timescales ({\it e.g.} Figure~\ref{fig:decline2rise}), peak bolometric luminosities ({\it e.g.} Figure~\ref{fig:peakbolo}) and other photometric properties. The correct interpretation of these distributions requires a well-defined selection function of our sample, the detailed simulation of which, however, is well beyond the scope of this paper. Further studies ({\it e.g.} Paper III) are required to confirm these distributions.

\section{Light Curve Parameters}
\label{sec:measure}
\subsection{Rise/decay time scales -- fast and slow SLSN-I events}
\label{subsec:Timescale}
Traditional rise times, defined as the interval between the explosion date and the peak, are notoriously difficult to measure, especially for distant transients such as SLSNe-I. Although the ZTF detection can go down to $20.5 - 21.0$\,magnitude, at the median redshift of our sample, it only probes an absolute magnitude of $\sim -20$\,mag. The upper limits before the first detection usually can not place strong constraints on the LC evolution in very early phases and only about one-third of our sample have high-quality early time data. In addition, \citet{Anderson2018} shows that a SLSN-I (SN\,2018bsz) can have a long, slowly rising “plateau” before a steeper, faster rise to the peak, which makes it harder to speculate explosion dates without enough deep, early data.

Instead of using the traditional rise time, we define $t_{\rm rise/decay,x}$ as the time interval between the peak and the flux when it is at a fraction $x$ of the peak value. Here the $x$ factor can be $10$\%\ ($\Delta$mag $=2.5$), or $1/e$\ ($\Delta$mag $\approx 1.09$). Time dilation is corrected for all time scale measurements. The derived $t_{\rm rise,10\%}$ and $t_{\rm rise/decay,1/e}$ are listed in Table~\ref{tab:LCproperties}. The poorly constrained time scales are not included in the following analysis.

\begin{figure}[htp]
\centering
\includegraphics[width=0.5\textwidth]{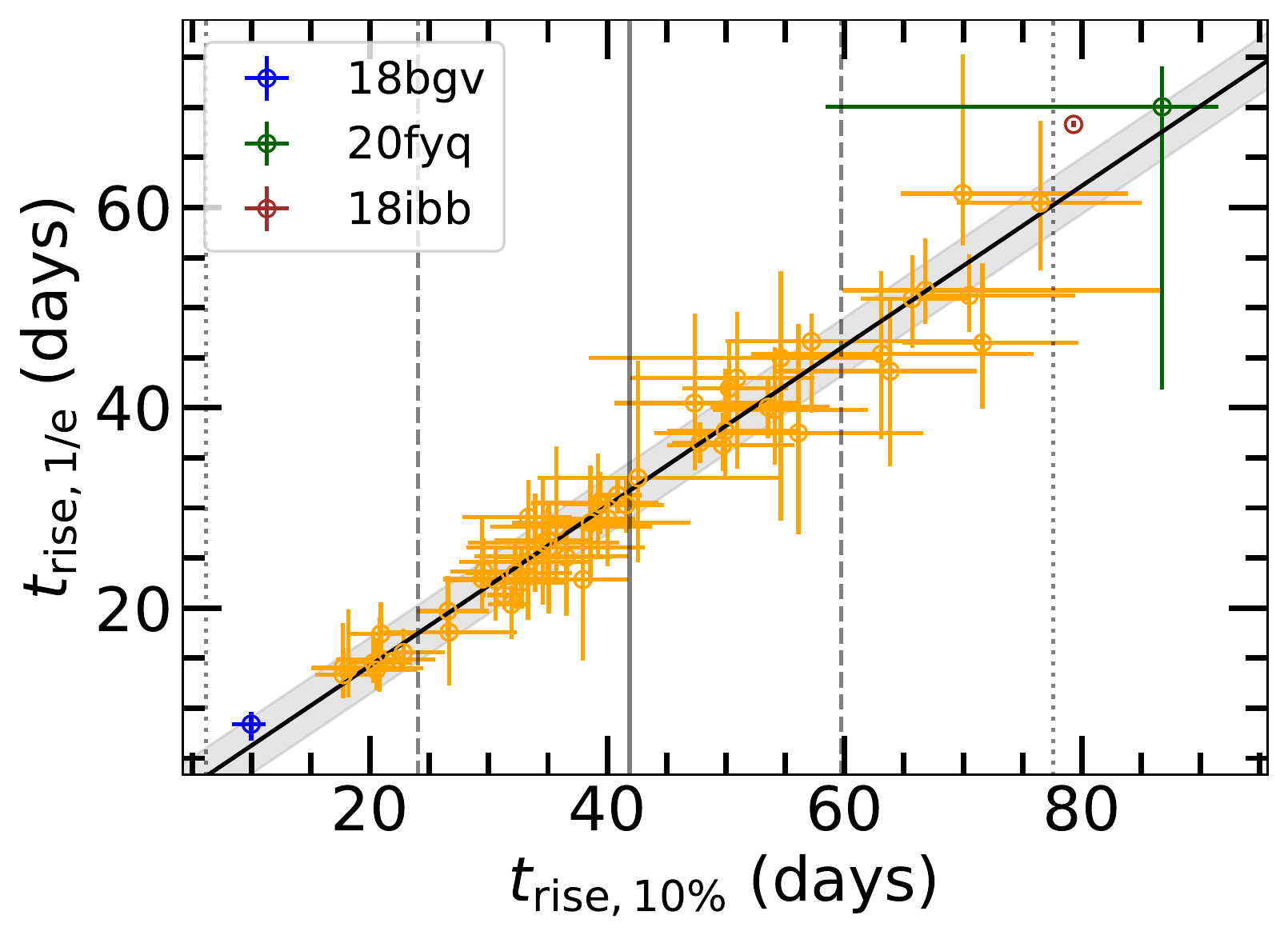}
\caption{Rise time measured at 10\% of the peak flux $t_{\rm rise,10\%}$ versus that measured at $1/e$ of the peak flux $t_{\rm rise,1/e}$. The vertical solid gray line marks the mean value of $t_{\rm rise,10\%}$, the dashed and dotted lines represent the $1\sigma$ and $2\sigma$ uncertainties, respectively. The black solid line and shaded area show the linear fit to the measured time scales ($t_{\rm rise,1/e} = 0.80~t_{\rm rise,10\%} - 1.73$\,days) and the $1\sigma$ dispersion. The events with extreme time scales are highlighted in different colors.
}
\label{fig:risetime}
\end{figure}

Figure~\ref{fig:risetime} shows a direct proportionality between $t_{\rm rise,1/e}$ and $t_{\rm rise,10\%}$ in the rest-frame $g$ band. Linear fit gives $t_{\rm rise,1/e} = 0.80~t_{\rm rise,10\%} - 1.73$\,days with $1\sigma$ uncertainty as small as $\sim 2.84$\,days. The small scatter implies that the LC rise rates at the very early phases do not have significant differences in our sample. The rise times $t_{\rm rise,10\%}$ cover a wide distribution, ranging from $10 - 90$\,days, with the mean (median) value and a standard deviation of $\overline{t_{\rm rise,10\%}} = 41.9 (38.3) \pm 17.8$\,days.

Of the total 56 SLSNe-I with $t_{\rm rise,10\%}$ measurements, the two slowest rising events SN\,2018ibb and ZTF20aapaecd (SN\,2020fyq) stand out in Figure~\ref{fig:risetime}, with time scales longer than $\overline{t_{\rm rise,10\%}} + 2\sigma = 78$\,days. 

For another slowly-evolving event ZTF20aadzbcf (SN\,2020fvm), its $t_{\rm rise,1/e}$ reaches 91 days, suggestive of an unusually long rise time, although $t_{\rm rise,10\%}$ is not well constrained. SN\,2020fvm has two LC peaks and we set the second, also the brighter one as its main peak. If the first one is set as the main peak, its $t_{\rm rise,1/e}$ is around 48 days. The fastest event in our sample is SN\,2018bgv \citep{Lunnan2020}, which rises to the peak in less than 10 days. Such a fast and luminous event is very rare and it may be associated with the Fast Blue Optical Transients \citep[FBOT,][]{Drout2014,Ho2021}. Compared with previous SLSN-I samples, our sample contains slightly more fast-evolving events. For instance, 9 out of 56 events \citep[9-26\%, calculated at a confidence level of 95\% based on][]{Gehrels_1986} are found to have $t_{\rm rise,1/e} \lesssim 15$\,days, while only 5 in 55 events (4-18\%) were reported in the previous samples.

Outside the $\pm 1\sigma$ range, we have 10 fast-evolving events with $t_{\rm rise,10\%} \leq \overline{t_{\rm rise,10\%}} - 1\sigma = 24$\,days and 10 slow events with $t_{\rm rise,10\%} \geq \overline{t_{\rm rise,10\%}} + 1\sigma = 60$\,days. The $t_{\rm rise,10\%}$ shows a continuous distribution and can not be divided into two separate fast and slow subclasses, as indicated by previous studies \citep{Nicholl2015b,DeCia2018}. 

\begin{figure}[htp]
\centering
\includegraphics[width=0.5\textwidth]{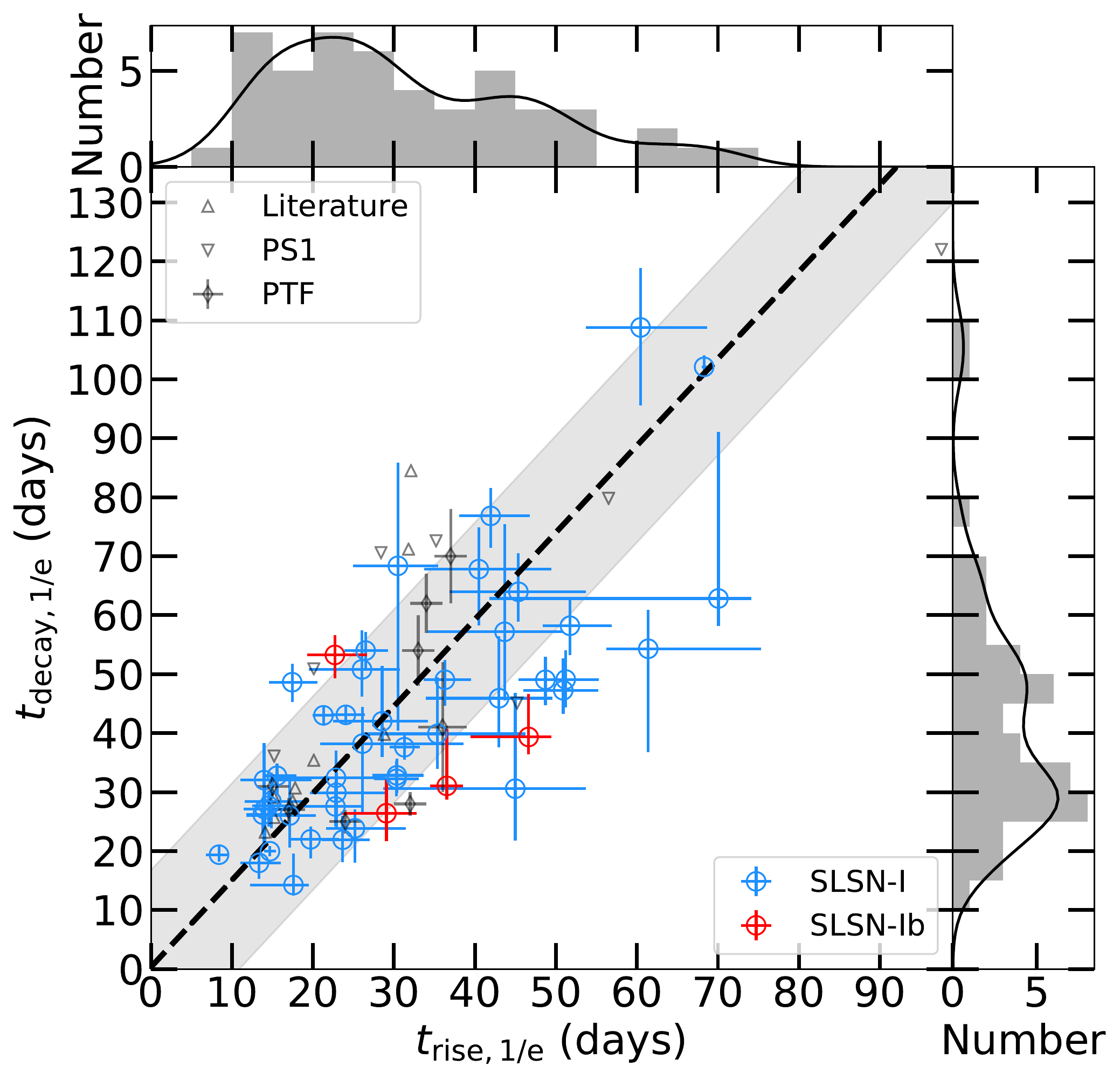}
\caption{
Time scales measured at $1/e$ of the peak flux for the rising and declining portions of LCs. 
Normal SLSNe-I are shown in blue circles while He-rich SLSNe-Ib are highlighted in red. The black dashed line and the shaded area show the best linear fit and 1-sigma error, $t_{\rm decay}=1.47~t_{\rm rise}+0.35$\,days.  The histograms along the horizontal and vertical axes show the distributions of the rise and decay time scales, respectively. The black solid lines show the kernel density estimation of the distributions. 
The time scales from other SLSN-I samples are plotted with smaller gray points for comparison. The data from PTF are measured at 1 mag below peak magnitude and is slightly shorter than the $1/e$-maximum (1.09 mag) time scales. And those from the literature \citep{Nicholl2015b} and the PS1 are measured from bolometric LCs and may have biases due to redshift and the SED.
}
\label{fig:decline2rise}
\end{figure}

Figure~\ref{fig:decline2rise} shows comparison of the rise and decay time scales measured at $1/e$ of the peak flux for 48 events. Both rise/decay time scales show a distribution centered at $\sim 25-30$\,days with an extended tail. For the whole sample, the rise and decay time scales show a strong positive correlation with a Spearman correlation coefficient of $\rho=0.73$ and null probability $p < 10^{-8}$, {\it i.e.} slowly-rising events tend to decay slowly. Applying a linear fit, we find $t_{\rm decay,1/e}=(1.47\pm0.07)~t_{\rm rise,1/e}+(0.35\pm2.50)$\,days, which is similar to what was found in previous studies \citep{Nicholl2015b, DeCia2018}. 


It is interesting to note that four of the five SLSNe-Ib in the \citet{Yan2020} sample, namely SN\,2018kyt, SN\,2019hge, ZTF19acgjpgh (SN\,2019unb) and ZTF20ablkuio (SN\,2020qef), have rise time scales which are 9 -- 32 days longer than that of the rise-decay time correlation. The last SLSN-Ib, ZTF19aamhhiz (SN\,2019kws), in the \citet{Yan2020} sample has a short rise time but much longer decay time. As noted in \citet{Yan2020}, these SLSNe-Ib may have He-rich CSM and CSM interaction may affect the LC evolution before or after the peak. Detailed modeling and discussions of their LCs are given in Paper II.

\begin{figure}[htp]
\centering
\includegraphics[width=0.5\textwidth]{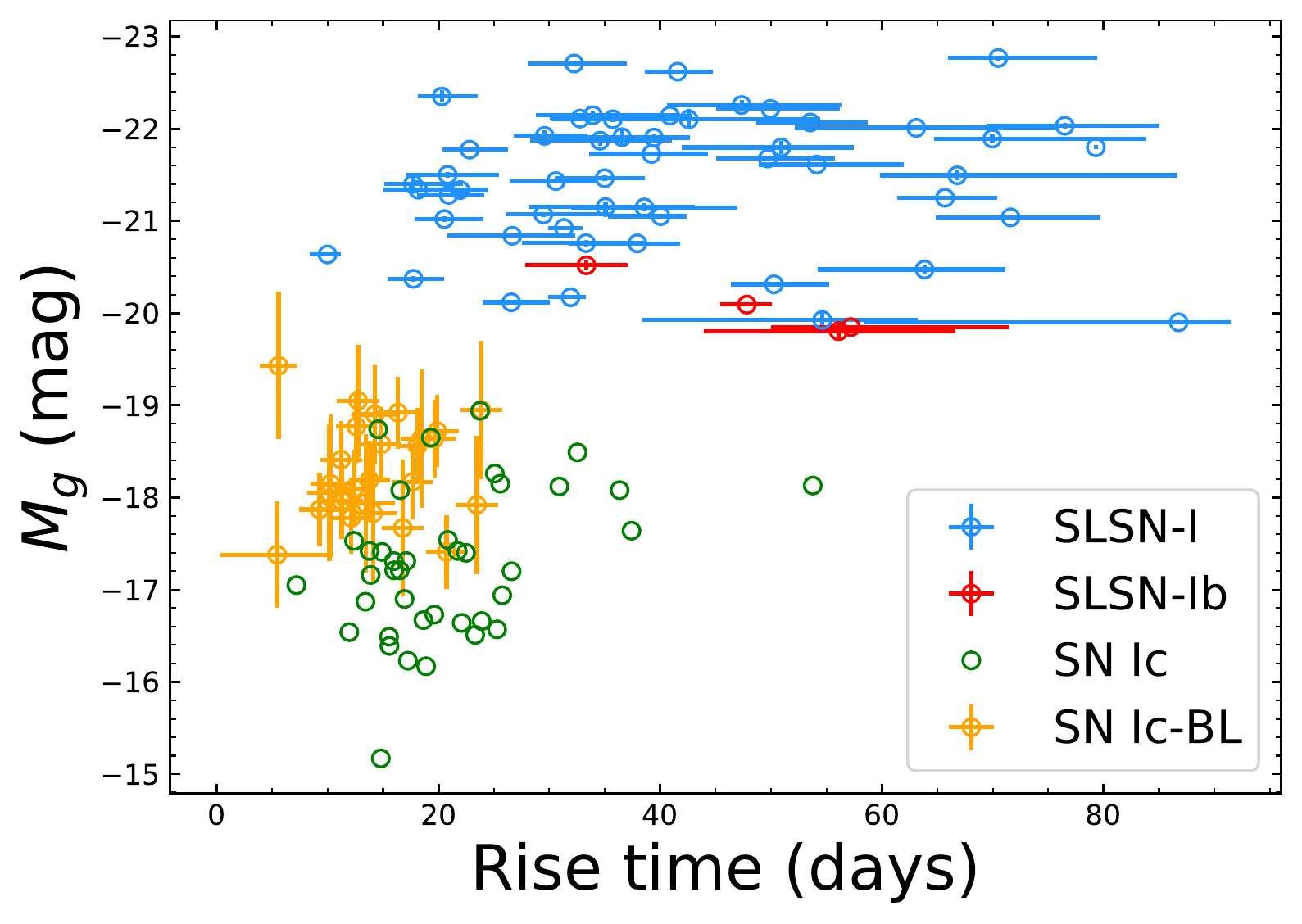}
\caption{
Rise times $t_{\rm rise,10\%}$ versus $g$-band absolute peak magnitudes $M_g$ for SLSNe-I (this paper), normal SNe\,Ic \citep[from][]{Barbarino_2021} and broad-lined SNe\,Ic \citep[SNe\,Ic-BL, from][]{Taddia_2019}. The $M_g$ of the SN\,Ic and SN\,Ic-BL samples is computed from the $r$-band magnitudes using a color correction of $\sim 0.36$\,mag \citep{Taddia2015,Prentice2016}. The rise times of SNe\,Ic and SNe\,Ic-BL are measured from the explosion date in $r$ band, which are slightly longer than $t_{\rm rise,10\%}$. The SLSNe-Ib in our sample are highlighted. 
}
\label{fig:magtrise}
\end{figure}

Although SNe\,Ic and SLSNe-I have similar post-peak spectra, they have very different peak luminosity and time scales. Figure~\ref{fig:magtrise} compares the peak $M_g$ and rise times between normal SNe\,Ic, broad-lined SNe\,Ic (SNe\,Ic-BL) and SLSNe-I. The SLSNe-I are from our sample, while the normal SN\,Ic sample and the SN\,Ic-BL sample are taken from \citet{Barbarino_2021} and \citet{Taddia_2019}, respectively. We exclude two bright SNe\,Ic from literature, iPTF12gty and iPTF15eov, which can be better classified as SLSNe-I as suggested by their authors. Both SN\,Ic samples are obtained from the (intermediate) PTF, which has similar observation depth to ZTF. The overall distribution in Figure~\ref{fig:magtrise} is similar to that shown in \citet{DeCia2018}. It should be noted that correction for observational selection biases and volumetric correction are required to interpret the distribution correctly. The peak $M_g$ of SLSNe-I is about 4 and 3\,magnitude brighter than those of normal SNe\,Ic and SNe\,Ic-BL, respectively. All SLSNe-Ib have low luminosities and moderate rise times compared with normal SLSNe-I. SLSNe-I have significantly longer rise times and wider dispersion compared to those of SNe\,Ic. First, this is physically related to the fact that they have larger ejecta masses and more massive progenitor stars (see Paper II). Second, SNe\,Ic are primarily powered by radioactive decay which has a constant time scale of energy injection. In the magnetar model proposed for SLSNe-I, the time scale of energy injection depends on magnetic field $B$ ($\propto B^{-2}$) and spin period $P$ ($\propto P^2$). The LC width expands with the decrease of $B$ \citep[see Figure~2 of ][]{Kasen2010}. The large spread in their rise times could suggest that the magnetic field strengths of the neutron stars can vary over a wide range as well.

\subsection{Color and blackbody temperature}
\label{subsec:temperature}

Understanding the variation and uniformity of the SLSN-I SEDs as a function of time should shed light on the physical nature of this population of stellar explosions. Although the complete characterization of the SLSN-I SEDs is beyond the scope of this paper, we can infer some basic properties by examining the distributions of $(g-r)$ colors and blackbody temperatures since both parameters are determined by the transient SEDs. 

\begin{figure*}[htp]
\center
\includegraphics[width=\textwidth]{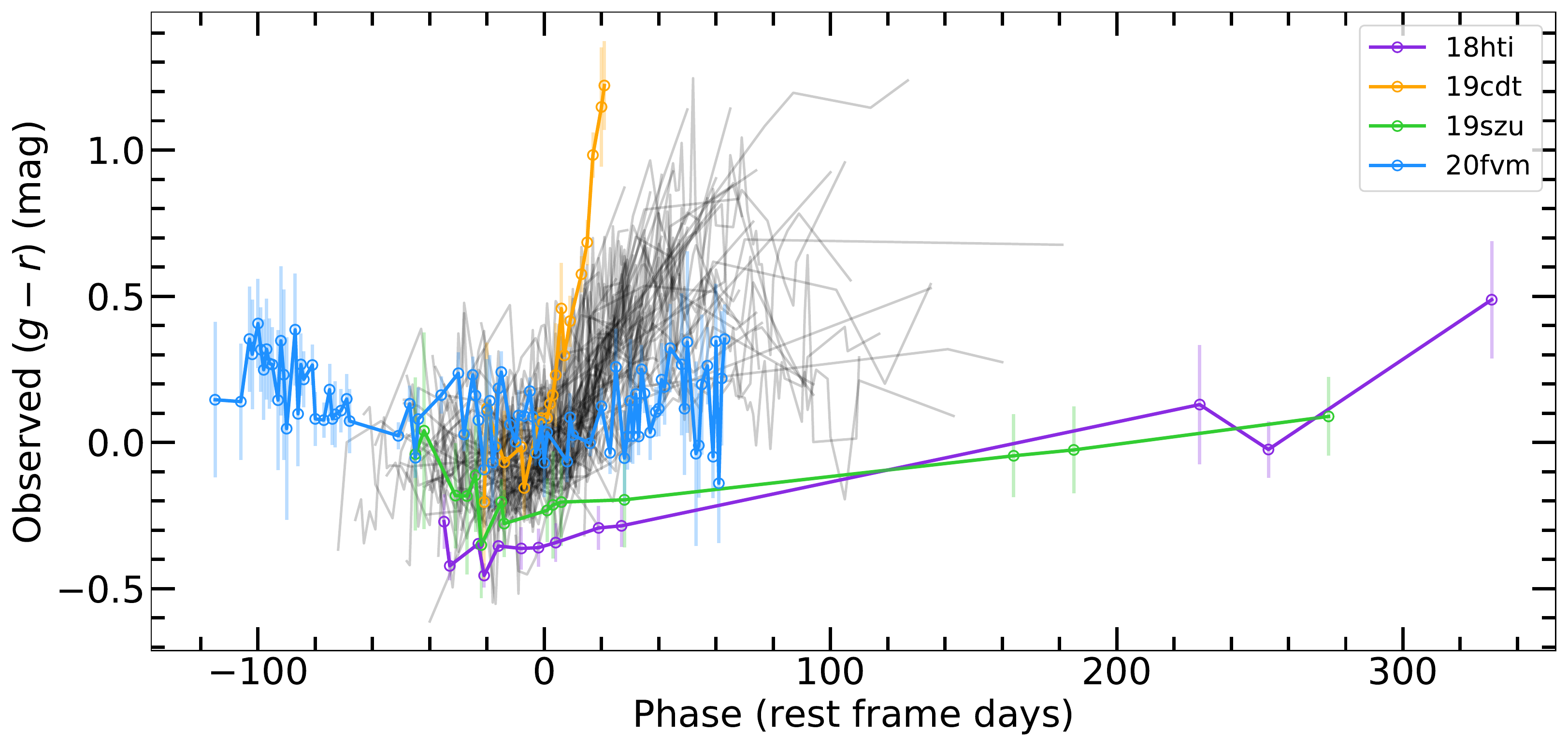}
\caption{The observed $(g-r)$ color evolution tracks with time. We highlight four events whose $(g-r)$ color tracks do not follow the general trend shown in gray.}
\label{fig:color}
\end{figure*}

\begin{figure}[htp]
\center
\includegraphics[width=0.5\textwidth]{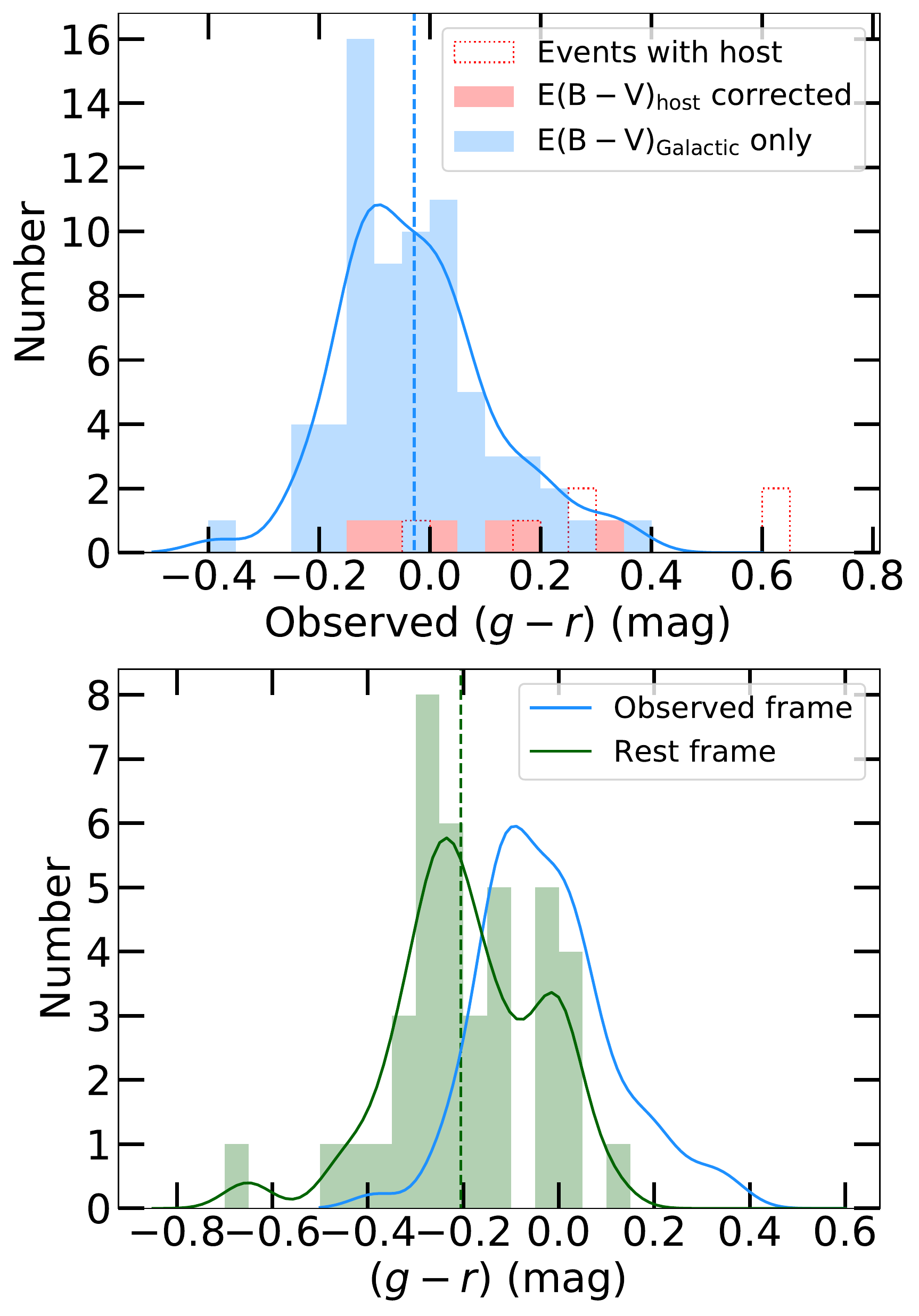}
\caption{The distribution of the $(g-r)$ colors at peak. \textbf{\emph{Top panel:}} The red dotted line represents the original colors of the events with host-galaxy reddening and the red area shows their corrected observed color. The blue area represents the observed colors of the events with only Galactic extinction correction. After host-galaxy reddening correction, the observed $(g-r)$ color at peak has a median value with the $1\sigma$ dispersion of $-0.03^{+0.12}_{-0.11}$\,mag, marked by the blue dashed line. The blue solid line shows the kernel density estimation of the distribution. \textbf{\emph{Bottom panel:}} The green area represents the rest-frame colors of the events with K-correction. The rest-frame $(g-r)$ color at peak has a median value with the $1\sigma$ dispersion of $-0.21^{+0.19}_{-0.12}$\,mag, which is represented by the green dashed line. The green and blue solid lines show the kernel density estimations of the rest-frame colors and the observed colors (normalized to the number of the rest-frame colors), respectively.}
\label{fig:peakcolor}
\end{figure}

Figure~\ref{fig:color} shows the observed color evolution as a function of time. All of the colors have been corrected for reddening but not for K-corrections, since that requires better spectral coverage than we have. The overall $(g-r)$ color trend evolves from blue ({\it i.e.} $\sim -0.3$\,mag, hotter temperatures) at early phases to red and reaches $(g-r) \sim +0.8$\,mag at about 2 months after the peak. In our sample, four events have peculiar color (temperature) evolution, and do not follow the general trend. We discuss these outliers below. We further examine the distribution of the observed $(g-r)$ color obtained at the maximum light in Figure~\ref{fig:peakcolor}. At the peak phases, the median observed color is close to zero, $(g-r)_{\rm med} = -0.03^{+0.12}_{-0.11}$\,mag.

Figures~\ref{fig:color} and \ref{fig:peakcolor} show large scatters. The observed colors are not easy to interpret because they sample different parts of the SEDs, depending on the transient redshifts. The proper measurements are the rest-frame color tracks, however, reliable results would require many more spectra than our sample has. Here we compute only the rest-frame $(g-r)$ colors at the peak phase using the color corrections computed from the near-peak spectra. The rest-frame ($g-r$) colors are corrected from the observed ($g-r$) for the events at $z \leq 0.17$ and ($i-r$) for $z>0.17$, if photometry data and spectra are available. All the peak colors are listed in Table~\ref{tab:peakcolor}. The green histogram in Figure~\ref{fig:peakcolor} shows that the median rest-frame $(g-r)_{\rm rest,med}$ is $-0.21^{+0.19}_{-0.12}$\,mag, which is consistent with $-0.27$\,mag calculated from the PTF SLSN-I sample \citep{DeCia2018}. The large scatter in the peak rest-frame $(g-r)$ indicates that the SEDs of SLSNe-I may show a diverse shape and hence a wide range of blackbody temperatures.

\begin{figure}[htp]
\center
\includegraphics[width=0.5\textwidth]{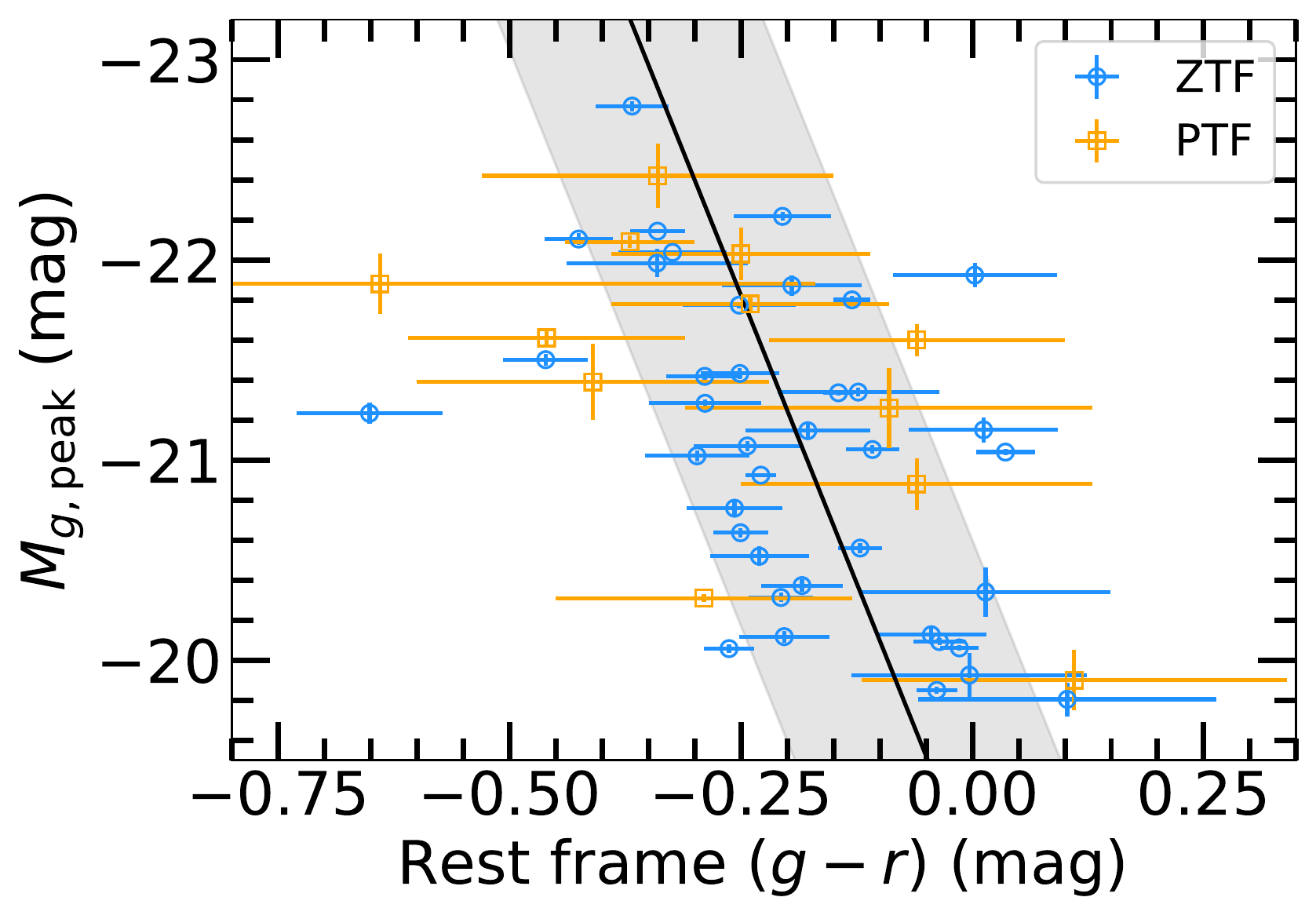}
\caption{The correlation between the $g$-band absolute peak magnitudes $M_{g,\rm peak}$ and the rest-frame ($g-r$) colors. The linear fit and $1\sigma$ error are shown by the black line and the shaded area. The SLSN-I sample from this paper is marked in blue while that from PTF \citep{DeCia2018} is in orange.}
\label{fig:Mg-color}
\end{figure}

Figure~\ref{fig:Mg-color} illustrates a moderate correlation ($\rho=0.52,~ p<10^{-3}$) between the peak rest-frame $(g-r)$ colors and the $g$-band absolute peak magnitudes $M_{g,\rm peak}$, {\it i.e.} brighter SLSNe-I tend to have bluer color, which was also previously found by \citet{Inserra2014} and \citet{DeCia2018} in smaller samples. If combined with the PTF sample from \citet{DeCia2018}, the correlation becomes stronger ($\rho=0.56,~ p\approx2\times10^{-5}$), and can be described by a linear function, $M_{g,\rm peak} = (11.5 \pm 2.6) \times (g-r) - (18.9 \pm 0.5)$\,mag, with $1\sigma$ error of 1.7 mag. This correlation can be used in cosmological searches of SLSNe, {\it e.g.} \citet{Inserra2021}. However, this is beyond the scope of this paper.

To fit the blackbody temperature, we adopt a modified blackbody function, defined by $f_\lambda = {\rm max}[ ~ 0, ~ 1 - A\times(1.0 - \lambda/3000.0) ~ ] \times B_\lambda$, with $B_\lambda$ being the Planck function, and $A$ being the scaling factor. The modified blackbody function aims to quantitatively capture the variations in the UV spectral suppression for different events, as shown by the HST UV spectra of SLSNe-I \citep{Yan2017a,Yan2018}. The scaling factor $A$ is derived from the fitting at $\lambda \leq 3000$~\AA~, set to zero at $\lambda \geq 3000$~\AA~ and fit in a range of 0 -- 3. Larger scaling factors represent stronger suppression in UV and $A=1$ represents the SED function used in \citet{Nicholl2017b}. The error associated with the temperature is estimated using the Markov Chain Monte Carlo (MCMC) method.

\begin{figure*}[htp]
\centering
\includegraphics[width=\textwidth]{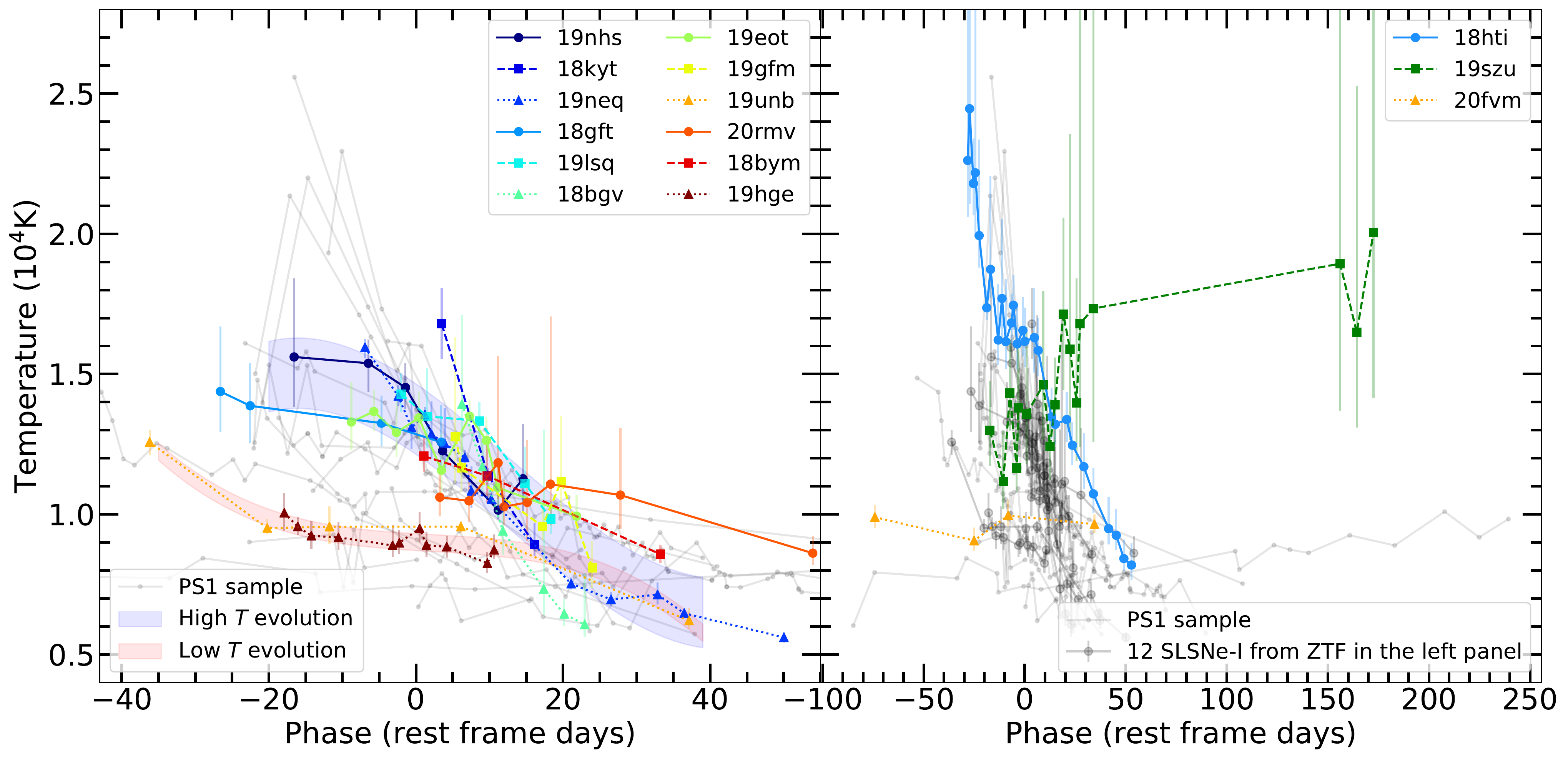}
\caption{Blackbody temperature evolution as a function of time. For comparison, we plot the temperature measurements from the PS1 sample in light gray. \textbf{\emph{Left panel:}} We include 12 of 15 events with at least 3 epochs of UV data. We apply third-order polynomial fits to the temperature evolution of two low-temperature events, SN\,2019hge and SN\,2019unb, and show the result with the $1\sigma$ error in the red shaded area. Similarly, the fit of 10 high-temperature events is shown by the blue shaded area. \textbf{\emph{Right panel:}} We highlight the three extraordinary events and plot the 12 normal events in dark gray for comparison.}
\label{fig:Temperature}
\end{figure*}

Of the 78 events in our sample, only 15 have at least 3 epochs of {\it Swift} UV photometry for properly computing the blackbody temperatures. In Figure~\ref{fig:Temperature}, the left panel shows the temperature evolution tracks for 12 events, and the right panel shows the rest three with peculiar color/temperature evolution. Although the statistics is not large, Figure~\ref{fig:Temperature} indicates the large temperature spread at any given phase, especially at pre-peak and peak phases.  In the left panel, some SLSNe-I have high temperatures at $\sim15000$\,K at $t \sim -10$\,days, and cool down to $\sim9000$\,K at $+20$\,days. Interestingly, there are also two cooler SLSNe-I ({\it i.e.} SN\,2019hge and SN\,2019unb) with the peak temperatures less than $\sim 10000$\,K.  Consistently, neither of them shows clear \ion{O}{2} absorption lines in their classification spectra, which require a high ionization temperature \citep[{\it i.e.} $T\sim 15000$\,K,][]{Quimby_2018}. 
We perform a polynomial fit to the temperature tracks and derive $T=0.090201\,t^3-3.4306\,t^2-180.30\,t+13395$\,K and $T=-0.052720\,t^3+0.3319\,t^2-15.84\,t+9007$\,K for the high and low temperature tracks, respectively. These are shown as blue and red shaded regions with the $\pm1\sigma$ uncertainty in Figure~\ref{fig:Temperature}. Consistent with the temperature evolution, the ($g-r$) colors of low-temperature events are $\sim 0.2 - 0.3$\,mag redder than those of high-temperature ones at the peak but become indistinguishable from that of the full sample after $+15$\,days after the peak. 

Moreover, the color distribution plots suggest that there are more low-temperature SLSNe-I. For instance, 10 out of 35 (29\%) events are found to have redder peak rest-frame colors than the two low-temperature events ($\sim 0.10$\,mag). However these 10 events have no UV photometry available, thus no temperature measurements. It is possible that these 10 red events could also have low peak temperatures.

\begin{figure}[htp]
\centering
\includegraphics[width=0.5\textwidth]{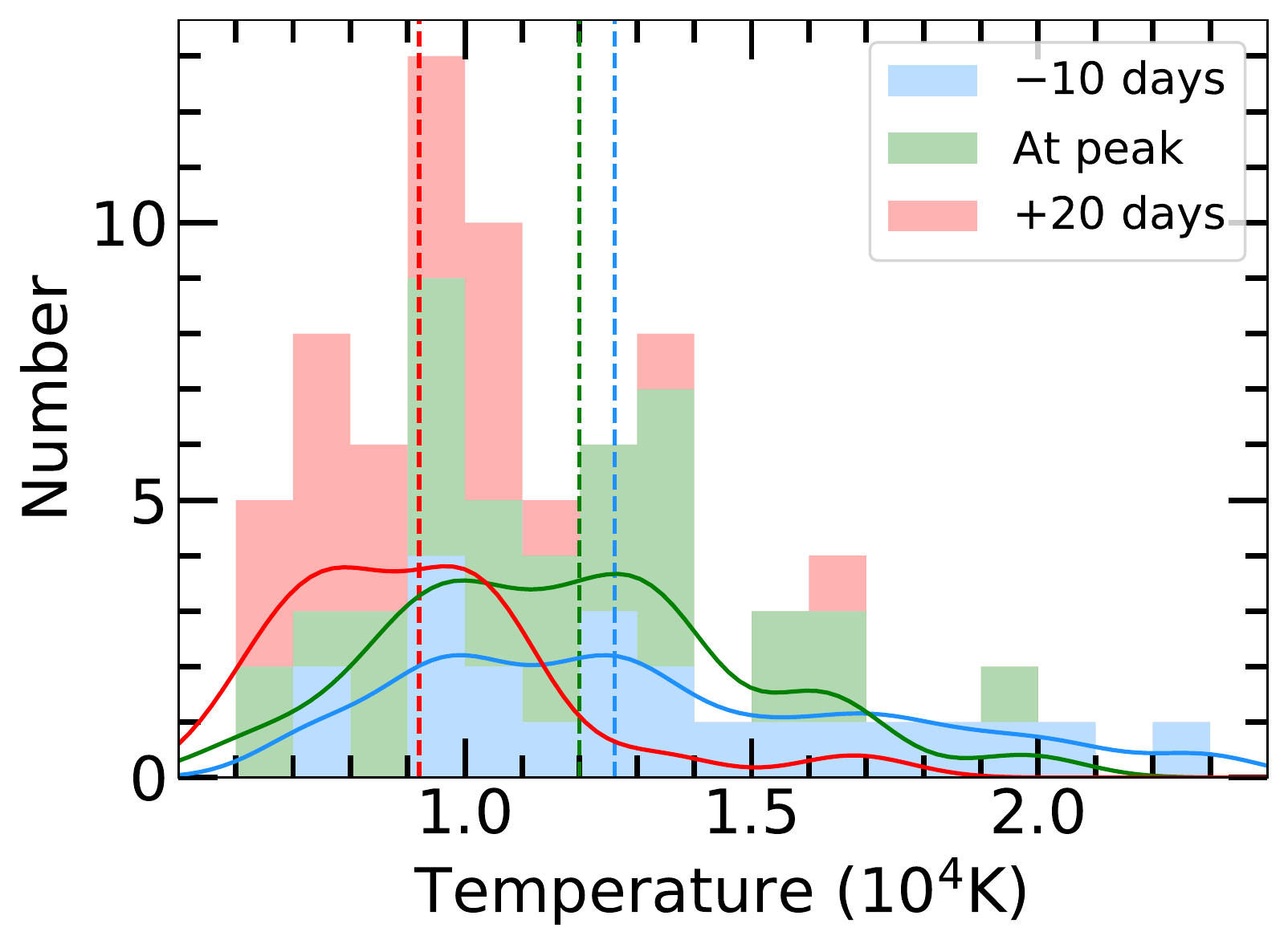}
\caption{The temperature distributions (including both ZTF and PS1 samples) at $-10$\,days, 0 days and $+20$\,days relative to the rest-frame $g$-band peak. The dashed lines mark the median temperature at different epochs, {\it i.e.} $T \sim 12600$\,K at $-10$\,days, $T \sim 12000$\,K at peak and $T \sim 9200$\,K at $+20$\,days. The solid lines show the kernel density estimation of the distribution.}
\label{fig:T_distribution}
\end{figure}

Combining the temperature measurements from the PS1 sample, we do see more low-temperature events. To better quantify the temperature distribution of SLSNe-I, we choose three typical epochs ({\it i.e.} $t \sim -10$\,days, 0 days and $+20$\,days relative to the rest-frame $g$-band peak) when sufficient data are available from the ZTF and PS1 sample. We show the temperature distribution at these epochs in Figure~\ref{fig:T_distribution}. 
At $t \sim -10$\,days, the temperature of SLSNe-I ranges from 7000 to 23000 K for different events, while this range is $6000 - 20000$\,K around the peak and $6000 - 12000$\,K (excluding one special event SN\,2019szu) at $t \sim +20$\,days. 
As SLSNe-I evolve, both the median value and the spread of temperature become smaller. This conclusion will still hold true if we expand the temporal range from $-20$ to $+40$\,days, according to the temperature trend shown in Figure~\ref{fig:Temperature}. Furthermore, the temperatures at any epochs in Figure~\ref{fig:T_distribution} are found to show flat and unimodal distributions, which indicates that SLSNe-I may have a continuous temperature distribution.

\begin{figure}[htp]
\centering
\includegraphics[width=0.5\textwidth]{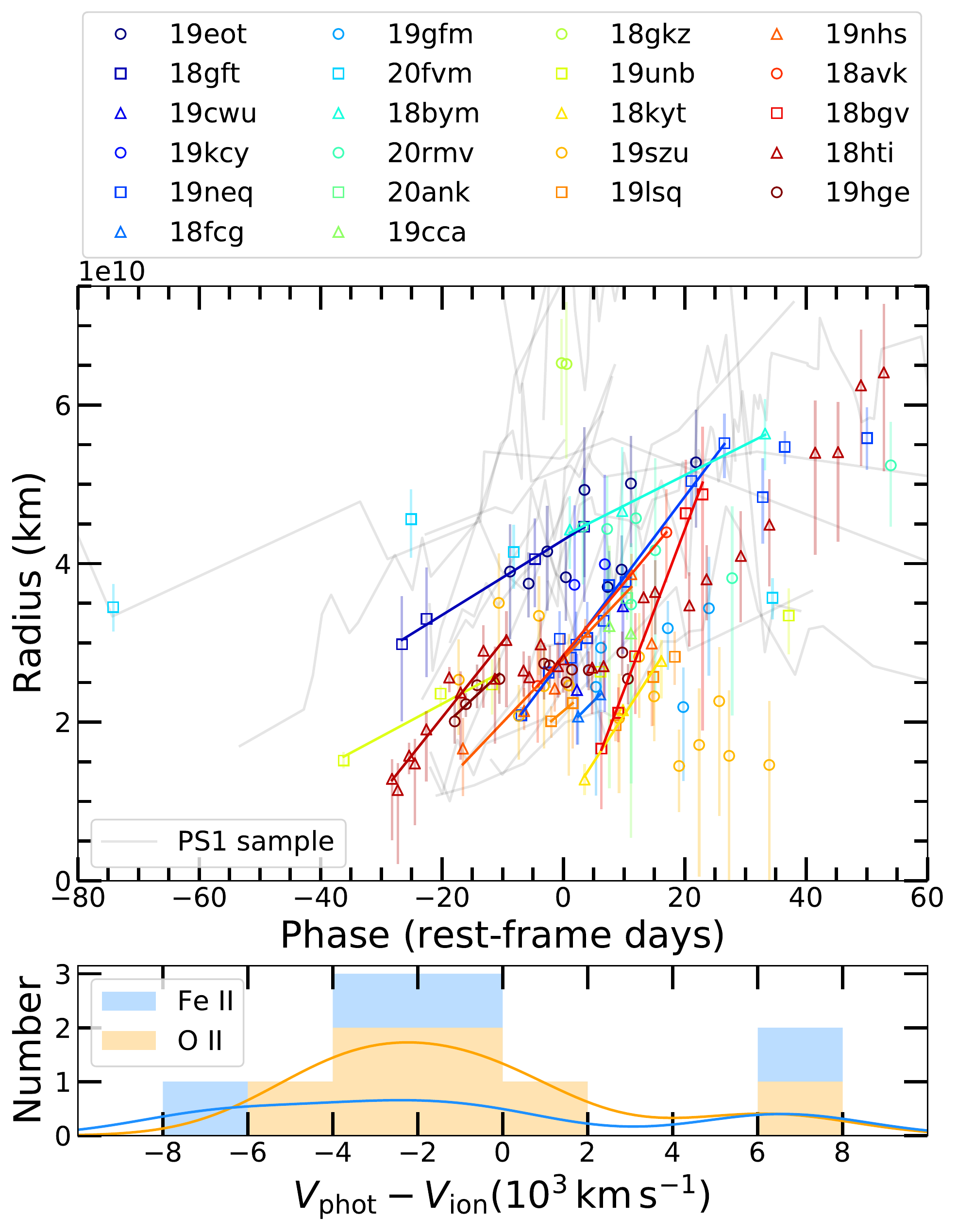}
\caption{\textbf{\emph{Top panel:}} The photosphere radius evolution as a function of time. Our measurements are plotted as open points while those for PS1 sample \citep{Lunnan2018} are shown in grey lines for comparison. For the events showing a linearly expanding photosphere before or at the peak, we apply a linear fit for them and plot the results with solid lines. 
\textbf{\emph{Bottom panel:}} The distribution of the velocity difference between those derived from photosphere radius ($V_{\rm phot}$) and those measured from the absorption lines in spectra ($V_{\rm ion}$). Blue and orange bars represent the value measured from \ion{Fe}{2} and \ion{O}{2}, respectively. The solid lines show the kernel density estimation of the distributions.}
\label{fig:radius}
\end{figure}

We also measure the blackbody radius of the photosphere. As shown in the top panel of Figure~\ref{fig:radius}, most events show a linearly expanding photosphere from explosion to $10 - 20$\,days around the peak. This is due to that the recession of the photosphere is negligible during this phase. We apply a linear fit to measure the photosphere velocity, $V_{\rm phot}$. In paper II, we measured the velocity from the \ion{Fe}{2} and \ion{O}{2} absorption lines in spectra. The velocity implied from the species $V_{\rm ion}$ are expected to be higher than $V_{\rm phot}$ since only the line features formed at higher velocity and lying outside the photosphere can be observed. As shown in the bottom panel of Figure~\ref{fig:radius}, the distribution of the $\Delta V = V_{\rm phot} - V_{\rm ion}$ proves that the most events have a negative $\Delta V$ and their $V_{\rm phot}$ is on average lower than $V_{\rm ion}$ by about $2000 - 3000 {\rm km\, s^{-1}}$ . Two outliers, SN\,2018bgv and SN\,2018kyt, are found to have significantly higher $V_{\rm phot}$ than $V_{\rm ion}$. SN\,2018bgv is the fastest-evolving event in our sample, while SN\,2018kyt also exhibits relatively fast-evolving behaviors. Thus the inconsistency between $V_{\rm phot}$ and $V_{\rm ion}$ may be due to that $V_{\rm phot}$ is the average velocity measured in a period of $10-20$\,days while $V_{\rm ion}$ is fast-evolving and measured at a single epoch. 
All of the measured temperature and radius values are listed in Table~\ref{tab:boloLC}. 

In summary, our measurements show that the SLSN-I have a wide range of temperatures, especially at early phases. 
While most SLSNe-I from ZTF have temperatures over 11000 K at the peak and cool down rapidly, there are also many lower-temperature and slowly-evolving events.  This indicates that the SEDs of SLSNe-I may have diverse shapes and different evolutionary tracks.

Finally, we discuss four peculiar events that have extraordinary color and temperature evolution as shown in Figures~\ref{fig:color} and \ref{fig:Temperature}.

SN\,2018hti is well sampled in UV and shows notably higher temperatures and much bluer $(g-r)$ colors compared with other events. Its LC can be well reproduced by a magnetar model \citep{Lin2020}. This event may represent some of the population with higher temperatures and bluer SEDs. However, we caution that this event has a very high Galactic extinction $E(B-V) = 0.4$\,mag. The dust extinction corrections in the UV-bands are highly uncertain. 

SN\,2020fvm has two almost equally bright LC peaks (at phase $\sim -60$ and $0$\,days, respectively) and the longest $1/e$-maxima rise time scale of 91 days.
Both its color and temperature evolutions are peculiar. The $(g-r)$ color around two peaks shows similar evolutionary trends, {\it i.e.} initially evolving from red to blue before the peak and then turning red after the peak. During the phase from $-75$\,days to $+35$\,days, its temperature remains almost constant of $\sim 10000$\,K, which could be due to the absence of sampling. Such a long time scale and double-peak evolution are unexpected for a simple radiative cooling or magnetar model. \citet{Dessart2019} examined the color evolution for various configurations of magnetar models, and found that almost all have fairly blue colors in the early phase and none shows the color change of $\sim 0.7$\,mag from red to blue. The ejecta interaction with an extended H-poor CSM could be a possible explanation. 

ZTF19aanesgt (SN\,2019cdt) has much redder color at +20\,days after the peak and it evolves much faster in comparison with other objects of our sample. No temperature evolution is computed due to the lack of UV data. This event is somewhat similar to SN\,2018bgv, both belonging to the fast-evolving and CSM-model-favored events (see LC modeling in Paper II), though it has redder $(g-r)$ color. Such a red color and fast LC evolution could also be consistent with a magnetar model with high kinetic energy, as shown in Figure~13 of \citet{Dessart2019}. Nevertheless, the \ion{Fe}{2} velocity and the kinetic energy of the ejecta derived for SN\,2019cdt are $\sim14400\,{\rm km\,s^{-1}}$ and $7.6\times10^{51}$\,erg, respectively, which are both higher than the average value (see Paper II). 

Finally, unlike any other SLSN-I, ZTF19acfwynw (SN\,2019szu) shows unusually high temperatures which are rising as the luminosity declines. At the early phases, its colors are blue, consistent with having high temperatures. Although the black-body temperatures at late time have large errors due to contamination of increasingly strong nebular emission lines, its spectral sequence has revealed a clear excess of UV continuum emission well past the peak phase. One possible explanation is CSM interaction, which could offer an additional heating source, boosting the emission at shorter wavelength. 
Another possibility is that the ejecta could be ionized by the ionizing flux from a long-lived central source \citep[{\it e.g.} a magnetar,][]{Margutti2017}. Ionization can increase the optical opacity dominated by the electron scattering and decrease the UV opacity dominated by line transitions of metals, which leads to a shift of the peak of the SED from optical to UV frequencies.

It is worth noting that SN\,2020fvm and SN\,2019cdt are poorly fit by simple magnetar models. It is possible that CSM interaction plays significant roles in these two systems. Additional modeling and analysis are presented in Paper II.

\subsection{Bolometric correction and bolometric LC}
\label{subsec:boloLC}
Bolometric light curve is an important indicator of the total radiative energy of a transient and sets constraints on the possible explosion mechanisms. Due to the high photospheric temperatures of SLSNe-I, UV radiation contributes to a large fraction of their emission \citep{Yan2017a,Yan2018}. Since only a small portion of the sample has UV data, we need to first derive an empirical bolometric correction (BC) relation, which can tie $g$- and $r$-band photometry to the bolometric luminosity, with ${\rm log}L_{\rm bol} = {\rm log}L_{gr} + \rm BC$ \footnote{Another definition of bolometric correction is $M_{\rm bolo} = M_K + {\rm BC} _K$ for $K$-band absolute magnitude $M_K$, where $M_{\rm bolo}$ is bolometric magnitude. Our correction is applied to two bands ($g$ and $r$), and we use the definition introduced in \S\ref{subsec:boloLC}.}. Here $L_{gr}$ is the sum of the $g$- and $r$-band luminosities. 

\citet{Lyman_2013} derived bolometric corrections for a sample of core-collapse SNe at low redshifts using their well-observed SEDs. We adopt a similar method and apply it to our sample. The basic concept is following. Bolometric correction is strongly influenced by the redshift and the temperature (slope of the spectra) at each epoch. As the observed $(g-r)$ color is also largely determined by the same two parameters, it is expected to see a correlation between the BC and the $(g-r)$ color. This correlation can serve as the basis of constructing bolometric LCs with only optical data.

We first compute the bolometric luminosity at the epochs when both UV and optical data are available, using the modified blackbody SED fit described in \S\ref{subsec:temperature}. The UV scaling factor $A$ is derived by fitting the UV to optical SEDs. For the epochs without UV data, we set $A=1.55$, which is the median value of the epochs with UV data. The bolometric luminosities are shown in Figure~\ref{fig:indboloLC_0} (only 15 events with at least three-epoch UV photometry) and listed in Table~\ref{tab:boloLC}.

\begin{figure}[htp]
\centering
\includegraphics[width=0.53\textwidth]{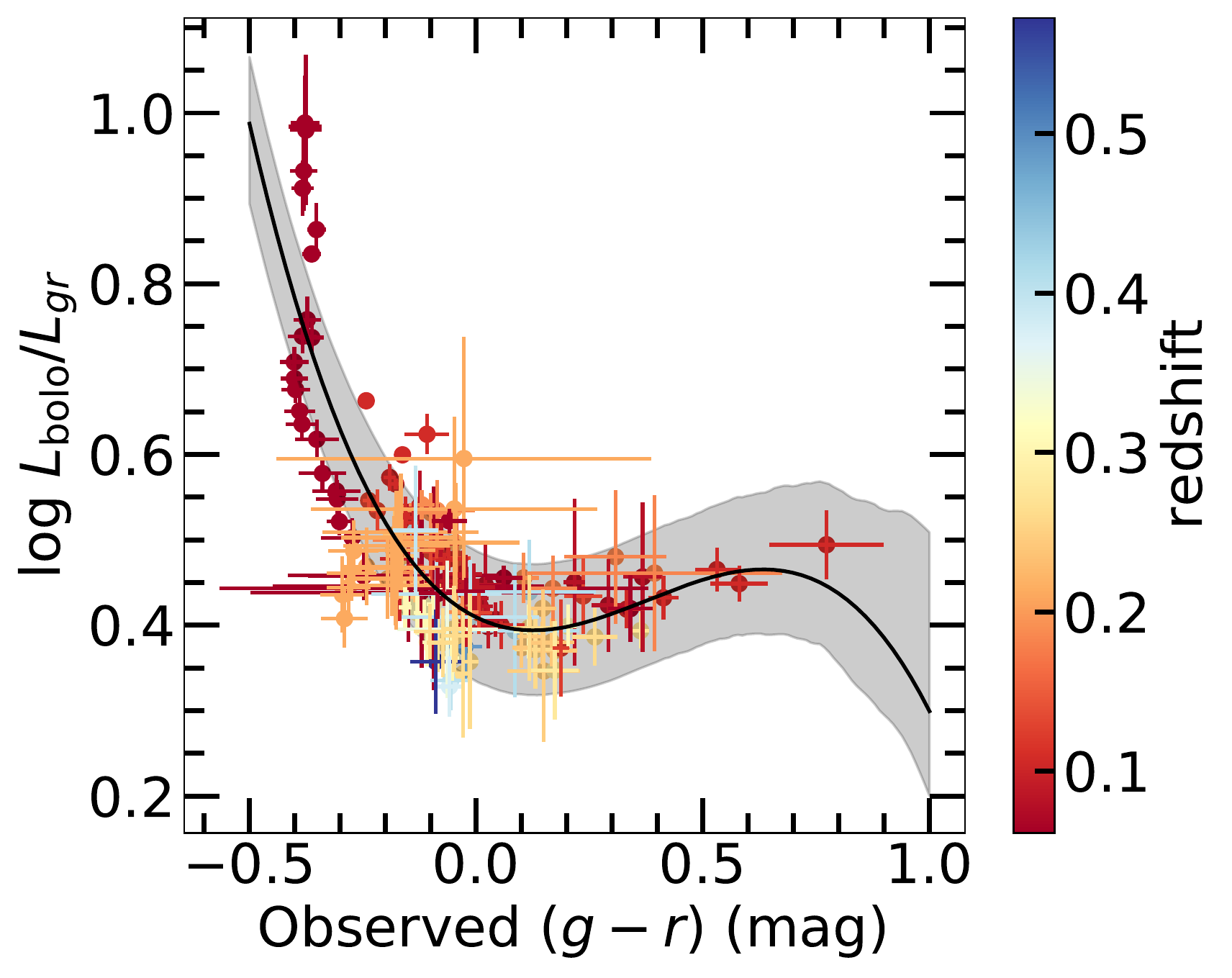}
\caption{Bolometric corrections (ratio between bolometric luminosity and the $gr$-band luminosity) derived from the events with both UV and $gr$-band photometry. X axis represents the $(g-r)$ color in the observed frame, where the Galactic and host-galaxy reddening have been corrected. The color bar on the right indicates the redshift of each data point. The black solid line shows the result of third-polynomial fit and the gray area shows $1\sigma$ uncertainty. 
}
\label{fig:bc}
\end{figure}

\begin{figure*}[htp]
\centering
\includegraphics[width=\textwidth]{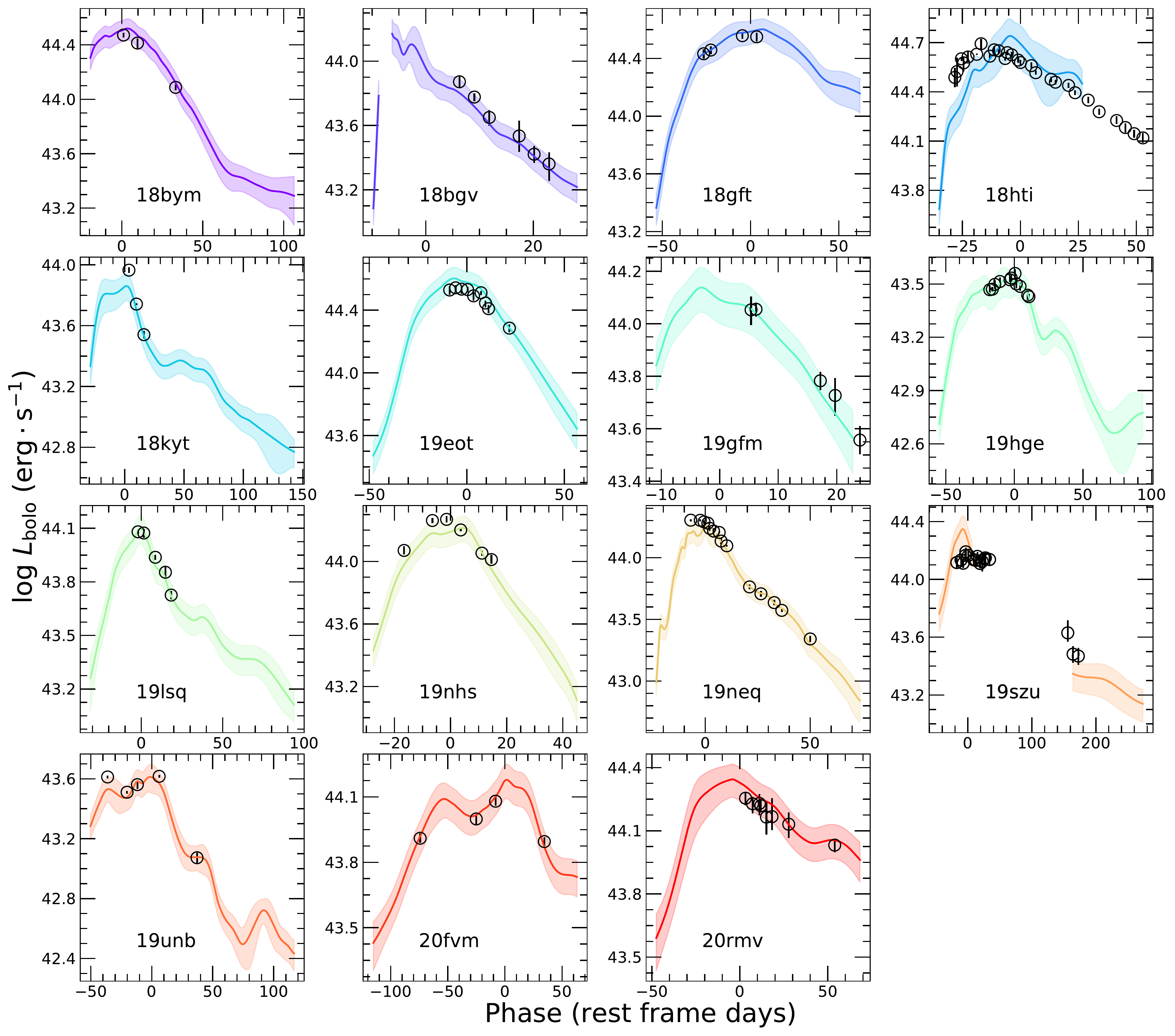}
\caption{Bolometric LCs for the events with at least three-epoch UV photometry. The black points present the bolometric luminosities constructed with both UV and optical photometry. The bolometric LCs derived from $g$- and $r$-band photometry are shown in solid lines with $1\sigma$ error marked by the shaded area. The breaks in LCs are due to the lack of data ({\it e.g.} SN\,2019szu) or that the ($g-r$) color is 0.1 mag beyond the range where the bolometric correction is reliable ({\it e.g.} SN\,2018bgv).}
\label{fig:indboloLC_0}
\end{figure*}

Figure~\ref{fig:bc} shows the observed $(g-r)$ color versus the bolometric correction, {\it i.e.} log$L_{\rm bol}/L_{gr}$, the luminosity ratio. We apply a third-order polynomial fit by optimizing a Gaussian likelihood function and derive an empirical relation as $\log (L_{\rm bolo} / L_{gr}) = -1.093\,x^3+1.244\,x^2-0.261\,x+0.410$, where $x=(g-r)$. Each point is weighted by the errors of both color and luminosity ratio. This equation can be used to compute the BC for low-$z$ SLSNe-I with similar $(g-r)$ color evolution as our sample. Note that our fit is applicable for $-0.4 < (g-r) < +0.9$\,mag, the phase range of $-74$\,days $<$ phase $< +173$\,days and $0.06 < z < 0.57$.  

As inferred from the residuals, the systematic error for the derived bolometric luminosity is about 19\%. The BC uncertainty  (shown as the shaded area in Figure~\ref{fig:bc}) is the combined errors from MCMC estimates and the systematic error. This error is the dominant one compared with other error sources like host-galaxy reddening and redshifts. In Figure~\ref{fig:indboloLC_0}, the bolometric luminosities constructed with both UV and optical photometry are consistent with the ones derived from $g$- and $r$-band photometry. This illustrates the reliability of our method.

\begin{figure*}[htp]
\centering
\includegraphics[width=\textwidth]{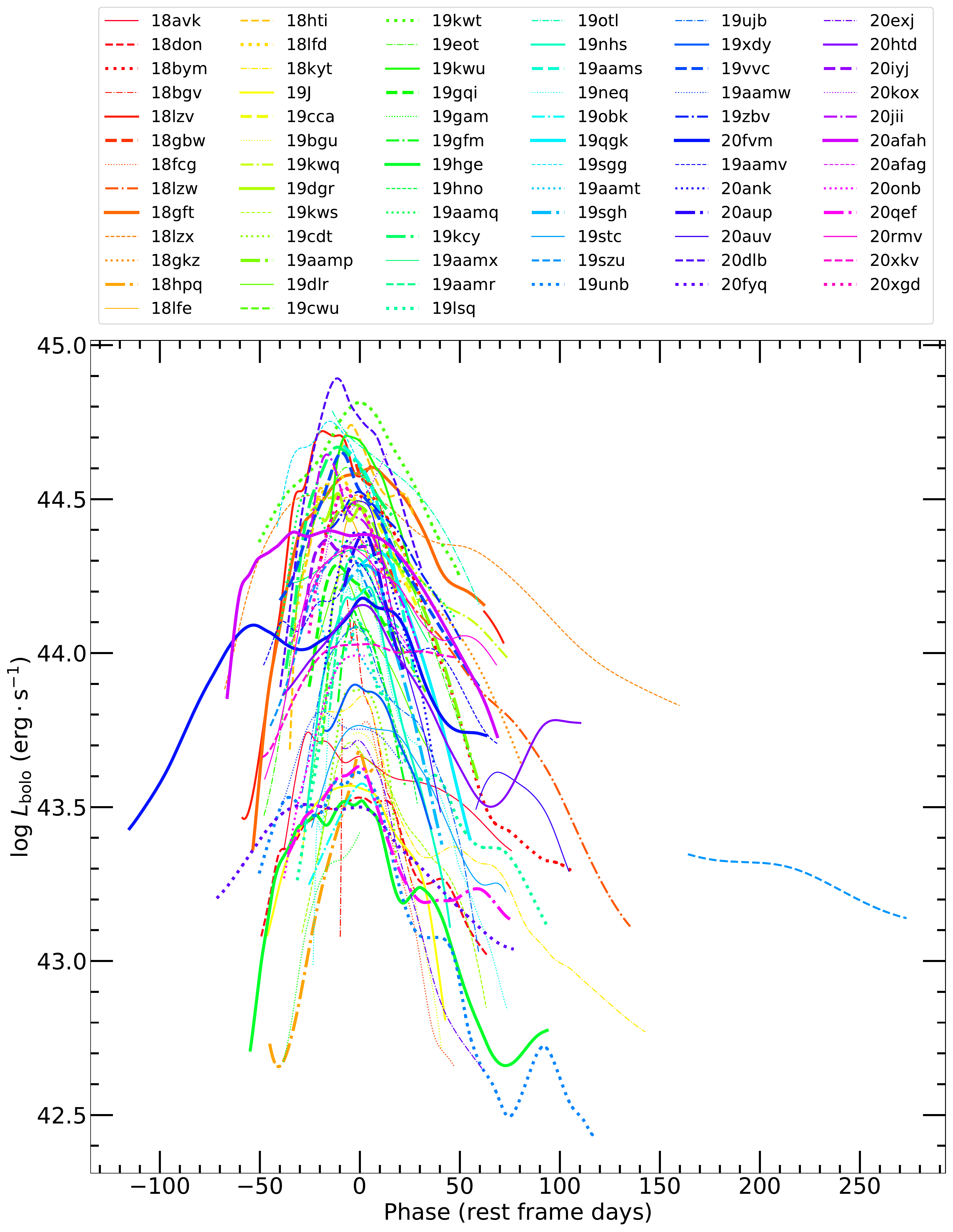}
\caption{The derived bolometric LCs from $g$- and $r$-band photometry for our whole sample. Due to the LC quality, SN\,2019fiy and SN\,2019aamu are excluded. The LCs are plotted in different colors, line styles and widths.}
\label{fig:boloLC}
\end{figure*}

The bolometric LCs derived from $g$- and $r$-band photometry are shown in Figure~\ref{fig:boloLC}. 
 The errors of both photometry and bolometric correction are combined together in quadrature to represent the error of the bolometric LCs. 
ZTF19aauiref (SN\,2019fiy) and SN\,2019aamu are excluded in this figure due to the bad photometry quality. Figure~\ref{fig:boloLC} illustrates a diversity in both peak luminosities and LC widths. Bumpy LCs are commonly seen and we perform detailed analysis in Paper II.

\begin{figure}[htp]
\centering
\includegraphics[width=0.5\textwidth]{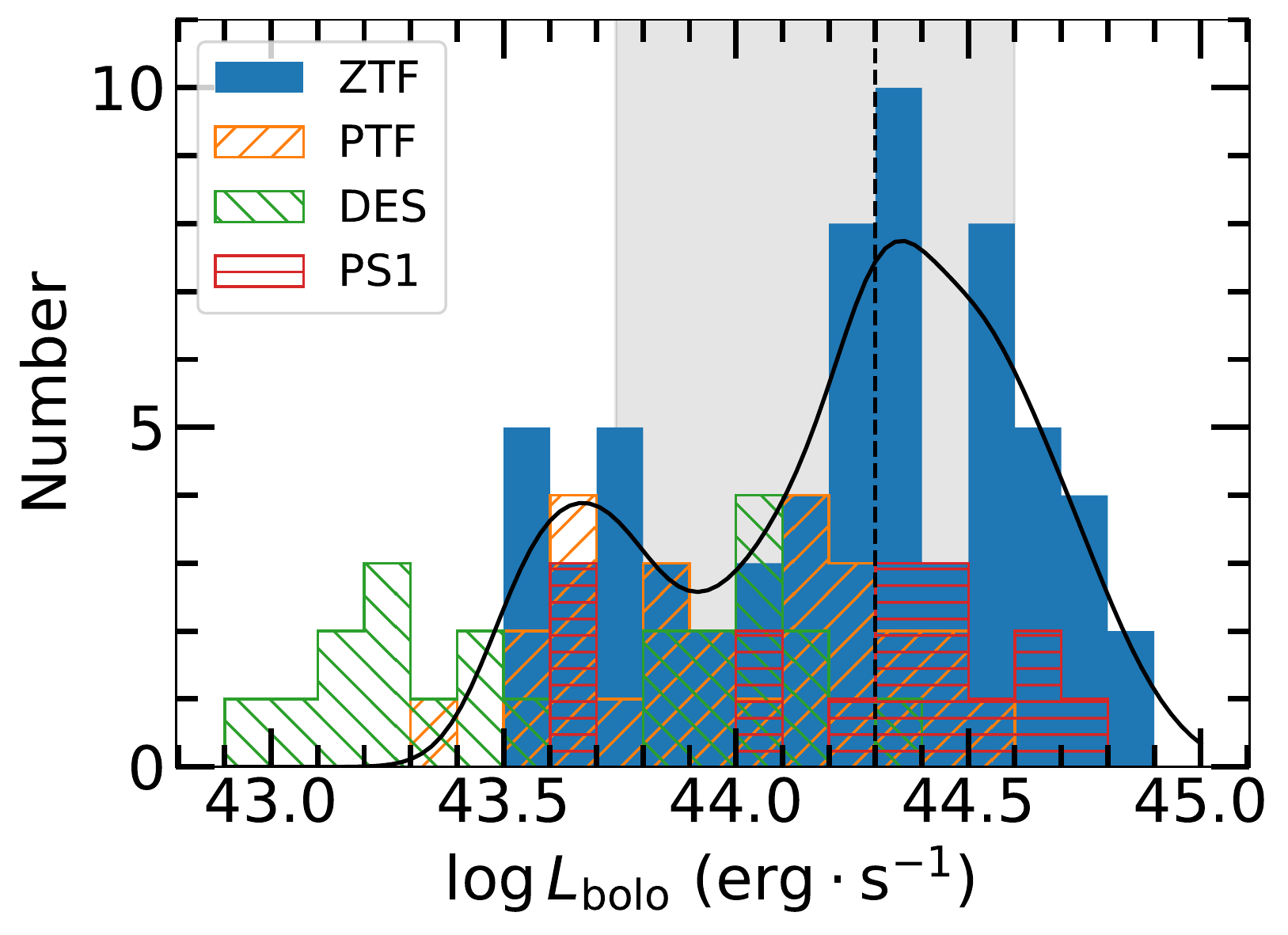}
\caption{The distribution of the peak bolometric luminosities. The black dashed line and the gray region show the median value and the $1\sigma$ dispersion of $2.00^{+1.97}_{-1.44} \times 10^{44}\,{\rm erg\, s^{-1}}$. The black solid line shows the kernel density estimation of the distribution. Other SLSN-I samples are plotted in different hatching patterns for comparison.}
\label{fig:peakbolo}
\end{figure}

The peak bolometric luminosities are tabulated in Table~\ref{tab:LCproperties} and the distribution is shown in Figure~\ref{fig:peakbolo}. The peak bolometric luminosities show a similar distribution as the absolute peak magnitudes. These are the raw distribution functions without any corrections for selection biases. Further studies are required to confirm such bimodal distributions. The peak bolometric luminosity spans from $3.22\times10^{43}$ to $7.80\times10^{44}\,{\rm erg\, s^{-1}}$, with a median value and the $1\sigma$ dispersion of $L_{\rm bolo,med} = 2.00^{+1.97}_{-1.44} \times 10^{44}\,{\rm erg\, s^{-1}}$. 
Our measurements are consistent with the results from the PTF and PS1 sample. Note that the DES sample tends to have lower luminosities, which could be due to that they are pseudo bolometric luminosities constructed from trapezoidal integration of photometry. 
Compared to the median peak luminosity of SNe\,Ib/c  \citep[$\sim 2\times10^{42}\,{\rm erg\, s^{-1}}$,][]{Prentice2016}, SLSNe-I are about 100 times more luminous, indicating different energy sources and/or explosion mechanisms from normal ccSNe.

\section{Conclusions}
\label{sec:summary}
The large sky coverage, high sensitivity and uniform cadence have made ZTF a very efficient discovery machine for superluminous supernovae. During the phase-I operation of ZTF survey, a total of 85 SLSNe-I were discovered. This paper presents 78 SLSNe-I whose LCs cover both pre- and post-peak phases. The other 7 SLSNe-I were still rising before October 30, 2021, and are not included in current analysis. 

This sample represents the largest sample of SLSNe-I at $z<1$ discovered from a single survey. 
Compared with the previous SLSN-I samples, our sample also has a better observing cadence on average. The large sample size and relative good observing cadence make it a good SLSN-I sample for statistical and detailed studies on SLSN-I light curves ({\it e.g.} precursor peak and post-peak bump features investigated in Paper II).

Based on this sample, we derive a bolometric correction relation for SLSNe-I, which allows a simple conversion between the optical $g$- and $r$-band photometry to the bolometric luminosity for low-$z$ SLSNe-I. The bolometric correction follows, 
$\log (L_{\rm bolo} / L_{gr}) = -1.093\,x^3+1.244\,x^2-0.261\,x+0.410$, where $x$ is the observed color $(g-r)$.


Our other findings are summarized as follows.

\begin{enumerate}

\item The rise time $t_{\rm rise,10\%}$ of our SLSN-I events has a mean value of $41.9\pm17.8$ days. Compared with SNe\,Ic, SLSNe-I have significantly longer rise times and wider dispersion \citep{Nicholl2015b,DeCia2018}. 
In our sample, there is one fast-evolving SLSN-I, SN 2018gbv, with $t_{\rm rise,10\%}$ shorter than 10 days; while there are only three slowly-evolving events with $t_{\rm rise,10\%} >$ 78\,days, which is about 5\% in our sample.
Compared with previous samples, the observed ratio of fast-rising ($t_{\rm rise,1/e} \lesssim 15$\,days) events in our sample is slightly higher, despite the target selection criteria disfavor the capture of fast-rising SNe. 

We confirm that the $t_{\rm rise,10\%}$ shows a continuous distribution and can not be divided into two clearly detached subclasses \citep{DeCia2018}. And as proven by many studies \citep{Nicholl2015b,Nicholl2017b,DeCia2018}, the $1/e$-maxima rise and decay time scales show a positive correlation, {\it i.e.} slowly-rising events tend to decay slowly, and the decay time scales are about $1.4-1.6$ times the rise time scales. 


\item The observed $(g-r)$ color at peak has a median value of $-0.03^{+0.12}_{-0.11}$\,mag and the rest-frame $(g-r)$ color $-0.21^{+0.19}_{-0.12}$\,mag. The majority of our SLSNe-I follows a wide color trend that evolves from blue ($(g-r) \sim -0.3$\,mag) at early phases to red, and reaches $(g-r) \sim +0.8$\,mag at about 2 months after the peak.  As \citet{Inserra2014} and \citet{DeCia2018} have proposed, we confirm that the peak rest-frame $(g-r)$ color is moderately correlated with the $g$-band absolute peak magnitudes, {\it i.e.} brighter SLSNe-I tend to have bluer color.


\item SLSNe-I have a wide range of temperatures at any given phase from $t \sim -20$ to $+40$\,days relative to the peak, especially at early phases, and the spread becomes smaller as SLSNe-I evolve. 
Based on the measurements from our sample and the PS1 sample, we suggest that the temperature of SLSNe-I at a given phase can have a continuous distribution.

\item We find four peculiar events that have extraordinary temperature and $(g-r)$ color evolution. SN\,2018hti shows a notably high temperature and blue color. SN\,2020fvm remains almost at a constant temperature of $10000\,K$ for over 100 days and has a double-peak color evolution like its LC. SN\,2019szu shows an unusual temperature evolution, which slowly rises from $13000$\,K at $-20$\,day to $20000$\,K at $+170$\,day. The color of SN\,2019cdt turns red much faster than that of any other event.

\item The absolute peak magnitudes of our SLSN-I sample are $-22.8\,{\rm mag} \leq M_g \leq -19.8$\,mag, with a median value and the $1\sigma$ error of $-21.48^{+1.13}_{-0.61}$\,mag. On average, the peak $M_g$ of SLSNe-I is around 4 mag and 3 mag brighter than normal SNe\,Ic and SNe\,Ic-BL, respectively. The peak bolometric luminosities of our sample distribute from $3.22\times10^{43}$ to $7.80\times10^{44}\,{\rm erg\, s^{-1}}$ and have a median value with the $1\sigma$ dispersion of $2.00^{+1.97}_{-1.44}\times10^{44}\,{\rm erg\, s^{-1}}$. 


\end{enumerate}

\acknowledgements

We acknowledge very helpful discussions about data processing of P200 photometry with Shengyu Yan from Tsinghua University. 

Based on observations obtained with the Samuel Oschin Telescope 48-inch and the 60-inch Telescope at the Palomar Observatory as part of the Zwicky Transient Facility project. ZTF is supported by the National Science Foundation under Grant No. AST-1440341 and a collaboration including Caltech, IPAC, the Weizmann Institute for Science, the Oskar Klein Center at Stockholm University, the University of Maryland, the University of Washington, Deutsches Elektronen-Synchrotron and Humboldt University, Los Alamos National Laboratories, the TANGO Consortium of Taiwan, the University of Wisconsin at Milwaukee, and Lawrence Berkeley National Laboratories. Operations are conducted by COO, IPAC, and UW. SED Machine is based upon work supported by the National Science Foundation under Grant No. 1106171. The ZTF forced-photometry service was funded under the Heising-Simons Foundation grant \#12540303 (PI: Graham). 

The Liverpool Telescope is operated on the island of La Palma by Liverpool John Moores University in the Spanish Observatorio del Roque de los Muchachos of the Instituto de Astrofisica de Canarias with financial support from the UK Science and Technology Facilities Council. The Nordic Optical Telescope is owned in collaboration by the University of Turku and Aarhus University, and operated jointly by Aarhus University, the University of Turku and the University of Oslo, representing Denmark, Finland and Norway, the University of Iceland and Stockholm University at the Observatorio del Roque de los Muchachos, La Palma, Spain, of the Instituto de Astrofisica de Canarias. This research has made use of data obtained through the High Energy Astrophysics Science Archive Research Center Online Service, provided by the NASA/Goddard Space Flight Center. This work was supported by the GROWTH project funded by the National Science Foundation under Grant No. 1545949. 

Z.~Chen acknowledges support from the China Scholarship Council. T.~K. acknowledges support from the Swedish National Space Agency and the Swedish Research Council. S.~Schulze acknowledges support from the G.R.E.A.T research environment, funded by {\em Vetenskapsr\aa det},  the Swedish Research Council, project number 2016-06012.  R.~L. acknowledges support from a Marie Sk\l{}odowska-Curie Individual Fellowship within the Horizon 2020 European Union (EU) Framework Programme for Research and Innovation (H2020-MSCA-IF-2017-794467). A.~Gal-Yam acknowledges support from the EU via ERC grant No. 725161, the ISF GW excellence center, an IMOS space infrastructure grant and BSF/Transformative and GIF grants, as well as the André Deloro Institute for Advanced Research in Space and Optics, the Schwartz/Reisman Collaborative Science Program and the Norman E Alexander Family M Foundation ULTRASAT Data Center Fund, Minerva and Yeda-Sela. The work of X.~Wang is supported by the National Natural Science Foundation of China (NSFC grants 12288102, 12033003 and 11633002), the Major State Basic Research Development Program (grant 2016YFA0400803), the Scholar Program of Beijing Academy of Science and Technology (DZ:BS202002), and the Tencent XPLORER Prize.

\facilities{PO:1.2m, PO:1.5m, PO:Hale, Liverpool:2m, NOT:2.56m, Keck:I, WHT:4.2m}

\software{Scikit-learn \citep{scikit-learn}, {\it FIREFLY} \citep{Wilkinson_2017}, {\tt george} \citep{Ambikasaran2015}, 
{\tt HEASoft} \citep[][\url{https://heasarc.gsfc.nasa.gov/lheasoft/download.html}]{HEAsoft2014}, 
SEDM pipeline \citep{Rigault2019}, 
pyraf-dbsp \citep{Bellm2016}, 
DBSP\_DRP \citep{Roberson2022}, 
LPipe \citep{Perley_2019}, 
Fpipe \citep{Fremling2016}, 
AutoPhOT \citep{Brennan2022}.
}
\hfill
\clearpage

\appendix

\newcounter{Atable}
\setcounter{Atable}{0}
\newcounter{Afigure}
\setcounter{Afigure}{0}

\renewcommand\thefigure{\Alph{section}\arabic{Afigure}} 
\renewcommand\thetable{\Alph{section}\arabic{Atable}} 

\section{Information on the ZTF SLSN-I sample}

\startlongtable
\addtocounter{Atable}{1}
\begin{deluxetable*}{lcccccccccc}
\tabletypesize{\scriptsize}
\tablecaption{The ZTF SLSN-I Sample}
\tablehead{\multirow{2}{*}[-4pt]{ZTF Name} &
\multirow{2}{*}[-4pt]{IAU Name} &
\multirow{2}{*}[-4pt]{RA$^a$} &
\multirow{2}{*}[-4pt]{DEC$^a$} &
\multirow{2}{*}[-4pt]{Redshift} &
\colhead{$E(B-V)^{b}$} &
\colhead{Discovery} &
\multirow{2}{*}[-4pt]{Template} &
\colhead{Spectrum} &
\colhead{Template} &
\colhead{Template} \\
&&&&&
(mag) &
group &
&
Phase$^c$ &
Phase$^c$ &
Type
} 

\startdata
ZTF18aaisyyp & SN\,2018avk & 13:11:27.72 & +65:38:16.7 & 0.132 & 0.011 & Gaia & PTF12gty & -13.5 & -7 & SLSN-I \\
ZTF18aajqcue & SN\,2018don & 13:55:08.65 & +58:29:42.0 & 0.0735 & 0.0089 & PS1 & SN2007gr & -2.1 & -7 & Ic \\
ZTF18aapgrxo & SN\,2018bym & 18:43:13.42 & +45:12:28.2 & 0.2744 & 0.0517 & ATLAS & PTF13ajg & +5.4 & +8 & SLSN-I \\
ZTF18aavrmcg & SN\,2018bgv & 11:02:30.29 & +55:35:55.8 & 0.0795 & 0.0074 & Gaia & SN2011ke & +17.8 & +20 & SLSN-I \\
ZTF18aazgrfl & SN\,2018lzv & 12:44:02.32 & +56:01:44.5 & 0.434 & 0.0081 & ZTF & PTF09atu & -19.1 & -20 & SLSN-I \\
ZTF18abjwagv & SN\,2018gbw & 15:55:38.02 & +28:21:38.0 & 0.3454 & 0.0386 & PS1 & PTF13ajg & -4.1 & -4 & SLSN-I \\
ZTF18abmasep & SN\,2018fcg & 21:09:36.78 & +33:28:59.6 & 0.1011 & 0.1435 & ZTF & PTF12dam & -4.4 & -22 & SLSN-I \\
ZTF18abrzcbp & SN\,2018lzw & 07:39:32.76 & +27:44:02.7 & 0.3198 & 0.0373 & ZTF & PTF10uhf & +27.6 & +15 & SLSN-I \\
ZTF18abshezu & SN\,2018gft & 23:57:17.95 & -15:37:53.3 & 0.2320 & 0.0256 & ZTF & PTF09cnd & -38.4 & -14 & SLSN-I \\
ZTF18abszecm & SN\,2018lzx & 22:29:27.24 & +13:10:39.8 & 0.4373 & 0.0529 & ZTF & PTF09cnd & +19.2 & +37 & SLSN-I \\
ZTF18abvgjyl & SN\,2018gkz & 07:58:11.54 & +19:31:07.9 & 0.2405 & 0.036 & ZTF & SN2007bi & +10.6 & +50 & SLSN-I \\
ZTF18acapyww & SN\,2018hpq & 18:28:41.24 & +75:48:47.3 & 0.124 & 0.0868 & Gaia & PTF10nmn & +25.0 & max & SLSN-I \\
ZTF18acenqto & SN\,2018ibb & 04:38:56.93 & -20:39:44.2 & 0.166 & 0.0275 & ATLAS & PTF12dam & -4.0 & +7 & SLSN-I \\
ZTF18acqyvag & SN\,2018lfe & 09:33:29.56 & +00:03:08.4 & 0.3505 & 0.0286 & PS1 & SN2011ke & +59.6 & +53 & SLSN-I \\
ZTF18acslpji & SN\,2018hti & 03:40:53.77 & +11:46:37.9 & 0.0613 & 0.3983 & ATLAS & PTF12dam & +21.2 & +7 & SLSN-I \\
ZTF18acxgqxq & SN\,2018lfd & 23:14:59.32 & +48:45:27.6 & 0.2686 & 0.1506 & ZTF & PTF12dam & -7.5 & -22 & SLSN-I \\
ZTF18acyxnyw & SN\,2018kyt & 12:27:56.23 & +56:23:35.6 & 0.1080 & 0.0091 & ZTF & PTF10hgi & +45.5 & +47 & SLSN-I \\
ZTF19aacxrab & SN\,2019J & 10:03:46.78 & +06:46:24.7 & 0.1346 & 0.0229 & PS1 & PTF10gvb & +8.8 & -6 & SLSN-I \\
ZTF19aajwogx & SN\,2019cca & 12:02:50.91 & -16:39:53.6 & 0.4103 & 0.0462 & ZTF & PTF10uhf & +28.5 & +15 & SLSN-I \\
ZTF19aaknqmp & SN\,2019bgu & 09:57:15.34 & +32:00:05.6 & 0.1480 & 0.0123 & ATLAS & PTF12gty & +16.6 & -7 & SLSN-I \\
ZTF19aalbrph & SN\,2019kwq & 17:07:58.84 & +58:42:03.9 & 0.49$\star$ & 0.0257 & ZTF & PTF09cnd & +68.9 & +37 & SLSN-I \\
ZTF19aamhast & SN\,2019dgr & 09:45:32.68 & +04:56:02.2 & 0.3815 & 0.0348 & ATLAS & PTF13ajg & +13.2 & +8 & SLSN-I \\
ZTF19aamhhiz & SN\,2019kws & 14:15:04.46 & +50:39:06.8 & 0.1977 & 0.0142 & ZTF & PTF12gty & +21.7 & -7 & SLSN-I \\
ZTF19aanesgt & SN\,2019cdt & 08:17:53.90 & +65:28:46.7 & 0.153 & 0.0441 & ZTF & PTF10nmn & +11.6 & +29 & SLSN-I \\
ZTF19aantokv & SN\,2019aamp & 14:37:49.27 & +20:18:16.6 & 0.4040 & 0.0237 & ZTF & PTF13ajg & +8.4 & +8 & SLSN-I \\
ZTF19aaohuwc & SN\,2019dlr & 11:17:34.18 & +00:30:02.6 & 0.26$\star$ & 0.0309 & ZTF & SN2004aw & +57.7 & +11 & Ic \\
ZTF19aapaeye & SN\,2019cwu & 14:51:37.29 & +48:59:13.7 & 0.32$\star$ & 0.0186 & ZTF & SN2011ke & +27.6 & +20 & SLSN-I \\
ZTF19aaqrime & SN\,2019kwt & 19:39:22.59 & +78:45:43.8 & 0.3562 & 0.0775 & ZTF & SN2007bi & +11.3 & +50 & SLSN-I \\
ZTF19aarphwc & SN\,2019eot & 18:00:29.95 & +50:17:43.3 & 0.3057 & 0.0366 & ZTF & PTF13ajg & -13.8 & -4 & SLSN-I \\
ZTF19aaruixj & SN\,2019kwu & 13:57:39.77 & +64:21:18.6 & 0.60$\star$ & 0.0138 & ZTF & PTF10vqv & +28.4 & +11 & SLSN-I \\
ZTF19aasdvfr & SN\,2019gqi & 14:21:11.98 & +28:54:05.9 & 0.3642 & 0.0129 & ATLAS & PTF11rks & +21.3 & +7 & SLSN-I \\
ZTF19aauiref & SN\,2019fiy & 14:05:46.73 & +33:27:38.3 & 0.67$\star$ & 0.0142 & PS1 & PTF09atu & +15.6 & +28 & SLSN-I \\
ZTF19aauvzyh & SN\,2019gam & 10:19:18.32 & +17:12:42.6 & 0.1235 & 0.0272 & ATLAS & SN2011ke & +17.6 & +29 & SLSN-I \\
ZTF19aavouyw & SN\,2019gfm & 15:35:46.59 & +24:03:45.0 & 0.1816 & 0.0465 & PS1 & PTF11rks & +16.8 & +18 & SLSN-I \\
ZTF19aawfbtg & SN\,2019hge & 22:24:21.20 & +24:47:17.1 & 0.0866 & 0.058 & ZTF & PTF12dam & -25.7 & +2 & SLSN-I \\
ZTF19aawsqsc & SN\,2019hno & 19:39:12.95 & +62:43:41.0 & 0.26 & 0.0583 & ATLAS & PTF09cnd & +1.6 & +9 & SLSN-I \\
ZTF19aayclnm & SN\,2019aamq & 20:55:36.14 & -08:40:31.4 & 0.386 & 0.0636 & ZTF & PTF10bjp & +3.8 & +8 & SLSN-I \\
ZTF19abaeyqw & SN\,2019kcy & 14:08:19.78 & +08:58:01.0 & 0.399 & 0.0227 & ZTF & PTF13ajg & -19.3 & -4 & SLSN-I \\
ZTF19abcvwrz & SN\,2019aamx & 15:57:48.27 & +27:28:03.5 & 0.41 & 0.0362 & ZTF & PTF09cwl & +13.0 & +22 & SLSN-I \\
ZTF19abdlzyq & SN\,2019aamr & 15:29:23.55 & +38:06:12.6 & 0.42$\star$ & 0.0118 & ZTF & PTF09cwl & +8.0 & +22 & SLSN-I \\
ZTF19abfvnns & SN\,2019lsq & 00:04:40.58 & +42:52:11.3 & 0.1295 & 0.0829 & ATLAS & PTF12dam & -3.3 & -22 & SLSN-I \\
ZTF19abkfshj & SN\,2019otl & 02:52:21.63 & -17:48:12.4 & 0.500 & 0.0241 & ZTF & PTF13ajg & +11.2 & -4 & SLSN-I \\
ZTF19abnacvf & SN\,2019nhs & 00:52:01.44 & +07:36:59.7 & 0.189 & 0.0514 & PS1 & PTF13ajg & -8.6 & -4 & SLSN-I \\
ZTF19abnqqdp & SN\,2019aams & 23:43:36.16 & +12:29:01.0 & 0.6360 & 0.0448 & ZTF & PTF13ajg & +0.8 & -4 & SLSN-I \\
ZTF19abpbopt & SN\,2019neq & 17:54:26.76 & +47:15:40.6 & 0.1060 & 0.0334 & ZTF & PTF11rks & +16.1 & +7 & SLSN-I \\
ZTF19abrbsvm & SN\,2019obk & 22:33:54.08 & -02:09:42.3 & 0.1656 & 0.048 & PS1 & PTF12gty & +12.8 & -7 & SLSN-I \\
ZTF19abuolvj & SN\,2019qgk & 22:29:57.55 & -04:06:02.2 & 0.3468 & 0.054 & ZTF & SN2010gx & +47.2 & +29 & SLSN-I \\
ZTF19abuyuwa & SN\,2019sgg & 01:01:11.77 & +14:01:35.4 & 0.5726 & 0.0373 & ZTF & PTF13ajg & -18.6 & -4 & SLSN-I \\
ZTF19abzoyeg & SN\,2019aamt & 21:15:08.00 & +32:43:01.3 & 0.2138 & 0.126 & ZTF & PTF12dam & +15.2 & +7 & SLSN-I \\
ZTF19abzqmau & SN\,2019sgh & 01:12:39.42 & +36:28:24.8 & 0.3436 & 0.0599 & ZTF & PTF10uhf & +39.8 & +15 & SLSN-I \\
ZTF19acbonaa & SN\,2019stc & 06:54:23.10 & +17:29:31.4 & 0.1178 & 0.072 & ZTF & PTF10hgi & +25.6 & +47 & SLSN-I \\
ZTF19acfwynw & SN\,2019szu & 00:10:13.14 & -19:41:32.4 & 0.2120 & 0.018 & ATLAS & PTF12dam & -15.2 & +2 & SLSN-I \\
ZTF19acgjpgh & SN\,2019unb & 09:47:57.02 & +00:49:36.0 & 0.0635 & 0.1045 & ZTF & PTF10hgi & -7.5 & +15 & SLSN-I \\
ZTF19ackjrru & SN\,2019ujb & 09:03:15.18 & +40:14:32.6 & 0.2008 & 0.0165 & ZTF & PTF11rks & +25.7 & +18 & SLSN-I \\
ZTF19acsajxn & SN\,2019xdy & 08:24:51.33 & +22:10:46.0 & 0.2206 & 0.0425 & ZTF & PTF10vwg & +31.6 & +22 & SLSN-I \\
ZTF19acucxij & SN\,2019vvc & 09:13:30.13 & +44:46:26.2 & 0.3314 & 0.0115 & ATLAS & PTF09cwl & +16.0 & +22 & SLSN-I \\
ZTF19acvxquk & SN\,2019aamu & 02:55:08.89 & +11:27:22.4 & 0.2590 & 0.1925 & ZTF & PTF12gty & +8.6 & -7 & SLSN-I \\
ZTF19adaivcf & SN\,2019zbv & 10:15:01.10 & +43:24:53.6 & 0.3785 & 0.0119 & ZTF & PTF13ajg & +10.4 & +8 & SLSN-I \\
ZTF20aahbfmf & SN\,2020ank & 08:16:14.65 & +04:19:26.9 & 0.2485 & 0.0193 & ZTF & PTF13ajg & -0.3 & -4 & SLSN-I \\
ZTF20aaifybu & SN\,2020auv & 16:34:12.51 & +13:05:51.9 & 0.280 & 0.0565 & ZTF & PTF13ajg & -2.4 & +8 & SLSN-I \\
ZTF19acujvsi & SN\,2019aamw & 23:48:54.54 & +24:59:59.8 & 0.22$\star$ & 0.0526 & ZTF & PTF10vwg & +137.6 & +147 & SLSN-I \\
ZTF20aadzbcf & SN\,2020fvm & 14:12:45.93 & +34:44:16.2 & 0.2428 & 0.0129 & ZTF & PTF10uhf & -1.4 & +15 & SLSN-I \\
ZTF20aagikvv & SN\,2019aamv & 12:45:01.65 & +33:33:14.1 & 0.3996 & 0.0127 & ZTF & PTF13ajg & -11.4 & +8 & SLSN-I \\
ZTF20aahrxgw & SN\,2020aup & 13:09:44.44 & +12:29:13.4 & 0.31$\star$ & 0.0208 & ZTF & PTF09cwl & +16.2 & +22 & SLSN-I \\
ZTF20aaoqwpo & SN\,2020dlb & 08:08:34.14 & +34:44:12.9 & 0.398 & 0.0327 & ZTF & SN2005ap & +0.2 & +4 & SLSN-I \\
ZTF20aapaecd & SN\,2020fyq & 14:46:10.44 & +23:48:02.0 & 0.1765 & 0.0325 & PS1 & SN2011ke & +27.5 & +53 & SLSN-I \\
ZTF20aattyuz & SN\,2020exj & 14:42:40.01 & +30:14:39.1 & 0.1216 & 0.0096 & ATLAS & PTF10hgi & +44.2 & +47 & SLSN-I \\
ZTF20aauoudz & SN\,2020htd & 17:44:17.28 & +38:55:30.4 & 0.3515 & 0.026 & PS1 & PTF09cnd & +7.4 & +13 & SLSN-I \\
ZTF20aavfbqz & SN\,2020iyj & 09:15:36.64 & +53:27:32.0 & 0.3690 & 0.0155 & ZTF & PTF09cwl & +1.9 & +1 & SLSN-I \\
ZTF20aavqrzc & SN\,2020kox & 11:06:04.97 & +26:17:28.7 & 0.456 & 0.0166 & ATLAS & PTF13ajg & +2.6 & -4 & SLSN-I \\
ZTF20aawfxlt & SN\,2020jii & 15:34:55.31 & +02:51:11.5 & 0.396 & 0.0427 & PS1 & PTF13ajg & -1.8 & -4 & SLSN-I \\
ZTF20aawkgxa & SN\,2020afah & 10:20:18.32 & +53:19:21.4 & 0.3754 & 0.0086 & ZTF & PTF09cnd & -41.8 & -16 & SLSN-I \\
ZTF20abisijg & SN\,2020afag & 00:15:46.25 & +47:00:08.5 & 0.3815 & 0.0898 & ZTF & PTF09cwl & +6.7 & +22 & SLSN-I \\
ZTF20abjwjrx & SN\,2020onb & 14:23:00.61 & +49:10:40.7 & 0.16$\star$ & 0.0208 & ZTF & SN2004aw & +24.8 & +4 & Ic \\
ZTF20ablkuio & SN\,2020qef & 22:56:10.53 & +28:45:53.3 & 0.1831 & 0.0459 & ATLAS & PTF12gty & +24.4 & -7 & SLSN-I \\
ZTF20abpuwxl & SN\,2020rmv & 00:40:00.19 & -14:35:25.1 & 0.2621 & 0.0185 & ATLAS & PTF10uhf & +13.2 & +15 & SLSN-I \\
ZTF20abzaacf & SN\,2020xkv & 22:37:46.00 & +23:31:37.4 & 0.2410 & 0.0303 & ATLAS & SN2011ke & -8.7 & +20 & SLSN-I \\
ZTF20aceqspy & SN\,2020xgd & 00:19:45.83 & +05:08:18.7 & 0.455 & 0.0145 & PS1 & PTF13ajg & -4.4 & -4 & SLSN-I \\
\enddata

\hspace*{\fill} \\
$^a$In J2000.\\
$^b$From \citet{Schlafly_2011}.\\
$^c$In rest-frame days.\\
$\star$ means the redshift is fit by \textit{superfit}.
\label{tab:basicinfo}
\end{deluxetable*}

\startlongtable
\addtocounter{Atable}{1}

\begin{deluxetable*}{lccc}
\tabletypesize{\scriptsize}
\tablecaption{Classification spectra}
\tablehead{
\colhead{Name} &
\colhead{Observing date} &
\colhead{Telescope + instrument} &
\colhead{Exposure time$^a$ (s)}}

\startdata
SN\,2018avk & 2018-05-04 & NOT+ALFOSC & $1800$ \\
SN\,2018don & 2018-06-09 & P200+DBSP & $2\times600/2\times600$ \\
SN\,2018bym & 2018-06-12 & P200+DBSP & $600/600$ \\
SN\,2018bgv & 2018-06-04 & NOT+ALFOSC & $2\times2400$ \\
SN\,2018lzv & 2018-07-14 & WHT+ISIS & $600$ \\
SN\,2018gbw & 2018-08-13 & P200+DBSP & $600/600$ \\
SN\,2018fcg & 2018-08-21 & P200+DBSP & $600/600$ \\
SN\,2018lzw & 2018-10-06 & P200+DBSP & $610/400$ \\
\enddata
\hspace*{\fill} \\
$^a$Slash-separated values indicate the exposure time of blue and red sides, respectively.\\
(This table is available in its entirety in machine-readable form.)
\label{tab:spec}
\end{deluxetable*}

\startlongtable
\addtocounter{Atable}{1}

\begin{deluxetable*}{lccccccc}
\tabletypesize{\scriptsize}
\tablecaption{Photometry data}
\tablehead{\multirow{2}{*}[-4pt]{Name} &
\colhead{MJD} &
\multirow{2}{*}[-4pt]{Filter} &
\colhead{Mag(AB)} &
\colhead{Error$_{\rm mag}$} &
\multirow{2}{*}[-4pt]{Label$^a$} &
\colhead{Flux$_{\rm ratio} \, ^b$} &
\multirow{2}{*}[-4pt]{Tel.+Ins.$^c$} \\
&
(days) &
&
(mag) &
(mag) &
&
($10^{-9}$) &
} 

\startdata
SN\,2018avk & 58202.30 & g & 20.48 & 0.21 & F & 6.40 & P48+ZTF \\
SN\,2018avk & 58202.32 & g & 20.64 & 0.18 & F & 5.53 & P48+ZTF \\
SN\,2018avk & 58202.32 & g & 20.74 & 0.16 & F & 5.04 & P48+ZTF \\
SN\,2018avk & 58202.34 & g & 20.78 & 0.19 & F & 4.86 & P48+ZTF \\
SN\,2018avk & 58202.36 & g & 20.92 & 0.14 & F & 4.27 & P48+ZTF \\
SN\,2018avk & 58202.36 & g & 20.80 & 0.21 & F & 4.77 & P48+ZTF \\
SN\,2018avk & 58203.35 & g & 20.54 & 0.15 & F & 6.06 & P48+ZTF \\
SN\,2018avk & 58204.39 & g & 20.10 & 0.19 & F & 9.09 & P48+ZTF \\
\enddata

\hspace*{\fill} \\
$^a$F means real detection and T means upper limit.\\
$^b$The ratio of observed flux and flux zero point.\\
$^c$Telescope+Instrument.\\
(This table is available in its entirety in machine-readable form.)
\label{tab:photometry}
\end{deluxetable*}

\startlongtable
\addtocounter{Atable}{1}

\begin{deluxetable*}{lccccccc}
\tabletypesize{\scriptsize}
\tablecaption{LC properties}
\tablehead{\multirow{2}{*}[-4pt]{Name} &
\colhead{K-correction$^a$} &
\colhead{$M_{\rm peak}$} &
\colhead{$L_{\rm peak}$} &
\colhead{$t_{\rm peak}$} &
\colhead{$t_{\rm rise,1/e}$} &
\colhead{$t_{\rm decay,1/e}$} &
\colhead{$t_{\rm rise,10\%}$} \\
&
(mag)  &
(mag) &
($10^{43}\,{\rm erg\, s^{-1}}$) &
(MJD) &
(days) &
(days) &
(days)
} 

\startdata
SN\,2018avk & -0.00 & $-20.27 \pm 0.03$ & $5.5 \pm 1.0$ & $58257.29^{+5.45}_{-4.35}$ & $41.94^{+4.84}_{-3.88}$ & $76.83^{+4.70}_{-5.44}$ & $50.27^{+4.99}_{-3.91}$ \\
SN\,2018don & 0.09 & $-20.05 \pm 0.04$ & $3.4 \pm 0.6$ & $58280.21^{+3.68}_{-3.49}$ & $48.70^{+3.55}_{-3.29}$ & $49.04^{+3.91}_{-4.31}$ & - \\
SN\,2018bym & -0.22 & $-22.03 \pm 0.03$ & $33.2 \pm 5.9$ & $58274.14^{+1.82}_{-1.87}$ & $24.05^{+2.33}_{-1.81}$ & $43.10^{+1.50}_{-1.46}$ & - \\
SN\,2018bgv & -0.09 & $-20.61 \pm 0.10$ & ($14.9 \pm 3.6$)$^b$ & $58253.74^{+1.30}_{-1.76}$ & $8.40^{+1.20}_{-1.63}$ & $19.39^{+1.66}_{-1.22}$ & $9.99^{+1.20}_{-1.63}$ \\
SN\,2018lzv & -0.39 & $-22.23 \pm 0.10$ & $52.6 \pm 16.1$ & $58340.35^{+12.83}_{-9.63}$ & $40.47^{+8.97}_{-6.75}$ & $67.81^{+7.07}_{-9.56}$ & $47.35^{+8.98}_{-6.75}$ \\
SN\,2018gbw & -0.33 & $-22.10 \pm 0.07$ & ($41.1 \pm 7.6$) & $58348.56^{+4.55}_{-3.23}$ & $22.58^{+3.53}_{-2.62}$ & - & $32.78^{+3.60}_{-2.77}$ \\
SN\,2018fcg & -0.13 & $-20.35 \pm 0.06$ & ($6.0 \pm 1.1$) & $58355.83^{+2.93}_{-2.54}$ & $13.33^{+2.67}_{-2.32}$ & $18.00^{+2.33}_{-2.68}$ & $17.75^{+2.76}_{-2.35}$ \\
SN\,2018lzw & -0.30$\star$ & ($-21.68 \pm 0.10$) & ($22.7 \pm 4.6$) & ($58360.51^{+3.57}_{-0}$) & - & ($56.57^{+1.40}_{-2.99}$) & - \\
SN\,2018gft & -0.18 & $-22.22 \pm 0.02$ & $40.1 \pm 7.2$ & $58420.27^{+7.66}_{-5.97}$ & $37.70^{+6.22}_{-4.85}$ & - & $49.95^{+6.26}_{-4.88}$ \\
SN\,2018lzx & -0.44 & $-22.01 \pm 0.04$ & $32.6 \pm 5.9$ & $58437.41^{+11.79}_{-9.64}$ & $60.45^{+8.24}_{-6.75}$ & $108.79^{+10.04}_{-13.22}$ & $76.50^{+8.55}_{-7.05}$ \\
SN\,2018gkz & -0.24 & ($-21.81 \pm 0.03$) & ($21.3 \pm 4.3$) & ($58383.86^{+9.92}_{-9.14}$) & - & ($86.65^{+8.11}_{-8.60}$) & - \\
SN\,2018hpq & -0.13$\star$ & $-20.10 \pm 0.03$ & $4.8 \pm 0.9$ & $58427.85^{+0.93}_{-1.31}$ & $20.37^{+1.88}_{-3.47}$ & - & $31.92^{+1.41}_{-1.99}$ \\
SN\,2018lfe & -0.33$\star$ & $-21.43 \pm 0.09$ & $27.4 \pm 9.1$ & $58465.45^{+3.90}_{-6.95}$ & $17.12^{+3.21}_{-5.34}$ & $26.03^{+6.25}_{-5.42}$ & - \\
SN\,2018hti & -0.09 & $-22.09 \pm 0.07$ & $55.1 \pm 14.9$ & $58460.47^{+8.51}_{-5.88}$ & $28.11^{+8.02}_{-5.55}$ & - & $35.73^{+8.03}_{-5.55}$ \\
SN\,2018lfd & -0.19 & $-21.98 \pm 0.11$ & $38.6 \pm 8.5$ & $58496.49^{+23.40}_{-9.15}$ & $35.45^{+18.61}_{-7.50}$ & - & - \\
SN\,2018kyt & -0.13 & $-20.51 \pm 0.08$ & $9.2 \pm 0.4\star\star$ & $58506.63^{+4.10}_{-6.14}$ & $29.07^{+3.75}_{-5.63}$ & $26.42^{+5.69}_{-4.76}$ & $33.35^{+3.70}_{-5.54}$ \\
SN\,2019J & -0.05 & $-19.86 \pm 0.13$ & $3.7 \pm 0.9$ & $58546.97^{+9.61}_{-18.26}$ & $44.98^{+8.70}_{-16.29}$ & $30.58^{+16.26}_{-8.77}$ & $54.62^{+8.60}_{-16.20}$ \\
SN\,2019cca & -0.37$\star$ & $-21.99 \pm 0.10$ & ($33.3 \pm 7.8$) & $58559.82^{+6.71}_{-10.28}$ & $35.36^{+10.88}_{-9.31}$ & $39.88^{+9.58}_{-5.97}$ & - \\
SN\,2019bgu & -0.14 & $-20.27 \pm 0.15$ & $5.5 \pm 1.4$ & $58559.92^{+5.53}_{-11.12}$ & - & $24.74^{+9.98}_{-5.20}$ & - \\
SN\,2019kwq & -0.43$\star$ & $-21.99 \pm 0.05$ & $32.2 \pm 5.8$ & $58563.33^{+5.87}_{-8.38}$ & $45.37^{+8.32}_{-8.52}$ & $63.96^{+6.57}_{-5.02}$ & $63.10^{+12.81}_{-10.97}$ \\
SN\,2019dgr & -0.37 & $-21.89 \pm 0.10$ & $33.0 \pm 9.8$ & $58581.83^{+5.11}_{-7.84}$ & $25.06^{+3.99}_{-5.84}$ & - & $36.56^{+3.87}_{-6.08}$ \\
SN\,2019kws & -0.31 & $-20.05 \pm 0.04$ & $5.1 \pm 0.9$ & $58573.97^{+4.64}_{-3.92}$ & $22.70^{+3.97}_{-3.40}$ & $53.26^{+3.36}_{-3.99}$ & - \\
SN\,2019cdt & -0.04 & $-20.85 \pm 0.05$ & $7.6 \pm 1.4$ & $58583.63^{+2.01}_{-6.10}$ & $17.59^{+1.89}_{-5.34}$ & $14.24^{+5.29}_{-1.77}$ & $26.66^{+5.70}_{-5.86}$ \\
SN\,2019aamp & -0.40 & $-21.92 \pm 0.04$ & $30.5 \pm 5.7$ & $58588.27^{+6.54}_{-6.84}$ & - & $43.93^{+5.98}_{-5.34}$ & - \\
SN\,2019dlr & -0.24 & $-21.11 \pm 0.07$ & $18.1 \pm 3.7$ & $58593.30^{+6.51}_{-7.17}$ & $28.54^{+5.67}_{-6.04}$ & $42.00^{+9.39}_{-6.08}$ & $38.58^{+8.43}_{-6.61}$ \\
SN\,2019cwu & -0.30 & $-21.27 \pm 0.04$ & $17.8 \pm 3.4$ & $58599.56^{+4.14}_{-3.73}$ & $17.43^{+3.15}_{-2.84}$ & $48.64^{+3.14}_{-3.32}$ & $20.92^{+3.20}_{-2.86}$ \\
SN\,2019kwt & -0.34 & $-22.74 \pm 0.05$ & $65.1 \pm 11.9$ & $58650.73^{+5.03}_{-4.48}$ & $51.21^{+4.08}_{-3.67}$ & $49.09^{+4.94}_{-4.72}$ & $70.50^{+8.91}_{-4.56}$ \\
SN\,2019eot & -0.28 & $-22.13 \pm 0.03$ & $40.1 \pm 7.1$ & $58644.97^{+2.46}_{-2.33}$ & $31.28^{+1.90}_{-1.80}$ & $37.61^{+2.04}_{-2.11}$ & $40.87^{+2.04}_{-1.90}$ \\
SN\,2019kwu & -0.51$\star$ & $-22.32 \pm 0.11$ & $50.8 \pm 9.4$ & $58620.64^{+5.06}_{-3.19}$ & $14.54^{+3.18}_{-2.02}$ & $27.67^{+2.62}_{-3.50}$ & $20.29^{+3.27}_{-2.19}$ \\
SN\,2019gqi & -0.34$\star$ & $-21.43 \pm 0.08$ & $19.3 \pm 4.8$ & $58636.92^{+4.27}_{-5.71}$ & $26.80^{+3.32}_{-4.30}$ & - & $34.99^{+3.61}_{-4.53}$ \\
SN\,2019fiy & -0.69 & $-22.71 \pm 0.03$ & - & $58640.96^{+5.24}_{-5.82}$ & $23.49^{+3.21}_{-3.54}$ & - & $32.24^{+4.75}_{-4.20}$ \\
SN\,2019gam & 0.32 & ($-19.88 \pm 0.12$) & ($2.6 \pm 0.7$) & ($58648.17^{+0}_{-4.25}$) & ($26.13^{+1.91}_{-4.49}$) & - & ($38.28^{+1.07}_{-3.92}$) \\
SN\,2019gfm & -0.02 & $-20.99 \pm 0.05$ & $13.7 \pm 2.7$ & $58645.17^{+3.58}_{-2.24}$ & $13.80^{+3.19}_{-2.03}$ & $26.18^{+2.32}_{-3.23}$ & $20.53^{+3.51}_{-2.68}$ \\
SN\,2019hge & -0.11 & $-19.82 \pm 0.05$ & $3.6 \pm 0.2\star\star$ & $58693.95^{+3.00}_{-7.76}$ & $46.63^{+2.79}_{-7.15}$ & $39.35^{+7.26}_{-2.92}$ & $57.21^{+14.29}_{-7.24}$ \\
SN\,2019hno & -0.21 & $-21.05 \pm 0.04$ & $13.4 \pm 2.5$ & $58663.92^{+7.75}_{-4.02}$ & $22.89^{+6.17}_{-3.24}$ & $29.89^{+3.51}_{-6.25}$ & $29.45^{+6.30}_{-3.30}$ \\
SN\,2019aamq & -0.36 & $-21.76 \pm 0.03$ & $32.9 \pm 11.8$ & $58690.78^{+11.60}_{-10.21}$ & $50.47^{+9.01}_{-7.92}$ & - & - \\
SN\,2019kcy & -0.41 & $-22.12 \pm 0.09$ & $46.6 \pm 11.8$ & $58693.95^{+16.31}_{-11.69}$ & $33.01^{+11.69}_{-8.40}$ & - & $42.58^{+11.82}_{-8.45}$ \\
SN\,2019aamx & -0.33 & $-21.80 \pm 0.07$ & $21.7 \pm 4.8$ & $58709.67^{+9.10}_{-12.60}$ & $42.97^{+6.58}_{-9.03}$ & $45.91^{+10.48}_{-8.31}$ & $50.92^{+6.51}_{-8.98}$ \\
SN\,2019aamr & -0.41 & $-21.36 \pm 0.10$ & $20.5 \pm 4.3$ & $58684.58^{+6.37}_{-3.66}$ & $14.04^{+4.50}_{-2.62}$ & $27.09^{+3.89}_{-5.38}$ & $17.70^{+4.50}_{-2.60}$ \\
SN\,2019lsq & -0.22 & $-20.87 \pm 0.04$ & $12.5 \pm 2.3$ & $58707.77^{+1.48}_{-1.44}$ & $21.29^{+1.36}_{-1.33}$ & $43.02^{+1.50}_{-1.80}$ & $31.31^{+1.65}_{-1.46}$ \\
SN\,2019otl & -0.46 & $-22.41 \pm 0.06$ & ($61.2 \pm 14.4$) & $58715.21^{+12.61}_{-13.82}$ & - & - & - \\
SN\,2019nhs & -0.05 & $-21.44 \pm 0.06$ & $18.6 \pm 0.9\star\star$ & $58732.19^{+3.96}_{-4.79}$ & $22.77^{+3.36}_{-4.06}$ & $27.57^{+4.16}_{-3.42}$ & $30.59^{+3.70}_{-4.18}$ \\
SN\,2019aams & -0.60 & $-22.11 \pm 0.08$ & ($47.2 \pm 13.6$) & $58730.76^{+8.27}_{-4.61}$ & $25.18^{+6.30}_{-3.55}$ & $23.86^{+3.23}_{-5.82}$ & $33.92^{+7.83}_{-5.12}$ \\
SN\,2019neq & -0.24 & $-21.30 \pm 0.03$ & $20.1 \pm 0.4\star\star$ & $58730.16^{+0.80}_{-0.83}$ & $14.69^{+0.73}_{-0.76}$ & $19.97^{+0.88}_{-0.83}$ & $21.97^{+0.74}_{-0.76}$ \\
SN\,2019obk & -0.10 & $-19.80 \pm 0.11$ & $3.8 \pm 1.1$ & $58735.13^{+10.43}_{-10.59}$ & $37.49^{+10.91}_{-10.12}$ & - & $56.10^{+10.52}_{-12.16}$ \\
SN\,2019qgk & -0.32$\star$ & ($-21.55 \pm 0.11$) & ($21.2 \pm 5.5$) & ($58744.36^{+9.70}_{-0}$) & - & ($43.96^{+3.91}_{-9.03}$) & - \\
SN\,2019sgg & -0.42 & ($-22.32 \pm 0.07$) & $56.6 \pm 12.0$ & ($58791.33^{+0}_{-8.53}$) & ($42.77^{+0.95}_{-5.51}$) & - & ($51.83^{+1.60}_{-5.61}$) \\
SN\,2019aamt & -0.24 & $-20.74 \pm 0.05$ & $17.0 \pm 7.1$ & $58773.51^{+6.45}_{-7.02}$ & $24.61^{+5.32}_{-5.79}$ & - & $33.31^{+5.37}_{-5.82}$ \\
SN\,2019sgh & -0.34 & $-21.32 \pm 0.05$ & $19.3 \pm 3.5$ & $58758.54^{+7.89}_{-3.86}$ & $13.96^{+5.89}_{-2.90}$ & $32.03^{+6.33}_{-10.59}$ & $18.16^{+6.34}_{-3.15}$ \\
SN\,2019stc & -0.02 & $-20.55 \pm 0.06$ & $5.8 \pm 1.0$ & $58783.42^{+12.67}_{-5.51}$ & $26.13^{+12.49}_{-5.24}$ & $38.22^{+6.25}_{-11.77}$ & - \\
SN\,2019szu & -0.06 & $-21.19 \pm 0.06$ & $22.3 \pm 5.2$ & $58829.39^{+6.34}_{-11.76}$ & $54.57^{+11.65}_{-11.07}$ & - & - \\
SN\,2019unb & -0.03 & $-20.13 \pm 0.07$ & $4.1 \pm 0.1\star\star$ & $58843.01^{+4.50}_{-4.87}$ & - & $20.80^{+4.64}_{-4.30}$ & - \\
SN\,2019ujb & -0.05 & $-21.13 \pm 0.08$ & $12.2 \pm 2.6$ & $58818.10^{+5.57}_{-7.89}$ & $26.06^{+4.67}_{-6.58}$ & $50.81^{+6.58}_{-4.66}$ & $35.07^{+8.06}_{-6.93}$ \\
SN\,2019xdy & -0.22$\star$ & $-20.72 \pm 0.06$ & $7.9 \pm 4.2$ & $58833.43^{+3.14}_{-5.67}$ & $22.84^{+6.08}_{-8.07}$ & $32.42^{+4.65}_{-2.59}$ & $37.94^{+3.82}_{-6.22}$ \\
SN\,2019vvc & -0.31$\star$ & $-22.06 \pm 0.02$ & $44.7 \pm 18.1$ & $58859.72^{+3.61}_{-3.91}$ & $40.06^{+2.86}_{-3.09}$ & - & $53.55^{+5.18}_{-4.85}$ \\
SN\,2019aamu & -0.25$\star$ & ($-21.64 \pm 0.07$) & - & ($58861.15^{+0}_{-7.55}$) & ($45.20^{+1.81}_{-6.33}$) & - & ($65.18^{+1.02}_{-6.05}$) \\
SN\,2019zbv & -0.39 & $-21.98 \pm 0.04$ & $33.4 \pm 6.0$ & $58866.72^{+3.55}_{-3.40}$ & $26.50^{+2.77}_{-2.60}$ & $54.00^{+3.13}_{-3.16}$ & - \\
SN\,2020ank & -0.22 & $-21.75 \pm 0.06$ & $24.6 \pm 4.5$ & $58893.40^{+2.91}_{-2.06}$ & $15.56^{+2.37}_{-1.69}$ & $32.76^{+2.06}_{-3.09}$ & $22.81^{+3.48}_{-2.41}$ \\
SN\,2020auv & -0.21 & $-21.85 \pm 0.09$ & $31.2 \pm 6.7$ & $58884.04^{+5.22}_{-6.59}$ & - & $26.77^{+5.23}_{-4.16}$ & - \\
SN\,2019aamw & -0.22$\star$ & $-20.49 \pm 0.07$ & $6.4 \pm 1.3$ & $58854.17^{+8.75}_{-11.51}$ & $43.66^{+7.27}_{-9.51}$ & $57.18^{+18.25}_{-11.62}$ & $63.85^{+7.30}_{-9.62}$ \\
SN\,2020fvm & -0.19 & $-21.39 \pm 0.05$ & $15.1 \pm 2.7$ & $58999.69^{+4.88}_{-2.82}$ & $91.12^{+4.02}_{-2.41}$$^c$ & - & - \\
SN\,2019aamv & -0.38 & $-21.37 \pm 0.06$ & $19.6 \pm 3.9$ & $58919.96^{+6.07}_{-3.72}$ & $51.73^{+5.17}_{-3.39}$ & $58.20^{+4.38}_{-5.00}$ & $66.80^{+19.86}_{-6.96}$ \\
SN\,2020aup & -0.28 & $-21.48 \pm 0.12$ & $24.6 \pm 4.8$ & $58882.76^{+5.31}_{-4.11}$ & $14.87^{+4.18}_{-3.27}$ & $28.44^{+4.82}_{-4.58}$ & $20.82^{+4.66}_{-3.70}$ \\
SN\,2020dlb & -0.36$\star$ & $-22.61 \pm 0.07$ & $78.0 \pm 23.9$ & $58948.76^{+3.86}_{-3.88}$ & $30.31^{+2.78}_{-2.80}$ & $32.26^{+3.08}_{-3.00}$ & $41.57^{+3.21}_{-2.93}$ \\
SN\,2020fyq & -0.18$\star$ & $-19.89 \pm 0.06$ & $3.2 \pm 0.6$ & $58965.65^{+4.40}_{-33.19}$ & $70.06^{+4.04}_{-28.25}$ & $62.83^{+28.28}_{-4.70}$ & $86.77^{+4.71}_{-28.37}$ \\
SN\,2020exj & -0.22 & $-20.08 \pm 0.07$ & $5.2 \pm 1.0$ & $58946.37^{+3.63}_{-2.43}$ & $19.70^{+3.51}_{-2.53}$ & $21.99^{+2.18}_{-3.24}$ & $26.57^{+3.44}_{-2.61}$ \\
SN\,2020htd & -0.33$\star$ & $-21.22 \pm 0.05$ & $14.3 \pm 2.6$ & $58987.98^{+4.85}_{-4.94}$ & $50.90^{+4.35}_{-4.93}$ & $47.26^{+5.44}_{-3.99}$ & $65.69^{+4.72}_{-4.30}$ \\
SN\,2020iyj & -0.35 & $-21.73 \pm 0.03$ & $23.3 \pm 4.3$ & $58995.41^{+6.73}_{-7.58}$ & $30.48^{+4.96}_{-5.57}$ & $68.36^{+17.54}_{-27.95}$ & $39.22^{+5.07}_{-5.65}$ \\
SN\,2020kox & -0.36 & $-21.78 \pm 0.11$ & ($44.6 \pm 12.7$) & $58994.23^{+9.02}_{-8.89}$ & $26.52^{+6.30}_{-6.20}$ & - & $34.57^{+6.46}_{-6.33}$ \\
SN\,2020jii & -0.40 & $-21.86 \pm 0.07$ & $44.2 \pm 13.9$ & $59000.44^{+4.02}_{-3.77}$ & $30.36^{+3.27}_{-3.00}$ & $32.84^{+2.83}_{-3.01}$ & $39.44^{+3.27}_{-3.22}$ \\
SN\,2020afah & -0.35$\star$ & $-21.84 \pm 0.05$ & $25.0 \pm 5.3$ & $59055.47^{+19.11}_{-7.13}$ & $61.40^{+13.90}_{-5.20}$ & $54.31^{+6.56}_{-17.50}$ & $69.96^{+13.91}_{-5.21}$ \\
SN\,2020afag & -0.37 & $-21.67 \pm 0.04$ & $23.6 \pm 4.3$ & $59065.68^{+3.48}_{-2.84}$ & $36.29^{+3.23}_{-2.59}$ & $49.05^{+3.39}_{-4.45}$ & $49.71^{+6.04}_{-4.63}$ \\
SN\,2020onb & -0.11 & $-21.07 \pm 0.12$ & $9.8 \pm 1.8$ & $59078.29^{+2.54}_{-5.48}$ & $28.91^{+2.25}_{-4.75}$ & - & $40.04^{+2.32}_{-4.78}$ \\
SN\,2020qef & -0.26 & $-20.03 \pm 0.05$ & $4.3 \pm 0.8$ & $59083.09^{+2.01}_{-2.18}$ & $36.52^{+1.99}_{-2.06}$ & $31.05^{+9.62}_{-2.30}$ & $47.82^{+2.28}_{-2.37}$ \\
SN\,2020rmv & -0.18 & $-21.60 \pm 0.05$ & $22.1 \pm 3.9$ & $59117.41^{+7.57}_{-6.28}$ & $39.81^{+6.23}_{-5.49}$ & - & $54.13^{+7.81}_{-5.25}$ \\
SN\,2020xkv & -0.36 & $-21.02 \pm 0.02$ & $10.7 \pm 2.0$ & $59144.80^{+9.81}_{-8.07}$ & $46.48^{+7.94}_{-6.55}$ & - & $71.62^{+8.08}_{-6.72}$ \\
SN\,2020xgd & -0.45 & $-21.83 \pm 0.09$ & $34.7 \pm 6.7$ & $59149.39^{+4.78}_{-3.70}$ & $23.67^{+3.31}_{-2.58}$ & $21.92^{+4.10}_{-3.77}$ & $29.58^{+3.85}_{-2.78}$ \\
\enddata

\hspace*{\fill} \\
$M_{\rm peak}$, $t_{\rm peak}$, $t_{\rm rise,1/e}$, $t_{\rm decay,1/e}$ and $t_{\rm rise,10\%}$ are measured in the rest-frame $g$-band LCs. \\
$^a$K-correction to rest-frame $g$-band at the peak. It equals $K_{g \rightarrow g}$ for $z \leq 0.17$ and $K_{r \rightarrow g}$ for $z>0.17$. \\
$^b$The value in parentheses means they are not well-constrained due to the lack of data. \\
$^c$SN\,2020fvm has two LC peaks and its $t_{\rm rise,1/e}$ is $47.52^{+6.66}_{-6.88}$\,days if the first (and the fainter) one is set as the main peak.\\
$\star$ means the K-correction is calculated by $-2.5\times {\rm log}(1+z)$.\\
$\star\star$ means the $L_{\rm peak}$ is calculated from multi-band data instead of corrected from $g$- and $r$-band flux.\\

\label{tab:LCproperties}
\end{deluxetable*}

\startlongtable
\addtocounter{Atable}{1}
\begin{deluxetable*}{lcc}
\tabletypesize{\scriptsize}
\tablecaption{($g-r$) colors at peak}
\tablehead{\multirow{2}{*}[-4pt]{Name} &
\colhead{Observed} &
\colhead{Rest frame} \\
&
(mag) &
(mag)
} 

\startdata
SN\,2018avk & $-0.03 \pm 0.03$ & $-0.21 \pm 0.03$ \\
SN\,2018don & $0.15 \pm 0.02$ & $-0.01 \pm 0.02$ \\
SN\,2018bym & $-0.11 \pm 0.01$ & $-0.32 \pm 0.06$ \\
SN\,2018bgv & $-0.23 \pm 0.03$ & $-0.25 \pm 0.03$ \\
SN\,2018lzv & $-0.03 \pm 0.07$ &  - \\
SN\,2018gbw & $-0.12 \pm 0.04$ &  - \\
SN\,2018fcg & $-0.16 \pm 0.04$ & $-0.18 \pm 0.04$ \\
SN\,2018gft & $-0.14 \pm 0.03$ & $-0.21 \pm 0.05$ \\
SN\,2018lzx & $0.02 \pm 0.04$ &  - \\
SN\,2018hpq & $-0.11 \pm 0.02$ &  - \\
\enddata

\hspace*{\fill} \\
(This table is available in its entirety in machine-readable form.)
\label{tab:peakcolor}
\end{deluxetable*}

\startlongtable
\addtocounter{Atable}{1}
\begin{deluxetable*}{lcccc}
\tabletypesize{\scriptsize}
\tablecaption{Temperatures and bolometric luminosities}
\tablehead{\multirow{2}{*}[-4pt]{Name} &
\colhead{Phase} &
\colhead{Temperature} &
\colhead{$L_{\rm bolo}$} &
\colhead{Radius}\\
&
(days) &
($10^4$\,K) &
($10^{43}\,{\rm erg\, s^{-1}}$) &
($10^{10}$\,km)
} 

\startdata
SN\,2018avk & $-4.23$ & $1.05^{+0.15}_{-0.11}$ & $5.14^{+0.58}_{-0.56}$ & $2.46^{+0.51}_{-0.71}$ \\
SN\,2018avk & $16.97$ & $0.74^{+0.05}_{-0.04}$ & $4.24^{+0.43}_{-0.40}$ & $4.39^{+0.54}_{-0.63}$ \\
SN\,2018bym & $1.06$ & $1.21^{+0.07}_{-0.06}$ & $29.67^{+1.13}_{-1.24}$ & $4.43^{+0.42}_{-0.50}$ \\
SN\,2018bym & $9.70$ & $1.14^{+0.13}_{-0.10}$ & $25.89^{+2.09}_{-2.29}$ & $4.66^{+0.81}_{-1.08}$ \\
SN\,2018bym & $33.24$ & $0.86^{+0.03}_{-0.03}$ & $12.23^{+0.58}_{-0.57}$ & $5.64^{+0.44}_{-0.46}$ \\
SN\,2018bgv & $6.26$ & $1.39^{+0.32}_{-0.19}$ & $7.44^{+0.68}_{-0.64}$ & $1.67^{+0.46}_{-0.77}$ \\
SN\,2018bgv & $9.04$ & $1.17^{+0.10}_{-0.09}$ & $5.98^{+0.34}_{-0.34}$ & $2.12^{+0.32}_{-0.37}$ \\
SN\,2018bgv & $11.82$ & $0.94^{+0.12}_{-0.09}$ & $4.47^{+0.50}_{-0.51}$ & $2.83^{+0.55}_{-0.73}$ \\
SN\,2018bgv & $17.38$ & $0.73^{+0.57}_{-0.09}$ & $3.43^{+0.84}_{-0.70}$ & $4.07^{+1.15}_{-6.33}$ \\
SN\,2018bgv & $20.16$ & $0.64^{+0.05}_{-0.04}$ & $2.64^{+0.34}_{-0.31}$ & $4.63^{+0.68}_{-0.82}$ \\
SN\,2018bgv & $22.94$ & $0.61^{+0.18}_{-0.05}$ & $2.29^{+0.42}_{-0.49}$ & $4.87^{+0.85}_{-2.98}$ \\
\enddata

\hspace*{\fill} \\
(This table is available in its entirety in machine-readable form.)
\label{tab:boloLC}
\end{deluxetable*}

\begin{figure*}[htp]
\addtocounter{Afigure}{1}
\includegraphics[width=\textwidth]{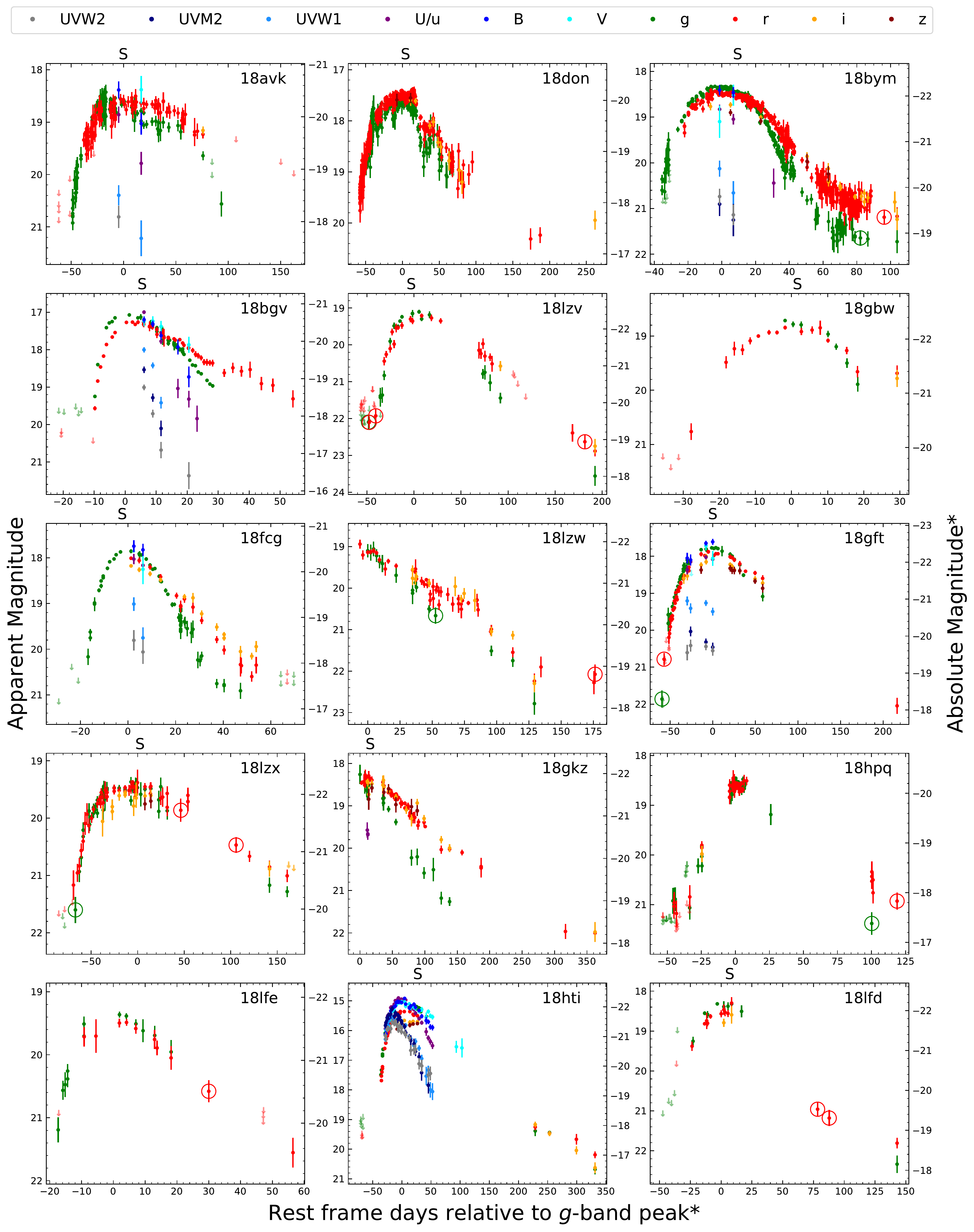}
\caption{The LCs of our sample. All magnitudes are in the AB system and have been corrected for Milky Way extinction. The absolute magnitude y-axis on the right side is calculated by assuming a constant K-correction of $-2.5\times {\rm log}(1+z)$. 
The rings mark the data measured from the combined image of multiple nearby images.
The symbol `S' at the top of each panel shows the epoch of the closest spectrum to the peak, which is used to calculate accurate K-corrections and color corrections. X-axis shows the rest-frame days relative to the observed $g$-band peak. SN\,2018gbw, SN\,2019cca, SN\,2019fiy, SN\,2019xdy, SN\,2019vvc, SN\,2019aamu and SN\,2020ank 
are relative to $r$-band peak.}
\label{fig:lc_0}
\end{figure*}

\begin{figure*}[htp]
\addtocounter{Afigure}{1}
\includegraphics[width=\textwidth]{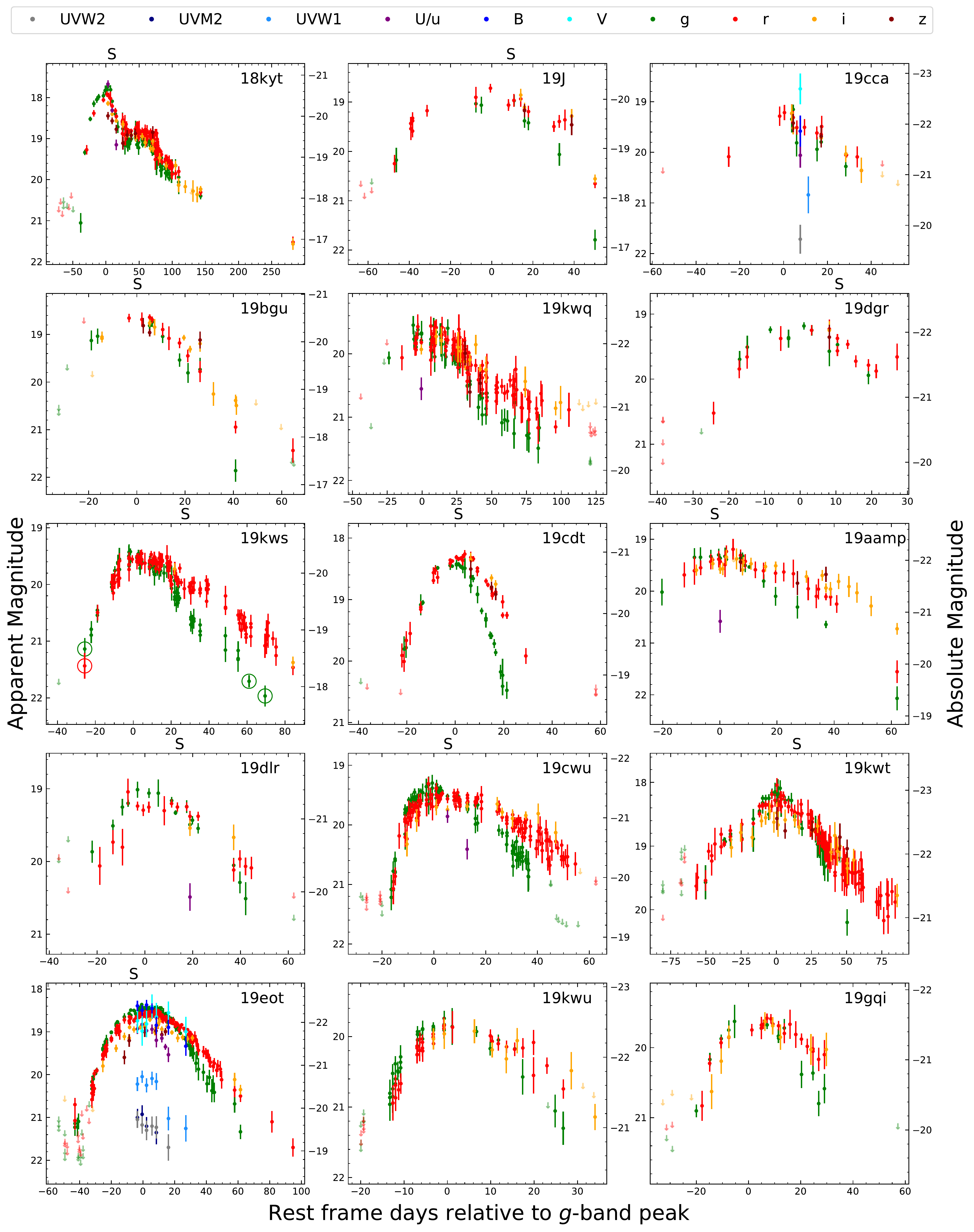}
\caption{Similar to Figure~\ref{fig:lc_0}.}
\label{fig:lc_1}
\end{figure*}

\begin{figure*}[htp]
\addtocounter{Afigure}{1}
\includegraphics[width=\textwidth]{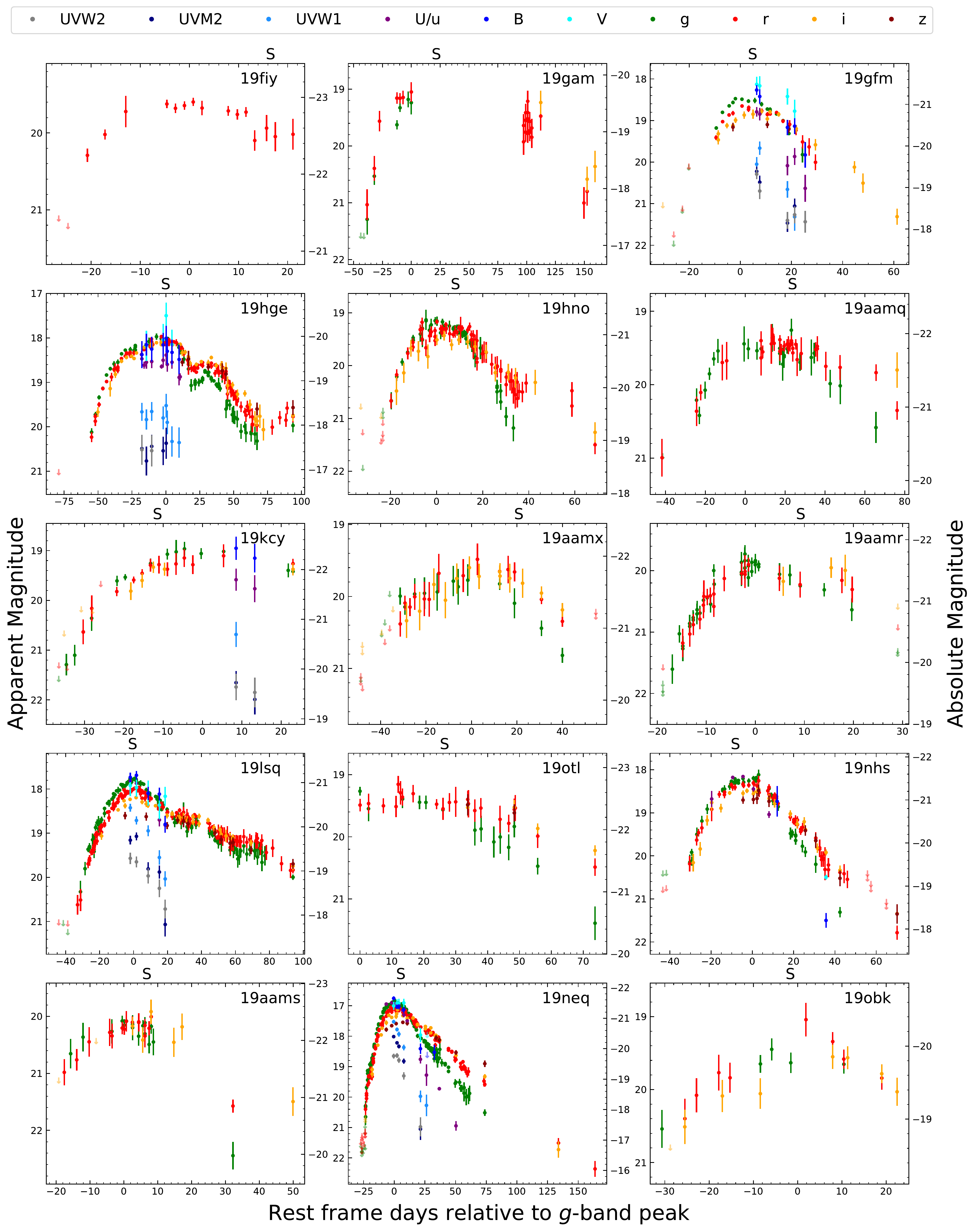}
\caption{Similar to Figure~\ref{fig:lc_0}.}
\label{fig:lc_2}
\end{figure*}

\begin{figure*}[htp]
\addtocounter{Afigure}{1}
\includegraphics[width=\textwidth]{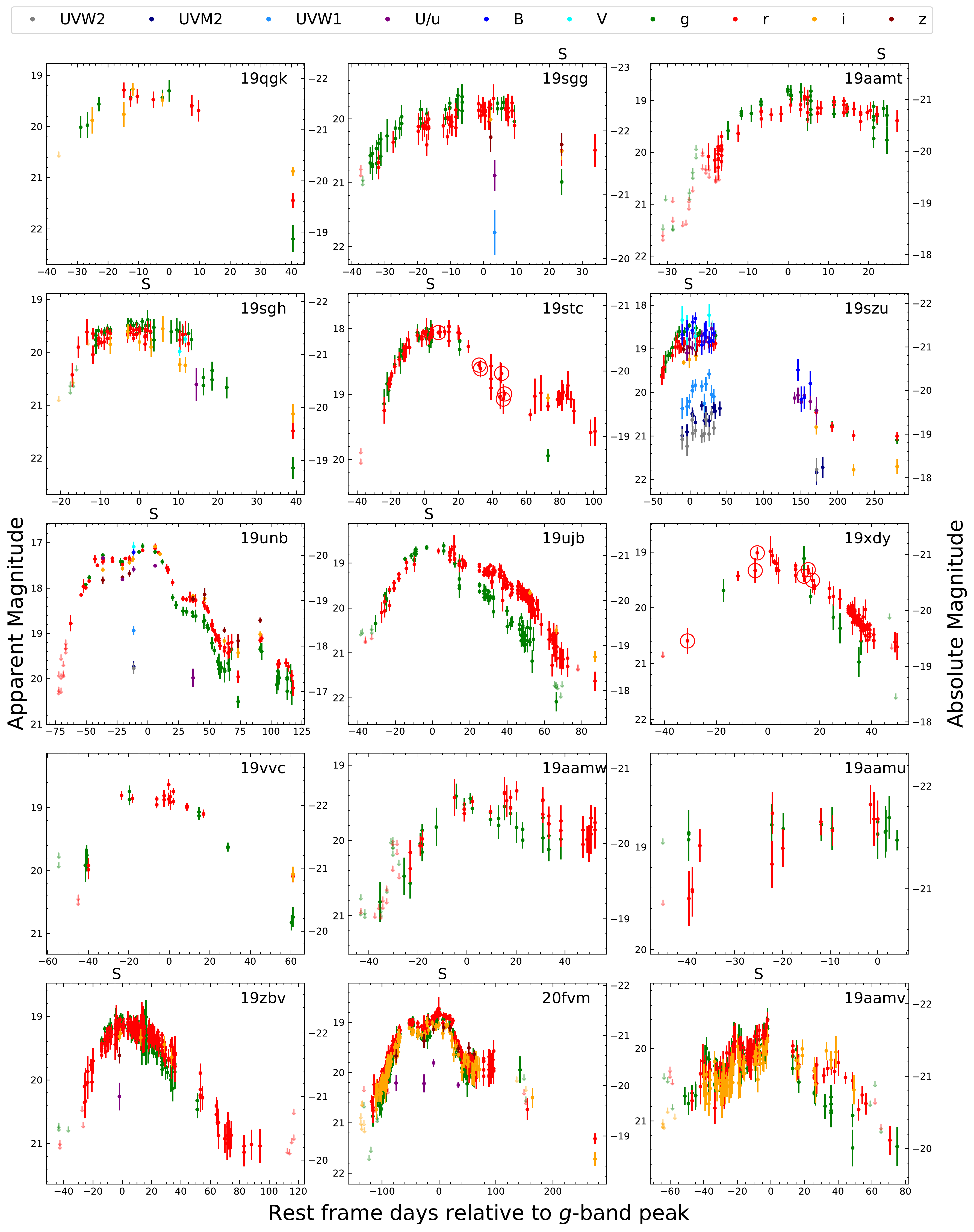}
\caption{Similar to Figure~\ref{fig:lc_0}.}
\label{fig:lc_3}
\end{figure*}

\begin{figure*}[htp]
\addtocounter{Afigure}{1}
\includegraphics[width=\textwidth]{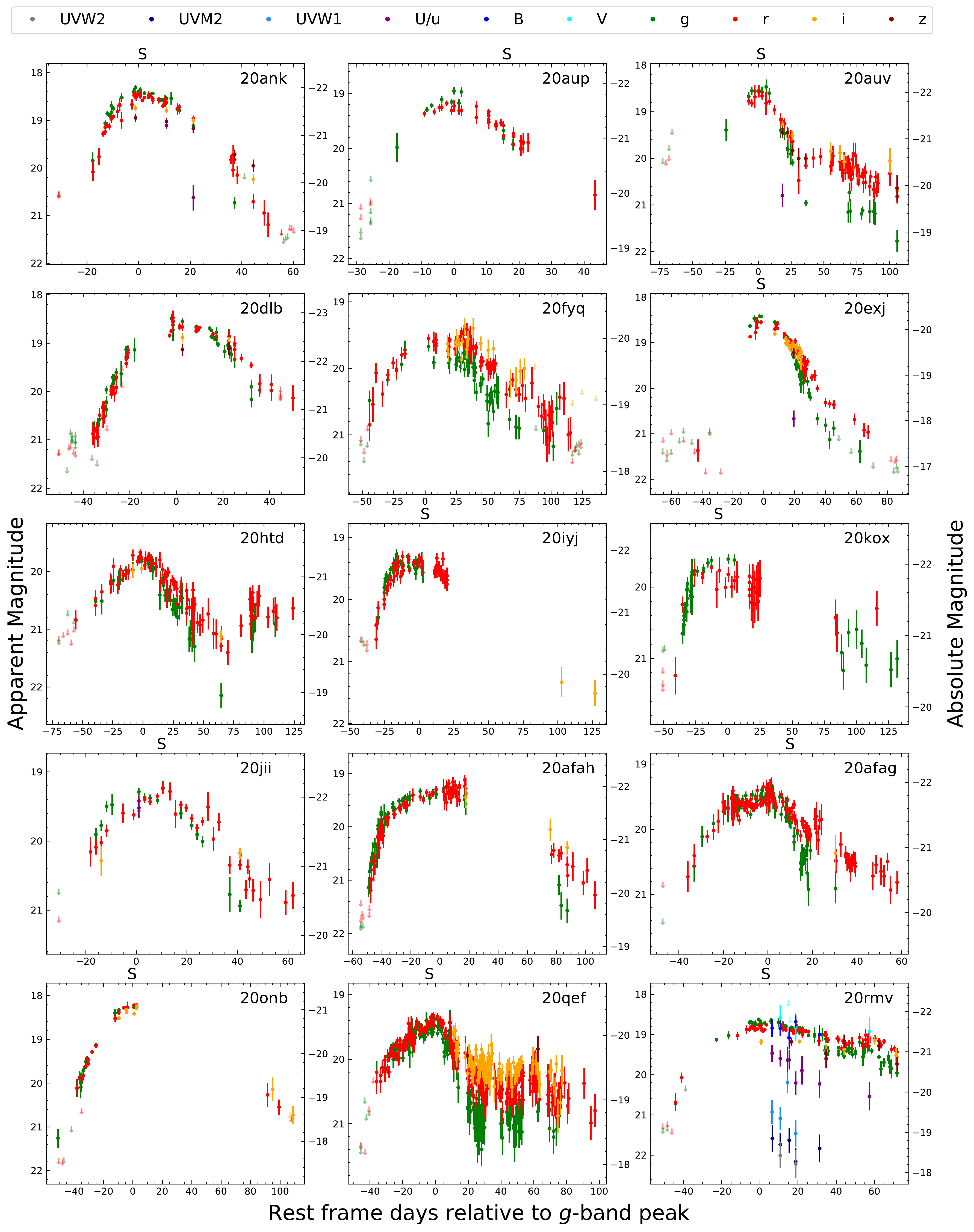}
\caption{Similar to Figure~\ref{fig:lc_0}.}
\label{fig:lc_4}
\end{figure*}

\begin{figure*}[htp]
\addtocounter{Afigure}{1}
\includegraphics[width=\textwidth]{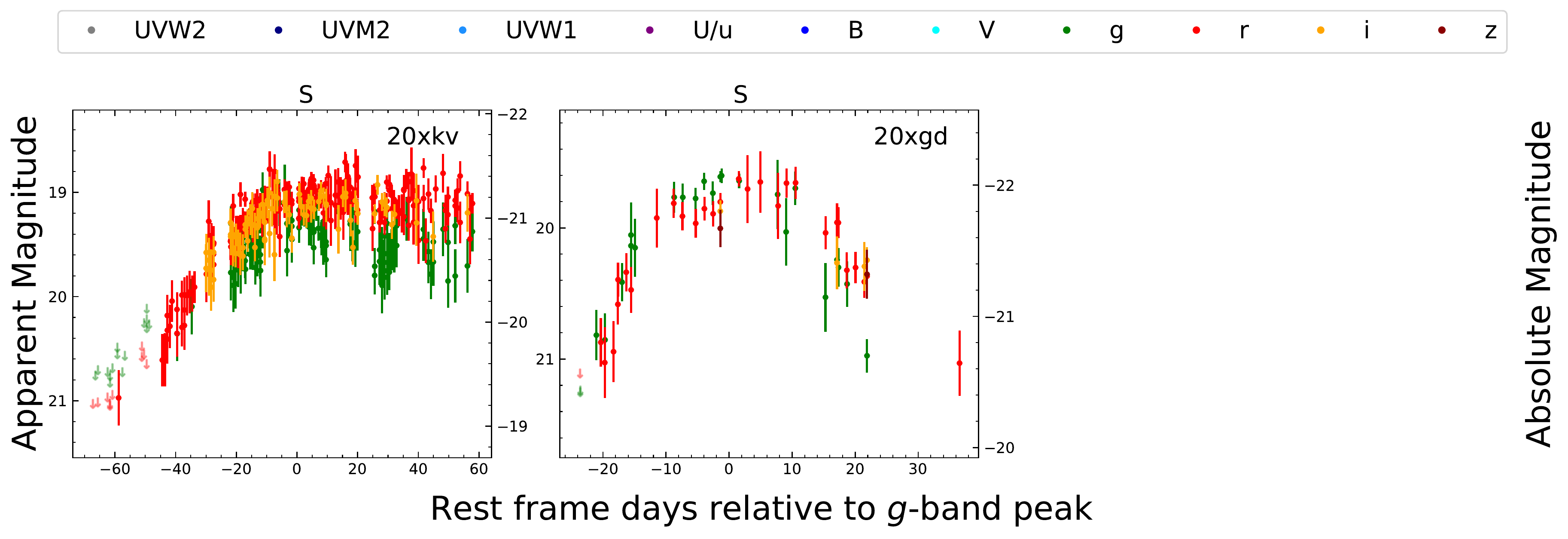}
\caption{Similar to Figure~\ref{fig:lc_0}.}
\label{fig:lc_5}
\end{figure*}

\begin{figure*}[htp]
\addtocounter{Afigure}{1}
\includegraphics[width=\textwidth]{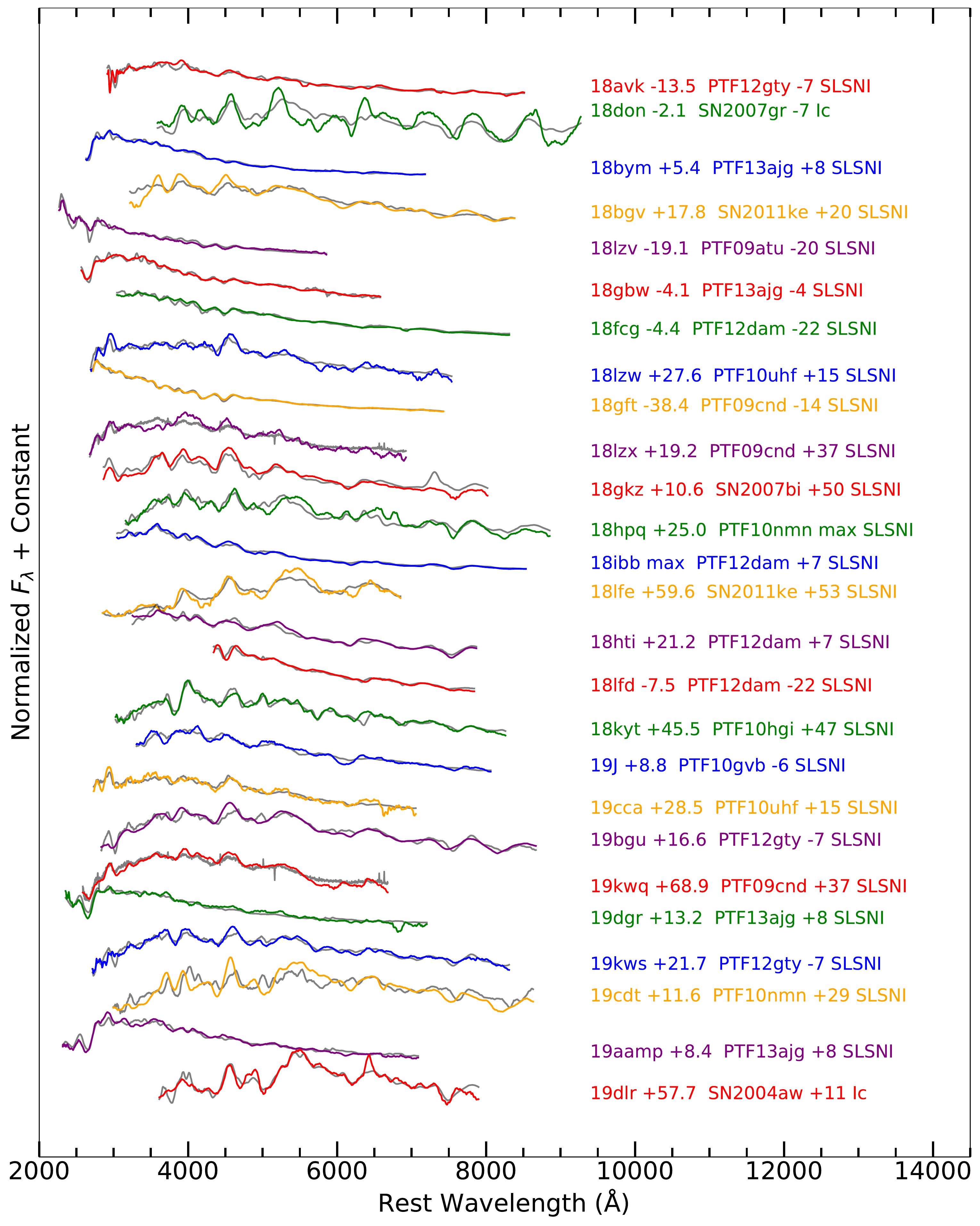}
\caption{Spectral classification. This figure shows the comparison between our spectrum and their best-matched templates (shown in gray) from \textit{superfit}. Event name, the phase of our spectrum (rest-frame days), the best-matched templates, the template phase and the template type are labeled after each spectrum in the same color.}
\label{fig:classification_0}
\end{figure*}

\begin{figure*}[htp]
\addtocounter{Afigure}{1}
\includegraphics[width=\textwidth]{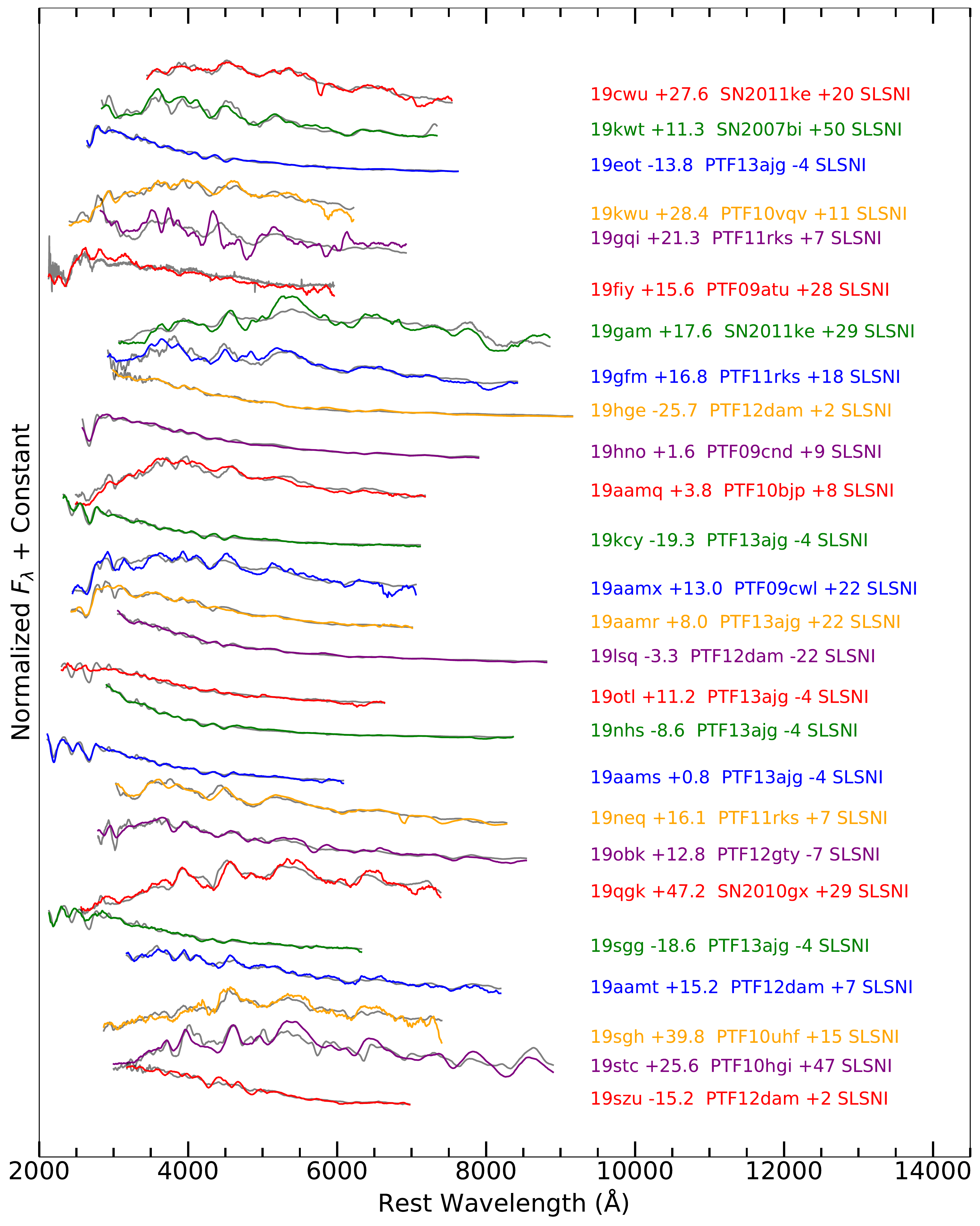}
\caption{Similar to Figure~ \ref{fig:classification_0}}
\label{fig:classification_1}
\end{figure*}

\begin{figure*}[htp]
\addtocounter{Afigure}{1}
\includegraphics[width=\textwidth]{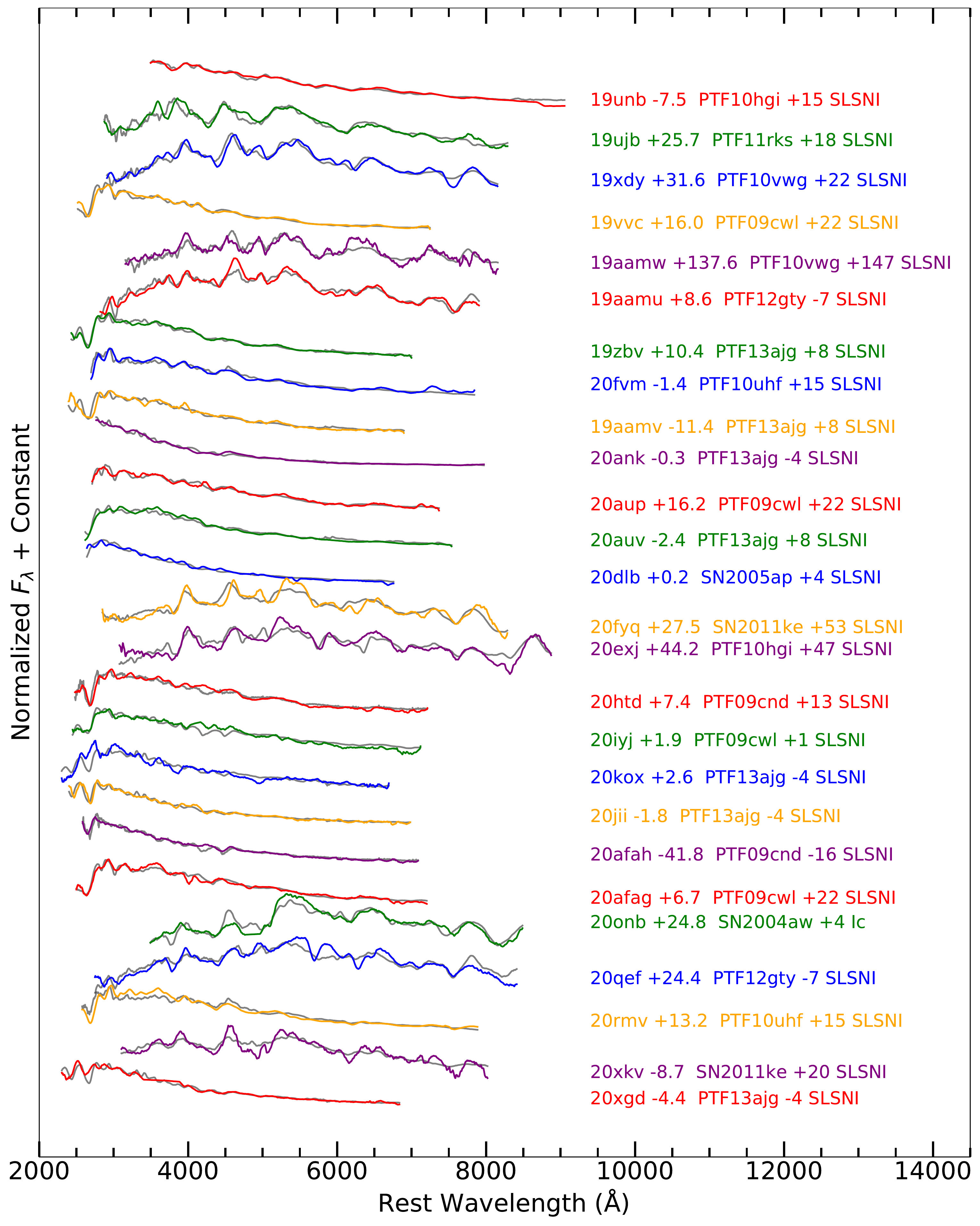}
\caption{Similar to Figure~ \ref{fig:classification_0}}
\label{fig:classification_2}
\end{figure*}

\clearpage
\bibliography{draft.bib}
\bibliographystyle{aasjournal}

\end{document}